\documentclass[prb,twocolumn,showpacs,amsmath,amssymb,superscriptaddress]{revtex4-1}
\usepackage{graphicx}
\usepackage{color}
\usepackage{braket}
\usepackage{amsmath}
\usepackage{amsfonts}

\usepackage[colorlinks = true,linkcolor = blue,urlcolor  = blue,     citecolor = blue,anchorcolor = blue]{hyperref}

\usepackage{amsmath,amssymb,mathrsfs}
\usepackage{bbm}
\usepackage[dvipsnames]{xcolor}

\date{\today}                  

\begin{document}

\title{Universal properties of boundary and interface charges in continuum models of one-dimensional insulators}

\author{Sebastian Miles}
\affiliation{Institut f\"ur Theorie der Statistischen Physik, RWTH Aachen, 
52056 Aachen, Germany and JARA - Fundamentals of Future Information Technology}

\author{Dante M. Kennes}
\affiliation{Institut f\"ur Theorie der Statistischen Physik, RWTH Aachen, 
52056 Aachen, Germany and JARA - Fundamentals of Future Information Technology}
\affiliation{Max Planck Institute for the Structure and Dynamics of Matter, Center for Free Electron Laser Science, 22761 Hamburg, Germany}

\author{Herbert Schoeller}
\affiliation{Institut f\"ur Theorie der Statistischen Physik, RWTH Aachen, 
52056 Aachen, Germany and JARA - Fundamentals of Future Information Technology}

\author{Mikhail Pletyukhov}
\email[Email: ]{pletmikh@physik.rwth-aachen.de}
\affiliation{Institut f\"ur Theorie der Statistischen Physik, RWTH Aachen, 
52056 Aachen, Germany and JARA - Fundamentals of Future Information Technology}

\begin{abstract}
We study single-channel continuum models of one-dimensional insulators induced by periodic potential modulations which are either terminated by a hard wall (the boundary model) or feature a single region of dislocations and/or impurity potentials breaking translational invariance (the interface model). We investigate the universal properties of excess charges accumulated near the boundary and the interface, respectively. We find a rigorous analytic proof for the earlier observed linear dependence of the boundary charge on the phase of the periodic potential modulation as well as extend these results to the interface model. 
The linear dependence on the phase shows a universal value for the slope, and is intersected by discontinuous jumps by plus or minus one electron charge at the phase points where localized states enter or leave a band of extended states. Both contributions add up such that the periodicity of the excess charge in the phase over a $2\pi$-cycle is maintained. While in the boundary model this property is usually associated with the bulk-boundary correspondence, in the interface model a correspondence of scattering state and localized state contributions to the total interface charge is unveiled on the basis of the so-called nearsightedness principle.

\end{abstract}

\maketitle

\section{Introduction}

The notion of topology has provided a powerful, novel viewpoint on a wide variety of phenomena in physics [\onlinecite{JackiwRebbi}-\onlinecite{hatsugai_1993}], successfully solidifying and expanding our understanding of condensed matter systems [\onlinecite{hasan_kane_2010}-\onlinecite{asboth_etal_16}]. 
Two fields that, e.g., have greatly profited from the paradigm of topology are the Quantum Hall effect (QHE) [\onlinecite{LaughlinQHE}] and the microscopic theory of polarization, now known as \textit{modern theory of polarization} (MTP) [\onlinecite{king_smith_vanderbilt_93}]. 

The field of the QHE significantly profited from the insight that the integer $n$ characterizing the Quantum Hall conductance $\sigma_{xy}=ne^2/h$ is linked to the topological invariant characterizing bands, now known as the TKNN invariant [\onlinecite{thouless_etal_prl_82}] and mathematically identified [\onlinecite{thouless_1983}] with the first Chern index $C_1$. The importance of phase-singular points of the Bloch states $|\psi_{k\alpha} \rangle$ (with the quasimomentum $k$ in the occupied bands $\alpha$) for both the conductance quantization and for the above mentioned identification of $n=C_1$ was quickly recognized  [\onlinecite{kohmoto_1985}]. This understanding has subsequently led Y. Hatsugai to the formulation [\onlinecite{hatsugai_1993}] of the seminal \textit{bulk-boundary correspondence}, which links the first Chern index (and thereby the topological characterization of the Bloch states in the Brillouin zone in terms of the Bloch wavefunction's phase vorticities) to the number of edge modes contained within the band gaps of the insulating system.

On the other hand, the MTP relies heavily on concepts of topology since it establishes a connection between localized charges $Q_B$ at boundaries and the Zak-Berry phase $\frac{\gamma_{\alpha}}{2 \pi}=i\int_{BZ} \frac{d k}{2 \pi} \langle \psi_{k\alpha} | \frac{d}{dk} \psi_{k\alpha} \rangle dk$ [\onlinecite{zak_89}] of the bulk Bloch bands even beyond any symmetry constraints [\onlinecite{king_smith_vanderbilt_93}]. This is the so-called \emph{surface charge theorem} [\onlinecite{king_smith_vanderbilt_surf_93},\onlinecite{marzari_etal_12}]. However, this connection is in general not unique but only determined up to an unknown integer [\onlinecite{vanderbilt_18}], since the MTP does not give a prescription which specific gauge of $| \psi_{k\alpha} \rangle$ should be used: 
Any nontrivial gauge transformation with a nonzero winding number changes this relation by an integer. 

Recently, systems have been proposed that capture both, the QHE and aspects of the MTP, simultaneously establishing an analogy between the two [\onlinecite{FracInDots},\onlinecite{thakurati_etal_18}]. Following the key idea of the mapping between a two-dimensional QHE model, which is translationally invariant in one dimension, and a one-dimensional (1D) model of an insulator with a modulated periodic potential, 
establishes the correspondence between the Hall conductance and the derivative of the boundary charge $Q_B$ with respect to the phase of the modulations of the 1D periodic potential. Studying a lattice model of a 1D insulator with quite a large number ($Z=10$) of sites per unit cell, 
reveals a linear dependence of $Q_B$ on the phase, which is accompanied by discontinuous jumps by plus/minus one electron charge at the phase points where an edge state enters/leaves the band. Thereby, 
$Q_B$ was elevated to be a relevant physical observable quantifying the spectral flow of the boundary eigenvalue problem. The slope of the linear dependence 
is universally given by the number of the occupied bands (in units of $2\pi/e$), suggesting that it should be identified not only with $n$, but with $C_1$ as well.

A wider class of generalized Aubry-Andr{\'e}-Harper models  [\onlinecite{aa_model},\onlinecite{harper_model}] has been later studied analytically [\onlinecite{pletyukhov_etal_prbr_20},\onlinecite{pletyukhov_etal_prb_20}] in the semi-infinite geometry with a hard-wall boundary condition at the origin.
The main conclusions of [\onlinecite{thakurati_etal_18}] have been confirmed, also proving that the slope of the linear dependence is indeed determined by the Chern index. In addition, it was shown that on top of the linear dependence for every finite $Z$ there is an additional non-universal $\frac{2\pi}{Z}$-periodic function, contributing to $Q_B$, which is expected to vanish at larger $Z$. Moreover, a distinguished gauge has been identified, in which the relation between the Zak-Berry phase and $Q_B$ holds exactly.
This gauge is fixed by choosing the last component of the Bloch vector to be real. As a consequence, the winding and the phase-singular points of the first Bloch vector component were shown to determine the universal values of the change $\Delta Q_B$ under a lattice shift by one site and for the jumps in $Q_B$, respectively.

In addition, in Ref.~[\onlinecite{weber_etal_prl_20}] quantum fluctuation of the boundary charge have been studied, and the so-called \textit{surface fluctuation theorem} has been formulated, relating the boundary charge fluctuations generically to the fluctuations of the bulk polarization. Furthermore, for one-dimensional systems, a universal $1/E_g$ low-energy scaling was established, with $E_g$ denoting the size of the gap in which the chemical potential lies. Other universal low-energy properties of the Wannier functions in these models have been recently addressed in Ref.~[\onlinecite{wannier_paper}].

An interface model combining the two semi-infinite lattice models with a mutual phase mismatch has been studied in Ref.~[\onlinecite{RatBCPlet}]. It has been shown that the interface charge $Q_I$ is given --- up to an integer --- by the dipole moment mismatch of the two subsystems. Furthermore, it was shown how to determine the integer for a specific type of the interface (realized by a hopping attenuation between the two subsystems at no phase mismatch) expressing it as a winding number of some complex function defined in the Brillouin zone. A principle stating that a local perturbation to a many-electron system may only lead to an integer-valued change in accumulated charges is called the \textit{nearsightedness principle} [\onlinecite{NearSightednessPrin}]. On its basis we expect an ubiquitous occurrence of winding number expressions for this kind of changes: A charge robustness with respect to numerous system parameters and its sharp quantization can be only guaranteed by a topological expression.

In this paper, we study single-channel continuum models of one-dimensional insulators. Mathematically they correspond to the limit $Z \to \infty$ (number of sites per unit cell), $a \to 0$ (lattice spacing) at the fixed $L=Za$ (unit cell length) of the previously studied lattice models. We investigate properties of both boundary and interface localized charges and reveal their universal features. Most of them are expected on the basis of our previous studies [\onlinecite{pletyukhov_etal_prbr_20}-\onlinecite{RatBCPlet}]: A linear phase dependence of $Q_B$ and $Q_I$, discontinuous jumps at the points where a localized state enters/leaves a band of extended states, both the surface charge and surface fluctuation theorems and the bulk-boundary correspondence. However, there are  additional features which are specific only to continuum models. In particular, the absence of the umklapp processes leads to the rigid values of the Chern indices ($=1$ for each band), as well as to additional constraints on numbers of edge states leaving and entering a specific band and on Bloch wavefunction's vorticity values. We find that these constraints stem from the properties of the Hill's equation which were thoroughly studied in the mathematical literature (see, e.g., Refs.~[\onlinecite{arscott}-\onlinecite{MagnusWinkler}]). As an additional motivation to study continuum models we literally quote the citation from Ref.~[\onlinecite{wang_troyer}] which we share without reserve: "There are several advantages to working with a continuum model instead of a tight-binding lattice model... Continuum models apply to a broader range of experimental situations...".

As for the continuum interface models, we establish that the quantized part of $Q_I$ is intimately related to analytic properties of the reflection coefficient.
This observation accords with the recent results of Ref.~[\onlinecite{nakata}].
In particular, extracting a function $\tilde{d}_{k \alpha}$, which is defined in the Brillouin zone for every band $\alpha$, from a scattering matrix denominator, we demonstrate that $\tilde{d}_{k \alpha}$ plays a role similar to that of the Bloch state phase in the boundary problem.

In sum, the present paper consolidates the findings of our previous papers [\onlinecite{pletyukhov_etal_prbr_20}-\onlinecite{RatBCPlet}] on the subject and firmly establishes the universal properties of both the boundary and the interface charges drawing a unified analogy between the two quantities.

The paper is organised as follows. In Sec. \ref{sec:kohn_method} a general construction of the Bloch states based on the approach of [\onlinecite{kohn_59}] for a generic but periodic 1D potential is discussed.  In Sec. \ref{sec:Boundary_Charge} we first determine eigenstates, both extended and localized, of the boundary problem with a hard-wall boundary introduced into the origin of the translationally invariant system. Next, we define the boundary charge which contains contributions from all eigenstates below the chemical potential put into the gap $\nu$. We rigorously prove the surface charge theorem, emphasizing the role of the gauge choosing the last  component of the Bloch state to be real. Then we establish the phase dependence of edge state dispersions, recover the bulk-boundary correspondence, and state the universal form for both the total boundary charge and the  band's individual contributions to it. In Sec. \ref{sec:Interface_Charge} we formulate the interface eigenvalue problem and construct  both its scattering and localized eigenstates. Next, we evaluate the total interface charge discussing individual contributions to it from  both of these types of eigenstates. This section is concluded with a novel expression for the phase dependence of the interface charge. All revealed properties are listed in the Summary Section \ref{sec:Summary}. The appendices provide detailed mathematical proofs of all intermediate statements and identities used in our derivations.

\begin{figure}[b]
    \centering
    \includegraphics[width=0.95\columnwidth]{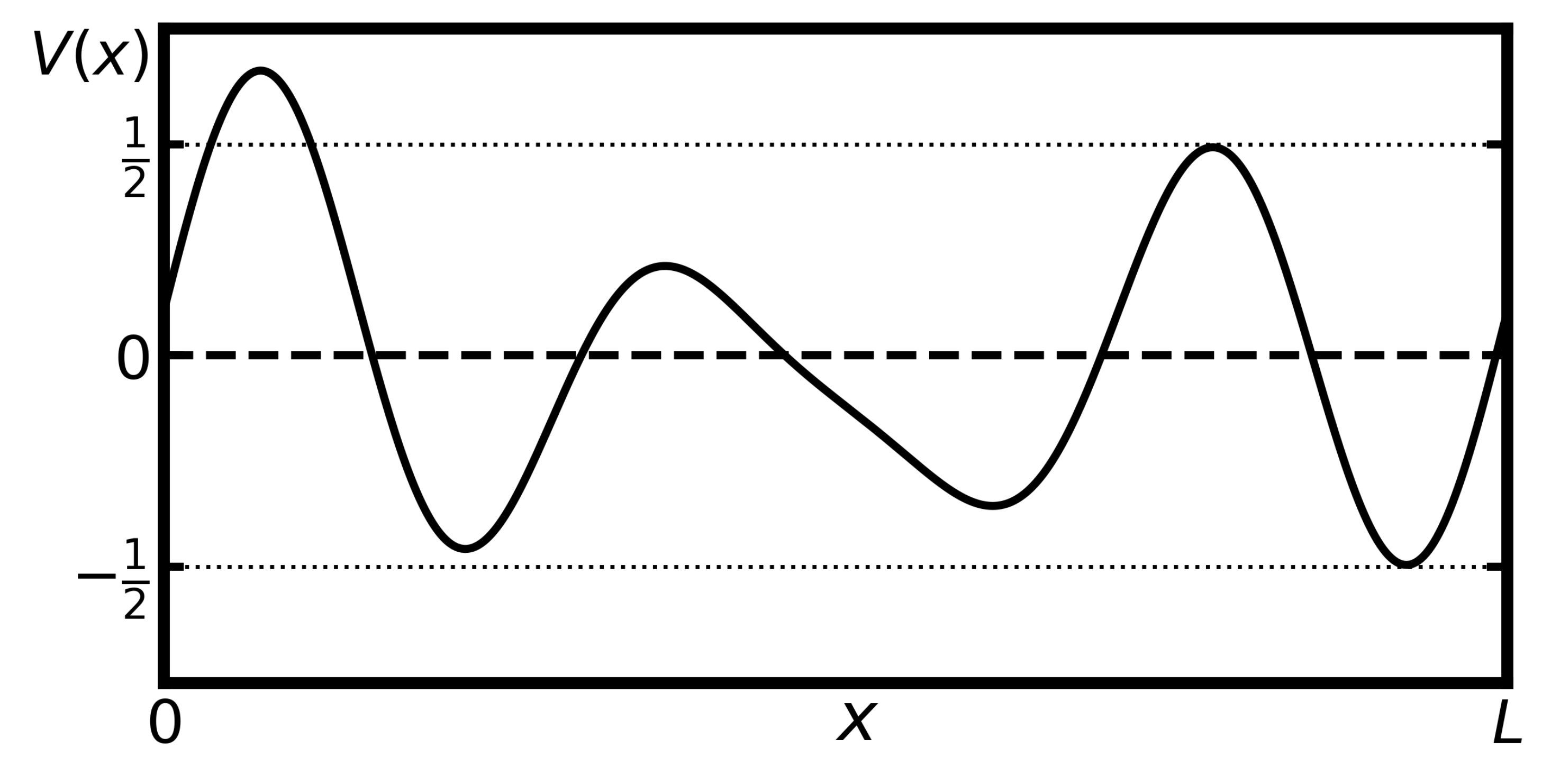}
    \caption{Potential landscape $V(x)=a_0 \cos \frac{2 \pi \omega_0}{L} x +a_1 \sin \frac{2 \pi \omega_1}{L} x +a_2 \sin \frac{2 \pi \omega_2}{L} x$ used in exemplary calculations. The parameters are $(a_0,a_1,a_2)=(0.1,0.4,0.2)$ and $(\omega_0,\omega_1,\omega_2)=(1,3,4)$.}
    \label{fig:BarePot}
\end{figure}

\begin{figure*}[t]
\centering
\includegraphics[width= 2\columnwidth]{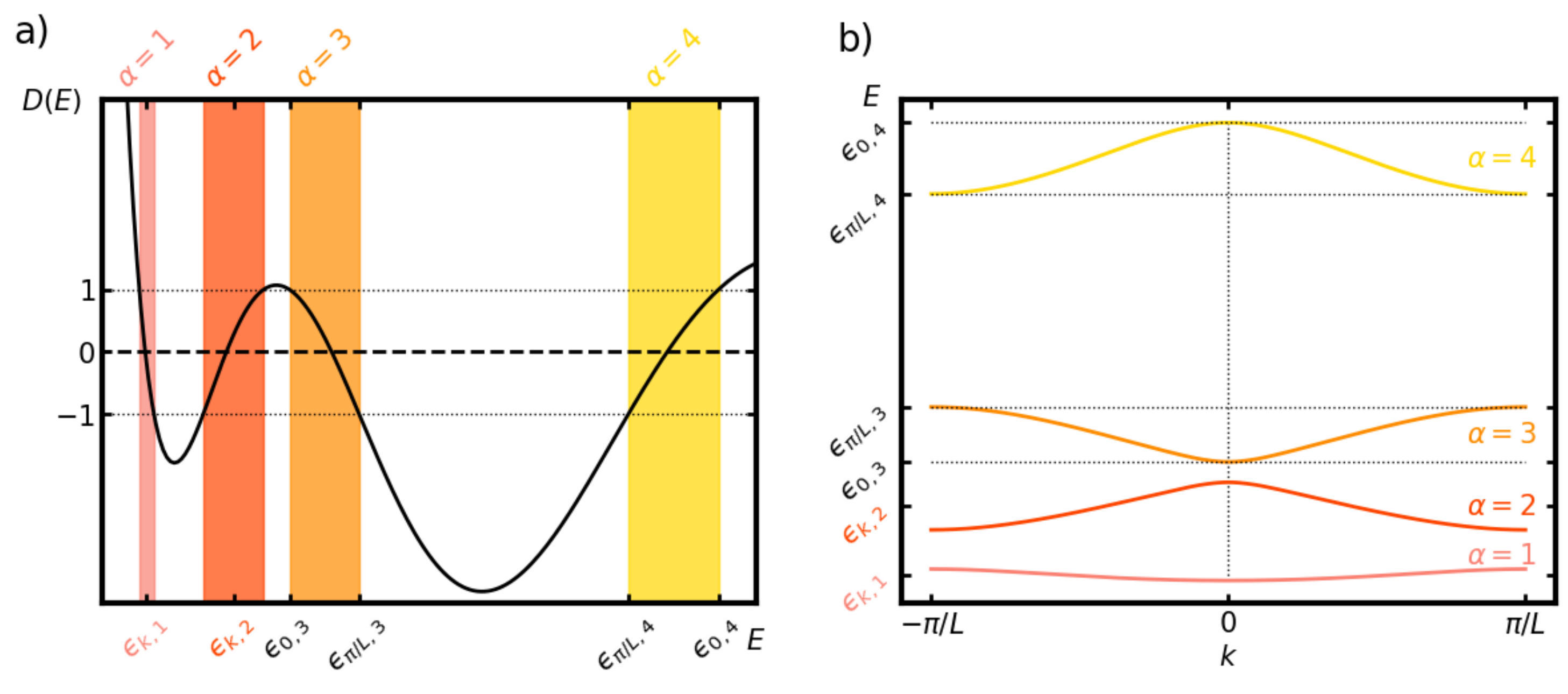} 
 \caption{a) The Lyapunov function $D(E)$ [Eq.~\eqref{dr}] for the potential choice depicted in Fig.~\ref{fig:BarePot}. In the calculation we set the mass $m=1$ and the unit cell length $L=10$. The bands given by the condition $|D(E)|\leq 1$ are indicated as the colored regions. Typically\cite{matveev} there are infinitely many bands (and band gaps), only the first four of them being shown in the figure.
 b) The band structure for the same model in dependence of $k$. 
 }
\label{fig:band}
\end{figure*}

\section{Bloch states construction for a periodic 1D potential}
\label{sec:kohn_method}

In this section we describe 
a construction of the Bloch states for a periodic one-dimensional  
potential following the ideas of  W.~Kohn [\onlinecite{kohn_59}]. In particular, we review properties of these eigenstates and thereby fix notations for
subsequent discussions. Throughout the paper we use the units $e= \hbar =1$.

The 1D Schr{\"o}dinger equation
\begin{align}
- \frac{1}{2 m} \psi'' (x) + V (x) \, \psi (x) = E \, \psi (x)
\label{schr_eq}
\end{align}
with a periodic potential $V(x) = V (x+L)$ is a particular form of the Hill's equation whose properties are comprehensively described in the mathematical literature (see, e.g., [\onlinecite{arscott}-\onlinecite{MagnusWinkler}]). For the following illustrations we arbitrarily choose a specific potential landscape shown in Fig.~\ref{fig:BarePot}, which does not have any internal symmetries. It will be used in all exemplary calculations of this paper.

At fixed $E$, a general solution of \eqref{schr_eq} can be expressed as combination of two linearly independent solutions, $\psi_1 (x,E) \equiv \psi_1 (x)$ and $\psi_2 (x,E) \equiv \psi_2 (x)$, that is
\begin{align}
\psi (x) = c_1 \, \psi_1 (x) + c_2 \, \psi_2 (x).
\label{lin_comb}
\end{align}

Choosing $\psi_1 (x)$ and $\psi_2 (x)$ to satisfy the initial conditions $\psi_1 (0)=1$, $\psi'_1 (0)=0$ and $\psi_2 (0)=0$, $\psi'_2 (0)=1$, respectively, we fix the fundamental matrix of the equation
\begin{align}
    \left( \begin{array}{cc} \psi_1 (x) & \psi_2 (x) \\ \psi'_1 (x) & \psi'_2 (x) \end{array} \right).
    \label{fund_sys}
\end{align}
Its determinant, the Wronskian,
\begin{align}
W (x) =\psi_1 (x) \, \psi'_2 (x) -\psi_2 (x) \, \psi'_1 (x) \equiv 1 \neq 0
\label{wron}
\end{align}
is a nonzero constant: It follows that $W' (x) =0$, and the specific value 1 results from the chosen initial conditions. This property guarantees linear independence of the vectors $\begin{pmatrix} \psi_1 (x) \\ \psi'_1 (x) \end{pmatrix} $ and $\begin{pmatrix} \psi_2 (x) \\ \psi'_2 (x) \end{pmatrix} $ at all $x$. Note that all elements of \eqref{fund_sys} are real.

According to the Bloch theorem, a smooth and bounded solution of \eqref{schr_eq} can be represented in the form
\begin{align}
\psi_k (x) = u_k (x) \, e^{i k x},
\label{eig_fun}
\end{align}
where $u_k (x)$ is a smooth and periodic function, $u_k (x) = u_k (x+L)$, and $k \in [ -\frac{\pi}{L}, \frac{\pi}{L})$. Thus, we obtain the boundary conditions for $\psi_k (x)$
\begin{align}
\psi_k (0) &= \psi_k (L) \, e^{-i k L} , \label{bc1} \\
\psi'_k (0) &= \psi'_k (L) \, e^{-i k L} . \label{bc2}
\end{align}
On the other hand, the representation \eqref{lin_comb} is still valid, implying a $k$ dependence of the coefficients,
\begin{align}
\psi_k (x) = c_{1,k} \, \psi_1 (x) + c_{2,k} \, \psi_2 (x).
\label{eig_fun1}
\end{align}
From the boundary conditions \eqref{bc1}, \eqref{bc2} we find
\begin{align}
0 &=  c_{1,k} \, [\psi_1 (L) - e^{i k L}] + c_{2,k} \, \psi_2 (L)  , \label{bc1a} \\
0 &=  c_{1,k} \, \psi'_1 (L) + c_{2,k} \, [\psi'_2 (L) -  e^{i k L}]. \label{bc2a}
\end{align}

To find nonzero $c_{1,k}$ and $c_{2,k}$, we claim
\begin{align}
\, [\psi_1 (L) - e^{i k L}] \, [\psi'_2 (L) -  e^{i k L}] -  \psi_2 (L) \, \psi'_1 (L) =0.
\end{align}
Taking into account \eqref{wron}, we establish the dispersion relation
\begin{align}
\cos k L = \frac12 \, [\psi_1 (L) + \psi'_2 (L)] \equiv D (E),
\label{dr}
\end{align}
where $D(E)$ is the so called Lyapunov function. Solving \eqref{dr}, one traditionally obtains the bands $\epsilon_{k\alpha}$ labelled by the band index $\alpha$ and possessing the property $\epsilon_{k\alpha}=\epsilon_{-k,\alpha}$ (see Fig.~\ref{fig:band}). As $E \to - \infty$, we can neglect $V (x)$ in \eqref{schr_eq} and thereby establish that asymptotically $ D(E) \approx \cosh (L \sqrt{2 m |E|}) \to + \infty$. Since $D (E)$ is a continuous function, this means that the bottom energy of the lowest band ($\alpha=1$) satisfies the equation $D(E)=+1$. In addition, within the band $\alpha$, the function $D (E)$ is monotonous [which follows, e.g., from the Schwarz inequality \eqref{schwarz1}], and
\begin{align}
     \text{sign} \left[\frac{\partial D (E)}{\partial E} \right] = (-1)^{\alpha}.
     \label{sign_dD}
\end{align}

Then, from \eqref{bc1a} we find
\begin{align}
\frac{c_{2,k}}{c_{1,k}}= - \frac{\psi_1 (L)-  e^{i k L}}{\psi_2 (L) },
\end{align}
and inserting this into \eqref{eig_fun1} we obtain
\begin{align}
& \psi_k (x) = c_{1,k} \left[ \psi_1 (x) -   \frac{\psi_1 (L)-  e^{i k L}} {\psi_2 (L) } \, \psi_2 (x) \right] \label{eig_fun2} \\
&= \frac{1}{\sqrt{N_k}} \left[ \psi_1 (x) \, \psi_2 (L) - \psi_1 (L) \, \psi_2 (x)  + \psi_2 (x) \, e^{i k L}\right] .
\label{eig_fun3}
\end{align}
Hereby we introduced the normalization $N_k$ by virtue of the relation
\begin{align}
     \int_0^L  dx \,\, |\psi_k (x)|^2 =1 .
     \label{normal}
\end{align}

Noticing that (see Appendix \ref{app:hill_eq} for details)
\begin{align}
  - \psi_2 (x-L) =  \psi_1 (x) \, \psi_2 (L) - \psi_1 (L) \, \psi_2 (x),
 \label{psi2L}
\end{align}
and explicitly writing the band index $\alpha$, we get the Bloch states in the form
\begin{align}
\psi_{k\alpha} (x) &= \frac{1}{\sqrt{N_{k\alpha}}} \left[ - \psi_2 (x-L)  + \psi_2 (x) \, e^{i k L}\right] ,
\label{eig_fun4}
\end{align}
along with
\begin{align}
u_{k\alpha} (x) &= \frac{e^{-i k x}}{\sqrt{N_{k\alpha}}} \left[ - \psi_2 (x-L)  + \psi_2 (x) \, e^{i k L}\right].
\label{u_fun}
\end{align}
Here $\psi_2 (x) = \psi_2 (x, \epsilon_{k \alpha})$ depends on $k$ and $\alpha$ via $ \epsilon_{k \alpha}$, but in the following we omit for brevity an explicit indication of this dependence.

Note that $\psi_{k\alpha} (x)= \psi_{k+ \frac{2 \pi}{L},\alpha} (x)$ and $u_{k \alpha} (x) = u_{k \alpha} (x+L)$, while the reciprocal relations are not valid, i.e. $u_{k\alpha} (x)\neq u_{k+ \frac{2 \pi}{L},\alpha} (x)$, $\psi_{k\alpha} (x) \neq \psi_{k\alpha} (x+L)$.

It is remarkable that
\begin{align}
    \psi_{k\alpha} (0) = u_{k\alpha} (0) = u_{k\alpha} (L) = \frac{\psi_2 (L)}{\sqrt{N_{k\alpha}}}   
    \label{gauge_real}
\end{align}
is real. This property fixes the gauge of the Bloch states which will prove to be advantageous for the calculation of the boundary charge in Sec. \ref{sec:Boundary_Charge}. 

In general, we also have the relations
\begin{align}
    \psi^*_{k\alpha} (x) &=  \psi_{-k,\alpha} (x), \\
    u^*_{k\alpha} (x) &=  u_{-k,\alpha} (x),
\end{align}
which can be realised in arbitrary gauge.

Finally, we introduce the Bloch states 
\begin{align}
    \Psi_{k\alpha} (x) = \sqrt{\frac{L}{2 \pi}} \, \psi_{k \alpha} (x)= \sqrt{\frac{L}{2 \pi}} \, u_{k \alpha} (x) \, e^{i k x},
    \label{Psi_basis}
\end{align}
which satisfy the following normalization and completeness relations
\begin{align}
    & \int d x \, \, \Psi_{k \alpha}^* (x) \, \Psi_{k' \alpha'} (x) = \delta (k-k') \, \delta_{\alpha \alpha'}, \label{normPsi} \\
    & \sum_{\alpha} \int_{-\pi/L}^{\pi/L} d k \, \, \Psi_{k \alpha} (x) \, \Psi_{k \alpha}^* (x') = \delta (x-x').
\end{align}
Note that the prefactor $\sqrt{L/(2\pi)}$ in \eqref{Psi_basis} is needed in order to reconcile \eqref{normPsi} with the normalization $\int_0^L d x \,  |u_{k \alpha} (x)|^2= 1$.

Later on the states \eqref{Psi_basis} are used for constructing physical observables. Some properties of  $N_{k \alpha}$, $\epsilon_{k \alpha}$, and $\psi_2 (L,\epsilon_{k \alpha})$, which are also useful for this purpose, are quoted in Appendix \ref{app:sign_psi2}.

\begin{figure*}[t]
\centering
\includegraphics[width= 2\columnwidth]{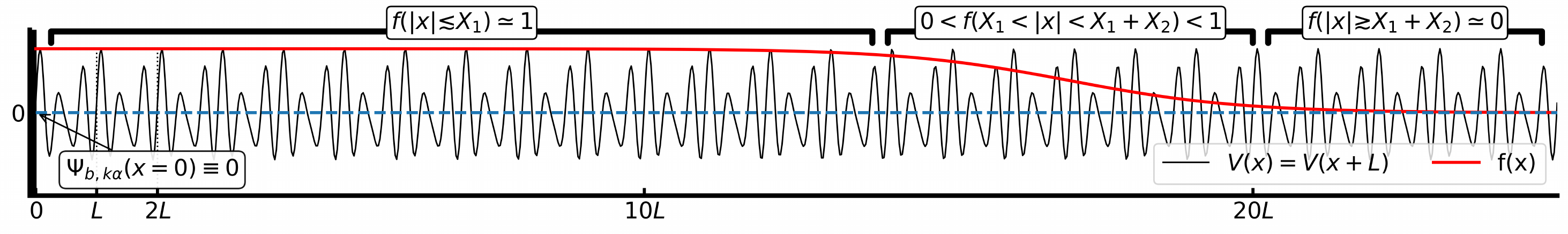} 
 \caption{Envelope function in the right half-infinite system: $f(x)=1$ for all unit cells lying to the left of $ X_1 $, while decaying smoothly to zero far from the boundary on a length scale $X_2$ much larger than the unit cell length $L$.
 }
\label{fig:EnvelopeFunction}
\end{figure*}

\section{Boundary charge} 
\label{sec:Boundary_Charge}

In this section we discuss solutions of the Schr{\"o}dinder equation \eqref{schr_eq} in the half-space $x>0$. This eigenvalue problem is equipped with the open boundary condition $\psi (x=0)=0$. Analogously to the case of 1D single-channel lattice models [\onlinecite{pletyukhov_etal_prb_20}], extended eigenstates of the present boundary problem can be constructed in terms of the Bloch states introduced in the previous section. In addition, we construct edge states which are exponentially localized near the boundary.

The central object of our study in this section is the boundary charge $Q_B$ accumulated near the boundary. For a system's filling determined by the chemical potential $\mu_{\nu}$ lying in the $\nu$th band gap, $Q_B$ receives contributions not only from an edge state eventually residing in this gap, but also from all occupied bands and the lower-gaps' edge states. We extend our previous lattice considerations [\onlinecite{pletyukhov_etal_prb_20}] to the case of continuum models and establish specific expressions for all contributions to $Q_B$. In particular, we provide a new rigorous proof of the surface charge theorem also fixing an expression for the integer part of $Q_B$ which is usually left unknown (cf., e.g., Ref.~[\onlinecite{vanderbilt_18}]).

Studying the dependence of $Q_B$ on the modulation phase of the potential (i.e., on the distance by which $V(x)$ is being shifted towards the boundary), we observe a manifestation of the bulk-boundary correspondence in the spirit of Ref.~[\onlinecite{hatsugai_1993}]. We also firmly establish the universal linear phase dependence of $Q_B$, with the slope being given by the Chern index of a band (bands). Thereby we provide a rigorous proof to the earlier numerical findings of Ref.~[\onlinecite{thakurati_etal_18}]. Moreover, we reveal that in continuum models the Chern index for each band is strictly equal to $+1$, and that the number of bounces of an edge state energy in a certain gap between two adjacent band edges during one pump cycle is given by the gap index.

Finally, we investigate the boundary charge fluctuations and conclude that in the narrow-gap limit they possess universal properties satisfying the surface fluctuation theorem which was stated in the previous works [\onlinecite{weber_etal_prl_20},\onlinecite{wannier_paper}].

\subsection{Eigenstates of the semi-infinite model}

In this subsection we construct the extended eigenstates of \eqref{schr_eq} in the right half-space $x>0$ and use them to evaluate each band's contribution to the boundary charge. 

The boundary problem eigenstates can be expressed in terms of the states \eqref{Psi_basis}:
\begin{align}
    \Psi_{\text{b},k\alpha} (x) = \Psi_{k\alpha} (x) - \Psi_{-k,\alpha} (x), \quad k \in \left[0, \frac{\pi}{L} \right],
    \label{eig_semi}
\end{align}
They have the eigenenergies $\epsilon_{k \alpha}$ (i.e. form the same bands as in the translationally invariant model) and satisfy the boundary condition
\begin{align}
    \Psi_{\text{b},k \alpha} (0) =0.
\end{align}
Their normalization reads
\begin{align}
    & \int_0^{\infty} d x \,\, \Psi^*_{\text{b},k \alpha} (x) \, \Psi_{\text{b},k'\alpha'} (x) = \delta (k-k') \, \delta_{\alpha \alpha'}.
\end{align}
Note that the set of all $\Psi_{\text{b},k\alpha} (x)$  is not complete, since localized edge states are also permitted. These edge states will be discussed in Sec. \ref{subsec:edge_states}.

The simplicity of the form of \eqref{eig_semi} is due to the special gauge choice which is fixed by the condition \eqref{gauge_real}. Using \eqref{eig_semi} we define the boundary charge [\onlinecite{thakurati_etal_18},\onlinecite{pletyukhov_etal_prb_20}] associated with the band $\alpha$
\begin{align}
    Q_{B,\alpha} &= \int_0^{\infty} d x \,\, [\rho_{\alpha} (x) - \bar{\rho}_{\alpha}] \, f (x)
    \label{eq:QBf}
\end{align}
in terms of the charge density
\begin{align}
    \rho_{\alpha} (x) &= \int_{0}^{\pi/L} d k \,\, |\Psi_{\text{b},k \alpha} (x)|^2 ,
    \end{align}
the average charge density in the bulk
\begin{align}
    \bar{\rho}_{\alpha} &= \frac{1}{L} \int_0^L d x  \int_{-\pi/L}^{\pi/L} d k \,\, |\Psi_{k \alpha} (x)|^2 = \frac{1}{L},
\end{align}
and the envelope function $f (x)$. The latter possesses the following properties: 1) $f (x) \approx 1$ in the range $|x| < X_1$; 2) in the range $X_1 < |x| < X_1 + X_2$ it smoothly crosses over to zero; 3) $f (x) \approx 0$ for $|x| >X_1 + X_2$ (see Fig. \ref{fig:EnvelopeFunction}). Here $X_{1,2} \gg \xi \gg L$ are scales large in comparison with the localization length $\xi= \kappa^{-1}$ (see Sec.~\ref{subsec:edge_states}
for the latter definition). 

\begin{figure*}[t]
\centering
\includegraphics[width= 2\columnwidth]{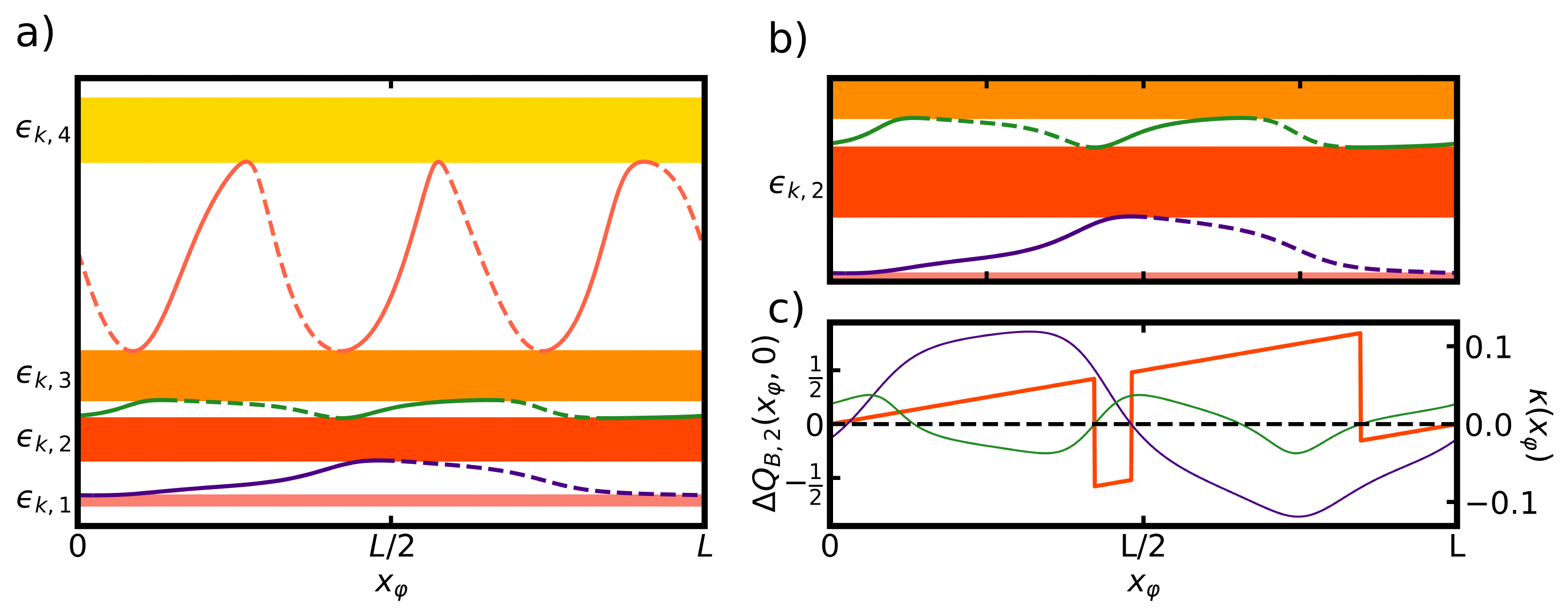} 
 \caption{a) Band structure including right/left (solid/dashed) edge state dispersions depending on $x_\varphi$ for $V(x)$ from Fig. \ref{fig:BarePot}. The number of touching points with adjacent bands equals the gap index implying a net difference of one touching point between leaving and entering edge states per band. By virtue of the bulk-boundary correspondence [see Sections \ref{SubSec:UnivBC} and \ref{SubSecBulkBoundary}] this fact reflects the constant band Chern index $C_{1,\alpha}=M_\alpha^{(-)}-M_\alpha^{(+)}= 1$. b) Close up of $\epsilon_{k,2}$. Edge states in the right system ($ \kappa > 0 $) and left system ($\kappa < 0 $) alternate upon traversing the touching points [see Eq.~\eqref{vort_bottom} and \eqref{vort_top}]. c) Boundary charge $\Delta Q_{B,2}(x_\varphi,0)=Q_{B,2} (x_{\varphi}) - Q_{B,2} (0)$ [Eq.~\eqref{eq:QBmain}]  and $\mathrm{Im}[k_{\text{e}} (x_\varphi)]=\kappa (x_\varphi)$ for the edge states adjacent to $\epsilon_{k,2}$. The universal linear growth is disrupted by discontinuous jumps at the positions of the touching points (= zeros of $\kappa (x_{\varphi})$). Since $ C_{1,\alpha} = 1 $ for each band  we regain periodicity under a shift $x_\varphi \rightarrow x_\varphi+L$.}
\label{fig:BoundaryCharge}
\end{figure*}

It can be shown [\onlinecite{pletyukhov_etal_prb_20}] that the boundary charge \eqref{eq:QBf} consists of the Friedel charge
\begin{align}
    Q_{F,\alpha} =& - \text{Re} \,  \int_0^{\infty} d x \int_{-\pi/L}^{\pi/L} d k  \,\, \Psi_{k\alpha}^2 (x)  \\
    =& - L \,\,  \text{Re} \,  \int_0^{L} d x \int_{-\pi/L}^{\pi/L} \frac{d k}{2 \pi}  \,\, \psi_{k\alpha}^2 (x)   \nonumber \\
    & \times \sum_{n=1}^{\infty} e^{2i (k L+i 0^+) (n-1) L} ,
    \label{QFsum}
\end{align}
which results from the charge density modulation near the boundary,
and the polarization charge 
\begin{align}
    Q_{P,\alpha} &=-\frac{1}{L} \int_0^L d x \,\, x \left( L \int_{-\pi/L}^{\pi/L} \frac{dk}{2 \pi} \,\, |\psi_{k\alpha} (x)|^2 - \frac{1}{L}\right) \label{eq:QP0} \\
    &= \frac12 - \int_0^L d x \,\,  x  \int_{-\pi/L}^{\pi/L} \frac{dk}{2 \pi} \,\, |u_{k \alpha} (x)|^2  ,
    \label{eq:QP}
\end{align}
which originates from the crossover region $X_1 \lesssim x \lesssim X_1+X_2$ of the envelope function $f (x)$, and it is expressed via the dipole moment of the unit cell (see below).

The Friedel charge \eqref{QFsum} is independent of the envelope function $f (x)$, since the corresponding Friedel charge density decays exponentially fast on the scale $\xi \ll X_1$. In turn, the polarization charge \eqref{eq:QP} relies on the properties of $f (x)$, and it is obtained from
\begin{align}
    Q_{P,\alpha} &= \int_0^{\infty} d x \,\, [\bar{\rho}_{\alpha} (x) - \bar{\rho}_{\alpha}] \, f (x) \\
    &= \sum_{n=0}^{\infty} \int_0^L d \bar{x} \,\,  [\bar{\rho}_{\alpha} (\bar{x}) - \bar{\rho}_{\alpha}] \, f (\bar{x} + n L), \label{eq:QPf} \\
    \bar{\rho}_{\alpha} (x) &=  \int_{-\pi/L}^{\pi/L} d k \,\, |\Psi_{k \alpha} (x)|^2 = \bar{\rho}_{\alpha} (x+L),
\end{align}
in the following way. Observing that the main contribution to \eqref{eq:QPf} comes from the range $X_1 < n L < X_1 + X_2$, we expand the envelope function $f (\bar{x} + n L) \approx f (n L) + \bar{x} f' (n L)$. The leading contribution identically vanishes after averaging over $\bar{x}$ (which is defined as $x \!\mod L$), and we get
\begin{align}
    Q_{P,\alpha} \approx \sum_{n=0}^{\infty}  f' (n L) \int_0^L d \bar{x} \,\, \bar{x} \, [\bar{\rho}_{\alpha} (\bar{x}) - \bar{\rho}_{\alpha}] .
\end{align}
Using the properties of the envelope function, we approximately evaluate the sum
\begin{align}
     \sum_{n=0}^{\infty}  f' (n L) \approx & \frac{1}{L} \int_0^{\infty} d y \,\, f' (y) \\
     =&  \frac{1}{L} [f (\infty) - f (0)] = - \frac{1}{L},
\end{align}
and thus reproduce \eqref{eq:QP0}.

Evaluating the sum in \eqref{QFsum} we express the Friedel charge in the form
\begin{align}
Q_{F,\alpha} &=   -L \int_0^L d x \int_{-\pi/L}^{\pi/L} \frac{d k}{2 \pi} \,\,\, \text{Re} \,  \left\{ \frac{ \psi_{k\alpha}^2 (x) \, e^{-i k L}}{- 2i \, \sin (k L+ i 0^+)} \right\}.
\end{align}
Note that a convergence factor $+i 0^+$ was added to the sum over $n$: We exchanged this summation with the integral over $k$, and this action had to be complemented by proper regularization.

Using a variant of the Sokhotski–Plemelj theorem
\begin{align}
 \frac{1}{\sin (k L+ i 0^+)} = \frac{1}{\sin k L} - i \pi \, \delta (k L)  + i \pi \, \delta (k L - \pi),
\end{align}
we obtain
\begin{align}
Q_{F,\alpha}^{\text{I}} =& -L \int_0^L d x \int_{-\pi/L}^{\pi/L} \frac{d k}{2 \pi} \,\, \text{Re}  \, \left\{  \frac{\psi_{k \alpha}^2 (x) \,  e^{-i k L}}{- 2i \, \sin k L} \right\}, \\
Q_{F,\alpha}^{\text{II}} =&   -\frac14  \int_0^L d x \int_{-\pi/L}^{\pi/L} d (k L) \nonumber \\
& \times \text{Re}  \left\{ \psi_{k \alpha}^2 (x) \,  e^{-i k L} \,\,
[\delta (k L)  -   \delta (k L - \pi)] \right\}  \nonumber \\
=& -\frac14  \int_0^L d x \,\, \text{Re}  \left\{  \psi_{0,\alpha}^2 (x)  +\psi_{\frac{\pi}{L},\alpha}^2 (x)   \right\} .
\end{align}
Since $ \psi_{0,\alpha} (x) $ and $\psi_{\frac{\pi}{L},\alpha} (x) $ are both real and normalized by \eqref{normal}, we get $Q_{F,\alpha}^{\text{II}}= - \frac12$. 

In turn,
\begin{align}
Q_{F,\alpha}^{\text{I}} =&  -L \int_0^L d x \int_{-\pi/L}^{\pi/L} \frac{d k}{2 \pi} \nonumber \\
& \times \text{Re}  \, \left\{ [\psi_{k\alpha} (x) \, e^{-i k L}  -\psi_{-k,\alpha} (x) \, e^{i k L}] \right. \nonumber \\
& \qquad \times \left. \psi_{k \alpha} (x) \, e^{-i k L} \, \frac{e^{i k L}}{- 2i \, \sin k L} \right\} \nonumber \\
&  + \frac{L}{2} \int_0^L d x \int_{-\pi/L}^{\pi/L} \frac{d k}{2 \pi} \,\, |\psi_{k \alpha} (x)|^2   \\
=&  -L \int_0^L d x \int_{-\pi/L}^{\pi/L} \frac{d k}{2 \pi} \,\, \frac{\psi_2 (x-L) }{N_{k\alpha}} \nonumber \\
& \times \left[   \psi_2 (x-L) - \psi_2 (x) \, \cos k L \right] +\frac12 .
\end{align}
So finally we get
\begin{align}
Q_{F,\alpha} =&  -L \int_0^L d x \int_{-\pi/L}^{\pi/L} \frac{d k}{2 \pi} \,\, \frac{\psi_2 (x-L) }{N_{k\alpha}}\nonumber \\
& \times \left[   \psi_2 (x-L) - \psi_2 (x) \, \cos k L \right].
\end{align}

\subsection{Berry phase expression for polarization}

In this subsection we extend the observation previously made for the lattice models in Ref.~[\onlinecite{pletyukhov_etal_prb_20}] that the boundary charge is given by the Zak-Berry phase [\onlinecite{zak_89}]. This statement is generally known as the surface charge theorem [\onlinecite{vanderbilt_18}]; however the role of the gauge of the Bloch states used for a construction of the Zak-Berry connection is usually not elucidated. Therefore this theorem is often stated up to an unknown integer which can be changed by a winding number of the phase of the gauge transformation.

The importance of the gauge \eqref{u_fun} (i.e., with the real last component of the Bloch vector $u_{k \alpha}$) was first recognized in Refs.~[\onlinecite{pletyukhov_etal_prbr_20},\onlinecite{pletyukhov_etal_prb_20}]. In particular, it was shown therein that while working in this specific gauge no additional integer contribution to $Q_B$ arises in lattice models. Below we show that this also holds true for continuum models.

Let us consider the expression
\begin{align}
 &  \text{Im} \,  \left[u_{k\alpha}^* (x) \, e^{i k (L-x)} \, \frac{d}{d k}  u_{k\alpha} (x) \, e^{-i k (L-x)}\right] \\
 =& \,\,  \text{Im} \,  \left[ \psi_{k \alpha}^* (x) \, e^{i k L} \, \frac{\partial }{\partial k}  \psi_{k \alpha} (x) \, e^{-i k L} \right] \nonumber \\
 &+ \frac{d \epsilon_{k\alpha}}{d k} \,\, \text{Im} \,  \left[\psi_{k \alpha}^* (x) \, \frac{\partial }{\partial E}  \psi_{k\alpha} (x) \right] \\
=&- \frac{L}{N_{k\alpha}}  \left[  \psi_2 (x-L)  - \psi_2 (x) \, \cos k L \right] \, \psi_2 (x-L) \nonumber \\
&+\frac{1}{N_{k\alpha}}   \frac{d \epsilon_{k\alpha}}{d k} \, \sin k L \,\,  F (x),
\end{align}
where
\begin{align}
F (x) =  \psi_2 (x) \, \frac{\partial \psi_2 (x-L)}{\partial E}- \psi_2 (x-L) \,  \frac{\partial \psi_2 (x)}{\partial E}  .
\end{align}

Due to the property 
\begin{align}
\int_0^L d x \, F (x) =0,
\label{psi_int}
\end{align}
whose proof is given in Appendix \ref{app:hill_eq}, we can write
\begin{align}
Q_{F,\alpha} =&    \int_0^L d x \int_{-\pi/L}^{\pi/L} \frac{d k}{2 \pi} \,\, \text{Im} \,  \left[ u_{k\alpha}^* (x) \, \frac{d}{d k}  u_{k \alpha} (x) \right] \nonumber \\
&-  \int_0^L d x  \,\, (L-x) \int_{-\pi/L}^{\pi/L} \frac{d k}{2 \pi} \,\, 
| u_{k \alpha} (x)|^2.
\end{align}
Combining this expression with \eqref{eq:QP}, we obtain
\begin{align}
Q_{B,\alpha}  &=  - \frac12 + \int_{-\pi/L}^{\pi/L} \frac{d k}{2 \pi} \,\, A_{k\alpha}, \label{QB_Berry}\\
A_{k\alpha} &= -i\int_0^L d x \,\, u_{k \alpha}^* (x) \, \frac{d}{d k} u_{k \alpha} (x),
\label{A_Berry}
\end{align}
where the last line expresses the Zak-Berry connection.

\subsection{Universal properties of the boundary charge} \label{SubSec:UnivBC}

The interesting behaviour of $Q_B$ under the shift of a potential towards the boundary was initially observed in numerical studies of Ref.~[\onlinecite{thakurati_etal_18}]. It was revealed that $Q_B$ linearly depends on the distance of the shift (i.e. on the potential modulation phase), the slope acquiring quantized universal values. In this subsection we provide an analytic proof of the linear dependence and discuss other universal properties of the boundary charge.

Suppose we shift our system towards the wall by $x_{\varphi}$. It means that we have to consider the boundary problem with a new potential
\begin{align}
    V^{(\varphi)} (x) = V (x+x_{\varphi}).
\end{align}
The new solution $\psi_{k \alpha}^{(\varphi)} (x)$ is related to the old one $\psi_{k \alpha} (x)$ by
\begin{align}
\psi_{k\alpha}^{(\varphi)} (x) &= \psi_{k\alpha} (x+ x_{\varphi}) \,\, e^{-i \Phi_{k \alpha} (x_{\varphi})}, \label{psi_shift} \\
u_{k\alpha}^{(\varphi)} (x) &= u_{k \alpha} (x+ x_{\varphi}) \,\,  e^{i k x_{\varphi}} \,\, e^{-i \Phi_{k \alpha} (x_{\varphi})}. \label{u_shift}
\end{align}
To enable the usage of \eqref{QB_Berry}, \eqref{A_Berry} in the shifted system, we must claim similarly to \eqref{gauge_real} that $\psi_{k\alpha}^{(\varphi)} (0) $ is real. This condition fixes (up to the irrelevant sign) the phase factor 
\begin{align}
    e^{ i \Phi_{k\alpha} (x_{\varphi})}  \sim \psi_{k\alpha}  (x_{\varphi}) = u_{k \alpha} (x_{\varphi}) \,\, e^{i k x_{\varphi}}.
\end{align}
Note that it is periodic in $k$ due to the analogous property of $\psi_{k\alpha} (x)$. In addition,
\begin{align}
    e^{ i \Phi_{k \alpha} (x_{\varphi}+L)} =e^{ i \Phi_{k \alpha} (x_{\varphi})} \,\, e^{i k L}.
    \label{Phi_x_per}
\end{align}

The boundary charge in the shifted system can be written with respect to the reference system value $Q_{B,\alpha} (0) \equiv Q_{B,\alpha}$ as
\begin{align}
     \Delta Q_{B,\alpha}(x_\varphi,0) &= Q_{B,\alpha} (x_{\varphi}) - Q_{B,\alpha} (0)\\
     &= \int_{-\pi/L}^{\pi/L} \frac{d k}{2 \pi} \,\, [A_{k\alpha}^{(\varphi)} - A_{k\alpha}] \\
    &= \frac{x_{\varphi}}{L} - \text{wn} \, [e^{i\Phi_{k\alpha} (x_{\varphi})}],
    \label{eq:QBmain}
\end{align}
where
\begin{align}
    &\text{wn} \, [e^{i \Phi_{k\alpha} (x_{\varphi})}] =- \text{wn} \, [e^{-i \Phi_{k\alpha} (x_{\varphi})}] \\
    &= -\frac{i}{2 \pi} \int_{-\pi/L}^{\pi/L} d k \,\, e^{-i \Phi_{k\alpha} (x_{\varphi})} \, \frac{d}{d k} e^{i \Phi_{k\alpha} (x_{\varphi})}
    \label{wn_def}
\end{align}
is the winding number of the corresponding phase factor. In this derivation we used the periodicity of $u_{k\alpha} (x)$ in $x$. Due to the property \eqref{Phi_x_per}, the boundary charge $Q_{B,\alpha} (x_{\varphi})$ is manifestly periodic under a shift $x_{\varphi} \to x_{\varphi} +L$, as it should be.

Piecewise, i.e. outside the points where the winding number jumps, the slope 
\begin{align} 
    L \, \frac{d Q_{B,\alpha} (x_{\varphi})}{d x_{\varphi}} = C_{1,\alpha} =1 \label{DQBtoC}
\end{align}
is universal and coincides with the first Chern index of the band $\alpha$ defined on the two-dimensional torus $(k, x_{\varphi})$:
\begin{align}
    C_{1,\alpha} =& - \int_{-\pi/L}^{\pi/L} 
    \frac{d k}{2 \pi} \int_0^L d x_{\varphi}  \nonumber \\
    & \times 2 \, \text{Im} \int_0^L d x \,\, \frac{d \psi_{k \alpha}^{(\varphi) \, *} (x)}{d k} \,\, \frac{d \psi_{k \alpha}^{(\varphi)} (x)}{d x_{\varphi}}.
    \label{chern_def}
\end{align}
This index is a sign-weighted count of the phase singular points across the band $\alpha$ [\onlinecite{kohmoto_1985},\onlinecite{pletyukhov_etal_prb_20}], thus giving (up to the sign) the net change of the winding number over one period in $x_{\varphi}$. The value of $C_{1,\alpha}$ must be synced --- as done in the first equality of \eqref{DQBtoC} --- with the slope of the linear part of $Q_{B,\alpha} (x_{\varphi})$ in order to maintain the periodicity $Q_{B,\alpha} (x_{\varphi})=Q_{B,\alpha} (x_{\varphi}+L)$.
In Appendix \ref{app:chern_eval} we show by explicit evaluation of \eqref{chern_def} that $C_{1,\alpha}=1$ for each band, i.e. prove the fulfillment of the second equality in \eqref{DQBtoC}.

The result \eqref{DQBtoC} differs to some extent from the findings of Ref.~[\onlinecite{pletyukhov_etal_prb_20}] for a  lattice model with $Z$ sites per unit cell. There, the slope (with respect to $\frac{\varphi}{2\pi}$) could take integer values $C_{1,\alpha} =1 +s_{\alpha} Z$ with some integer $s_\alpha$. The limit $Z \to \infty$ should bring us to the case of the presently studied continuum models. In order to get finite Chern indices, it is necessary to ensure that upon taking this limit the potential as a function of the phase variable does not make any kinks between the lattice sites.  This procedure ensures $s_{\alpha} =0$ and yields the result $C_{1,\alpha} = 1$ which is consistent with that of our present consideration. 

The bulk-boundary correspondence [\onlinecite{hatsugai_1993}] implies that the found value of $C_{1,\alpha}$ should be supported by the same value for the difference between numbers of edge states entering and leaving the band $\alpha$ on the interval $0 \leq x_{\varphi} < L$. We inspect this property in the next subsection.

To complete the discussion of the boundary charge we also outline expressions for the boundary charge in the left subsystem
\begin{align}
    Q_{B,\alpha}^L  &=\int_{-\infty}^0 d x \,\, [\rho_{\alpha} (x) - \bar{\rho}_{\alpha}] \, f (x) , \label{eq:QBfL} \\
    &= Q_{F,\alpha}^L  + Q_{P,\alpha}^L = -1 -Q_{B,\alpha},
    \label{eq:QBLQBR}
\end{align}
where
\begin{align}
    Q_{F,\alpha}^L =& - \int_{-\infty}^{0} d x \int_{-\pi/L}^{\pi/L} d k   \,\, \text{Re} \,  \Psi_{k \alpha}^2 (x)  \\
=& -L \,\, \text{Re} \, \int_{0}^{L} d x \int_{-\pi/L}^{\pi/L} \frac{d k}{2 \pi}  \,\,  \psi_{k \alpha}^2 (x) \nonumber  \\
& \times \sum_{n=-\infty}^0 e^{2 i (kL-i 0^+) (n-1)} \\
=& -1 - Q_{F,\alpha} , \label{eq:QFL} \\
     Q_{P,\alpha}^L =& - Q_{P,\alpha} . \label{eq:QPL}
\end{align}
For simplicity we choose $f (x) = f(-x)$ in \eqref{eq:QBfL}. In general, the values of $X_{1,2}$ may however differ for each half-space, but this does not affect the results \eqref{eq:QFL} and \eqref{eq:QPL}.

\subsection{Edge states}
\label{subsec:edge_states}

Besides the eigenstates \eqref{eig_semi} the semi-infinite model with the open boundary condition possesses edge localized states. We notice that the function \eqref{eig_fun3} vanishes at $x=0$, if the condition 
\begin{align}
    \psi_2 (L) = 0
    \label{edge_cond}
\end{align}
is fulfilled. This can not happen within the energy range of any band, since
according to the property \eqref{sign_psi2} the function $\psi_2 (L)$ is sign-definite there. Therefore, the condition \eqref{edge_cond} can be only fulfilled in band gaps. For this to happen we should allow for complex valued $k$.

The other condition \eqref{dr} relating $k$ and $E$ to each other contains the function $D(E)$, which is real at every $E$. This imposes the constraint that complex-valued $k$ can only be of the form either $k = \pm \pi/L + i \kappa$ or $k = i \kappa$, both with real $\kappa$.

The condition \eqref{wron} is also fulfilled at every $E$. Combining it with \eqref{edge_cond}, we obtain
\begin{align}
    \psi_1 (L) = \frac{1}{\psi'_2 (L)}.
\end{align}
Furthermore, from \eqref{dr} it follows
\begin{align}
    (-1)^s \, \cosh \kappa L = \frac12 \left( \psi_1 (L) + \frac{1}{\psi_1 (L)} \right).
    \label{dr_es}
\end{align}
Here $s=0$ corresponds to the choice $\text{Re}\, k =0$, and $s=1$ corresponds to the choice $\text{Re} \, k =\pm \pi/L$.

Equation \eqref{dr_es} has two solutions: $\psi_1 (L) =(-1)^s \, e^{\kappa L}$ and $\psi_1 (L) =(-1)^s \, e^{-\kappa L}$. They give the following eigenfunctions of edge states
\begin{align}
    \psi_{\text{e}} (x) & \sim [-\psi_1 (L) + e^{i k L}] \,\, \psi_2 (x) \nonumber \\
    & =-2 \, (-1)^s \, \sinh \kappa L \,\, \psi_2 (x), \\
    \psi_{\text{e}} (x) & \sim [-\psi_1 (L) + e^{i k L}] \,\, \psi_2 (x) \equiv 0,
    \label{triv_sol}
\end{align}
respectively. Thus, a nontrivial solution has
\begin{align}
    e^{\kappa L} = (-1)^s \, \psi_1 (L) >0.
    \label{nontriv_es}
\end{align}

Labelling each band gap by the index $\alpha$ of the band which lies beneath it, we search for the relation between $s$ and $\alpha$.  Comparing \eqref{pr2} and \eqref{sign_D_dpsi2}, we find that at the edge state energy $E= \epsilon_{\text{e},\alpha}$ it holds
\begin{align}
    \text{sign} \,\, \psi_1 (L,\epsilon_{\text{e},\alpha}) = \text{sign}\,  \frac{\partial \psi_2 (L,\epsilon_{\text{e}})}{\partial E} = \text{sign}\,\, D (\epsilon_{\text{e},\alpha}). 
\end{align}
But $\text{sign}\,  D (\epsilon_{\text{e},\alpha}) =(-1)^{\alpha}$ (see Fig.~\ref{fig:band}), and therefore
\begin{align}
    (-1)^s = (-1)^{\alpha},
\end{align}
i.e. edge states residing in odd gaps have $\text{Re} \, k =\pm \pi/L$, while edge states residing in even gaps have $\text{Re} \, k =0$.

Being primarily interested in the right subsystem ($x>0$), we impose the condition $\kappa >0$ in order to have a normalizable (decaying) solution. From \eqref{nontriv_es} it follows that a nontrivial edge state solution in the right subsystem exists, if
\begin{align}
    |\psi_1 (L)|> 1 \quad \Rightarrow \quad \kappa = \frac{1}{L} \, \ln |\psi_1 (L)| >0.
\end{align}
In turn,  a nontrivial edge state solution in the left subsystem exists, if
\begin{align}
    |\psi_1 (L)|< 1 \quad \Rightarrow \quad \kappa = \frac{1}{L} \, \ln |\psi_1 (L)| <0.
\end{align}
In Appendix \ref{app:one_per_gap} we prove that in each band gap there is precisely one edge state, corresponding either to the right subsystem ($x>0$, $\kappa>0$) or to the left subsystem ($x<0$, $\kappa<0$). Its normalized wavefunction reads
\begin{align}
 \Psi_{\text{e}, \alpha} (x) =  \frac{\Theta( x \,\,  \text{sign} \, \kappa)}{\sqrt{N_{\text{e}, \alpha}}} \, \psi_2 (x),
\end{align}
with the corresponding normalizations
\begin{align}
N_{\text{e}, \alpha} &= \int_0^{\infty} d x \,\, \psi_2^2 (x) =\sum_{n=0}^{\infty} \int_0^{L} d x \,\, \psi_2^2 (x+n L)\nonumber \\
&=\frac{(-1)^{\alpha}}{4m} \,  \frac{\partial \psi_2 (L)}{\partial E} \, \frac{1}{\sinh \kappa L}, \quad \kappa>0, \label{norm1} \\
N_{\text{e}, \alpha} &= \int_{-\infty}^0 d x \,\, \psi_2^2 (x) =\sum_{n=0}^{\infty} \int_{-L}^0 d x \,\, \psi_2^2 (x-n L) \nonumber \\
&=- \frac{(-1)^{\alpha}}{4m} \, \frac{\partial \psi_2 (L)}{\partial E} \, \frac{1}{\sinh  \kappa L} , \quad \kappa< 0. \label{norm2}
\end{align}
These expressions are derived with help of \eqref{psi2bL}, \eqref{pr1} and \eqref{psi2L}, \eqref{pr2}. We remark that at the edge state energy $E = \epsilon_{\text{e}, \alpha}$ it always holds $(-1)^{\alpha} \, \frac{\partial \psi_2 (L)}{\partial E} > 0 $ (see the discussion in Appendix \ref{app:one_per_gap}).

\subsection{Phase dependence of edge states}

If we shift the potential $V (x)$ towards the hard wall by $x_{\varphi}$, the edge state energy $\epsilon_{\text{e}, \alpha}$ will change its position within the band gap $\alpha$. At some values of $x_{\varphi}$ the function $\epsilon_{\text{e}, \alpha} (x_{\varphi})$ can touch either lower or upper bands adjacent to the corresponding band gap. In the following we study the edge state energy dependence on $x_{\varphi}$.

We start with expressing the fundamental system $\{ \psi_1^{(\varphi)} (x), \psi_2^{\varphi} (x)\}$ of the shifted model with $V^{(\varphi)} (x) = V (x+x_{\varphi})$ in terms of the original reference system $\{ \psi_1 (x), \psi_2 (x)\}$:
\begin{align}
    \psi_1^{(\varphi)} (x) &= c_{11} \, \psi_1 (x+x_{\varphi}) + c_{12} \, \psi_2 (x+x_{\varphi}), \label{psi1_shift} \\
    \psi_2^{(\varphi)} (x) &= c_{21} \, \psi_1 (x+x_{\varphi}) + c_{22} \, \psi_2 (x+x_{\varphi}). \label{psi2_shift}
\end{align}
Applying the defining relations
\begin{align}
    \psi_1^{(\varphi)} (0) &=1, \quad \psi_1^{(\varphi)\, \prime} (0) =0, \\ \psi_2^{(\varphi)} (0) &=0, \quad \psi_2^{(\varphi)\, \prime} (0) =1,
\end{align}
we obtain the equations
\begin{align}
    1 &= c_{11} \, \psi_1 (x_{\varphi}) +c_{12} \, \psi_2 (x_{\varphi}), \\
    0 &= c_{21} \, \psi'_1 (x_{\varphi}) +c_{22} \, \psi'_2 (x_{\varphi}), 
\end{align}
and
\begin{align}
    0 &= c_{21} \, \psi_1 (x_{\varphi}) +c_{22} \, \psi_2 (x_{\varphi}), \\
    1 &= c_{21} \, \psi'_1 (x_{\varphi}) +c_{22} \, \psi'_2 (x_{\varphi}).
\end{align}
Solving them we get
\begin{align}
    \left( \begin{array}{cc} c_{11} & c_{12} \\ c_{21} & c_{22} \end{array} \right) = \left( \begin{array}{cc} \psi'_2 (x_{\varphi}) & - \psi'_1 (x_{\varphi}) \\ - \psi_2 (x_{\varphi}) & \psi_1 (x_{\varphi}) \end{array} \right).
    \label{c_matr}
\end{align}

An immediate consequence of this solution is that the function $D (E)$ is independent of $x_{\varphi}$: Considering
\begin{align}
    D^{(\varphi)} (E) =& \frac12 \left\{ \psi_1^{(\varphi)} (L) + \psi_2^{(\varphi) \, \prime} (L) \right\}  \label{DE_phi} \\
    =& \frac12 \left\{ \psi'_2 (x_{\varphi}) \, \psi_1 (L+ x_{\varphi})-\psi'_1 (x_{\varphi}) \, \psi_2 (L+ x_{\varphi})\right. \nonumber \\
    & \left. -\psi_2 (x_{\varphi}) \, \psi'_1 (L+ x_{\varphi}) + \psi_1 (x_{\varphi}) \, \psi'_2 (L+ x_{\varphi}) \right\}, \nonumber
\end{align}
and employing \eqref{psi1bL}, \eqref{psi2bL}, and \eqref{wron} we obtain
\begin{align}
    D^{(\varphi)} (E) = \frac12 \left\{ \psi_1 (L) + \psi'_2 (L) \right\} = D (E).
    \label{DE_phi_indep}
\end{align}
Due to this property all band edges are flat with respect to $x_{\varphi}$ in Fig.~\ref{fig:BoundaryCharge}.
We also note that the relation \eqref{psi_shift} implies the equality
\begin{align}
    \frac{\psi_{k \alpha}^{(\varphi)\, \prime} (0)}{\psi_{k \alpha}^{(\varphi)} (0)} = \frac{\psi'_{k \alpha} (x_{\varphi})}{\psi_{k \alpha} (x_{\varphi})}.
    \label{phase_elim_rel}
\end{align}
With help of \eqref{psi1_shift}, \eqref{psi2_shift}, \eqref{c_matr} and \eqref{LminusL1}-\eqref{LminusL4} we verify that it actually holds for all energies.

The edge state conditions in the shifted system 
\begin{align}
    \psi_2^{(\varphi)} (L) &=0, \\
    \psi_1^{(\varphi)} (L) &= (-1)^{\alpha} \, e^{\kappa (x_{\varphi} ) L}
\end{align}
are expressed with help of the above relations  as
\begin{align}
  \frac{\psi_1 (L+x_{\varphi})}{\psi_1 (x_{\varphi})} = \frac{\psi_2 (L+x_{\varphi})}{ \psi_2 (x_{\varphi})}= (-1)^{\alpha} \, e^{\kappa (x_{\varphi} ) L}.
  \label{es_shift_cond}
\end{align}
Solving them we find the edge state dispersions $\epsilon_{\text{e},\alpha} (x_{\varphi})$ and showcase them in Fig.~\ref{fig:BoundaryCharge}(a,b).

\subsection{Band-touching points and vorticities}
\label{subsec:band_touch}

At the band-touching point $x^*_{\varphi}$ the condition
\begin{align}
    \kappa (x_{\varphi}^*) =0
    \label{touch_first}
\end{align}
is fulfilled.
Following Ref.~[\onlinecite{hatsugai_1993}], we assign to this point the vorticity
\begin{align}
    \text{sign} \, \frac{d \kappa (x_{\varphi}^*)}{d x_{\varphi}}.
    \label{vort_def}
\end{align}
Using in the following the right system ($x>0$) as the reference one, we observe that the vorticity $-1$ counts an edge state entering the band (change of $\kappa$ from positive to negative), while the vorticity $+1$ counts an edge state leaving the band (change of $\kappa$ from negative to positive).

Evaluating \eqref{vort_def} on the basis of \eqref{es_shift_cond} and \eqref{psi1bL}, \eqref{psi2bL} we obtain
\begin{align}
    \text{sign} \, \frac{d \kappa (x_{\varphi}^*)}{d x_{\varphi}} &= (-1)^{\alpha} \, \text{sign} \, \left[ \psi'_1 (L) \, \frac{d}{d x_{\varphi }} \frac{\psi_2 (x_{\varphi}^*)}{\psi_1 (x_{\varphi}^*)} \right] \nonumber \\
    &= (-1)^{\alpha} \, \text{sign} \, \left[ \psi'_1 (L) \right]
    \label{vort_sign1}
\end{align}
and
\begin{align}
    \text{sign} \, \frac{d \kappa (x_{\varphi}^*)}{d x_{\varphi}} &= (-1)^{\alpha} \, \text{sign} \, \left[ \psi_2 (L) \, \frac{d}{d x_{\varphi }} \frac{\psi_1 (x_{\varphi}^*)}{\psi_2 (x_{\varphi}^*)} \right] \nonumber \\
    &=- (-1)^{\alpha} \, \text{sign} \, \left[ \psi_2 (L) \right].
    \label{vort_sign2}
\end{align}
From these expressions (which are equivalent, see below) we make an important conclusion: The vorticity value is a property of the band edge, i.e. it has the same value for all touching points at the same band edge. Distinguishing between the top edge of the band $\alpha$ and the bottom edge of the band $(\alpha+1)$, both facing the same band gap $\alpha$, and using \eqref{sign_psi2} we establish
\begin{align}
    \text{sign} \, \frac{d \kappa (x_{\varphi}^*)}{d x_{\varphi}} \bigg|_{\alpha+1, \, \text{bottom}} &= -1, \label{vort_bottom}\\
    \text{sign} \, \frac{d \kappa (x_{\varphi}^*)}{d x_{\varphi}} \bigg|_{\alpha, \, \text{top}} &= +1. \label{vort_top}
\end{align}
Due to the continuity of $\epsilon_{\text{e}, \alpha} (x_{\varphi})$ in $x_{\varphi}$ the edge state trajectory for the right system originates in the lower band (at the top of the band $\alpha$) and terminates in the upper band (at the bottom of the band $\alpha+1$).

To demonstrate the equivalence of \eqref{vort_sign1} and \eqref{vort_sign2}, we relate the functions $\psi'_1 (L)$ and $\psi_2 (L)$ at the band edge, where $D(E) =(-1)^{\alpha}$, via 
\begin{align}
    \psi'_1 (L) \, \psi_2 (L) &= -1 + \psi'_2 (L) \, \psi_1 (L)  \nonumber \\
    &= -1 + [2 (-1)^{\alpha} - \psi_1 (L)] \,\, \psi_1 (L) \nonumber \\
    &= -[(-1)^{\alpha} - \psi_1 (L)]^2 \, \leq  \, 0.
\end{align}
This means that $\text{sign} \, [ \psi'_1 (L)] = - \text{sign} \, [ \psi_2 (L)] $ (unless we have a touching point at $x_{\varphi}^* =0$, which can be always excluded by choosing a different reference system).

The observation about the vorticity values made above allows us to conclude
that
\begin{align} \label{MatoMap1}
    M_{\alpha}^{(-)} = M_{\alpha+1}^{(+)},
\end{align}
where $M_{\alpha}^{(\pm)}$ are the numbers of edge states which enter/leave the band $\alpha$ (labels correspond to vorticities $\mp 1$). These properties can be recognized in Fig.~\ref{fig:BoundaryCharge}, where, within a gap, the edge dispersions touch the adjacent bands the same number of times thus demonstrating the property \eqref{MatoMap1}.

\subsection{Bulk-boundary correspondence} 
\label{SubSecBulkBoundary}

To prove the bulk-boundary correspondence, we have to show that
\begin{align}
    M_{\alpha}^{(-)} - M_{\alpha}^{(+)} = C_{1,\alpha} = +1.
    \label{bbc}
\end{align}

Let us prove that
\begin{align}
    M_{\alpha}^{(-)} & = \alpha, \label{Mm} \\
    M_{\alpha}^{(+)} & = \alpha -1. \label{Mp}
\end{align}
These relations automatically imply \eqref{bbc}.

First, we summarize the conditions satisfied at the touching points. Combining \eqref{es_shift_cond} and \eqref{touch_first} as well as using \eqref{psi1bL} and \eqref{psi2bL} we derive
\begin{align}
    \psi_2 (L) \, \psi_1 (x^*_{\varphi}) + \psi_2 (x^*_{\varphi}) \, \psi'_2 (L) &= (-1)^{\alpha} \, \psi_2 (x^*_{\varphi}).
\end{align}
Eliminating $\psi'_2 (L) =2 (-1)^{\alpha} - \psi_1 (L)$,
we express this condition in the form
\begin{align}
    \psi_1 (x_{\varphi}^*) \, \psi_2 (L) -[\psi_1 (L) - (-1)^{\alpha} ] \, \psi_2 (x_{\varphi}^*)=0.
   \label{psi_s0}
\end{align}
Hereby one can recognize an equation for the roots of the periodic (even $\alpha$) and antiperiodic (odd $\alpha$) solutions \eqref{eig_fun3} of Eq. \eqref{schr_eq}. 

A problem of counting zeros of periodic and antiperiodic solutions of a linear differential equation with periodic coefficients has a long tradition in the mathematical literature [\onlinecite{arscott,ince,mckean}]. Adapting the existing statements and their proofs initially formulated in Ref.~[\onlinecite{ince}] to our present notations (see Appendix \ref{app:zeros} for details), we eventually conclude that 
\begin{align}
    M^{(-)}_{\alpha} = M_{\alpha+1}^{(+)} = \alpha .
    \label{Mmp_alpha}
\end{align}
This relation is equivalent to \eqref{Mm} and \eqref{Mp}, and the bulk-boundary correspondence \eqref{bbc} is confirmed. 

The discussed  property is demonstrated in Fig.~\ref{fig:BoundaryCharge}(c): In order to ensure the periodicity of $Q_{B,\alpha} (x_{\varphi})$ which grows linearly in $x_{\varphi}$ with the slope $\frac{1}{L}$ on the interval $L$, there must be an overall discontinuous jump by $-1$. The jumps occur only when an edge state either leaves or enters the band. Thus, the bulk-boundary correspondence \eqref{bbc} expresses the necessary charge balance during a pump cycle.  

\subsection{Total boundary charge}
\begin{figure}[t]
    \centering
    \includegraphics[width=\columnwidth]{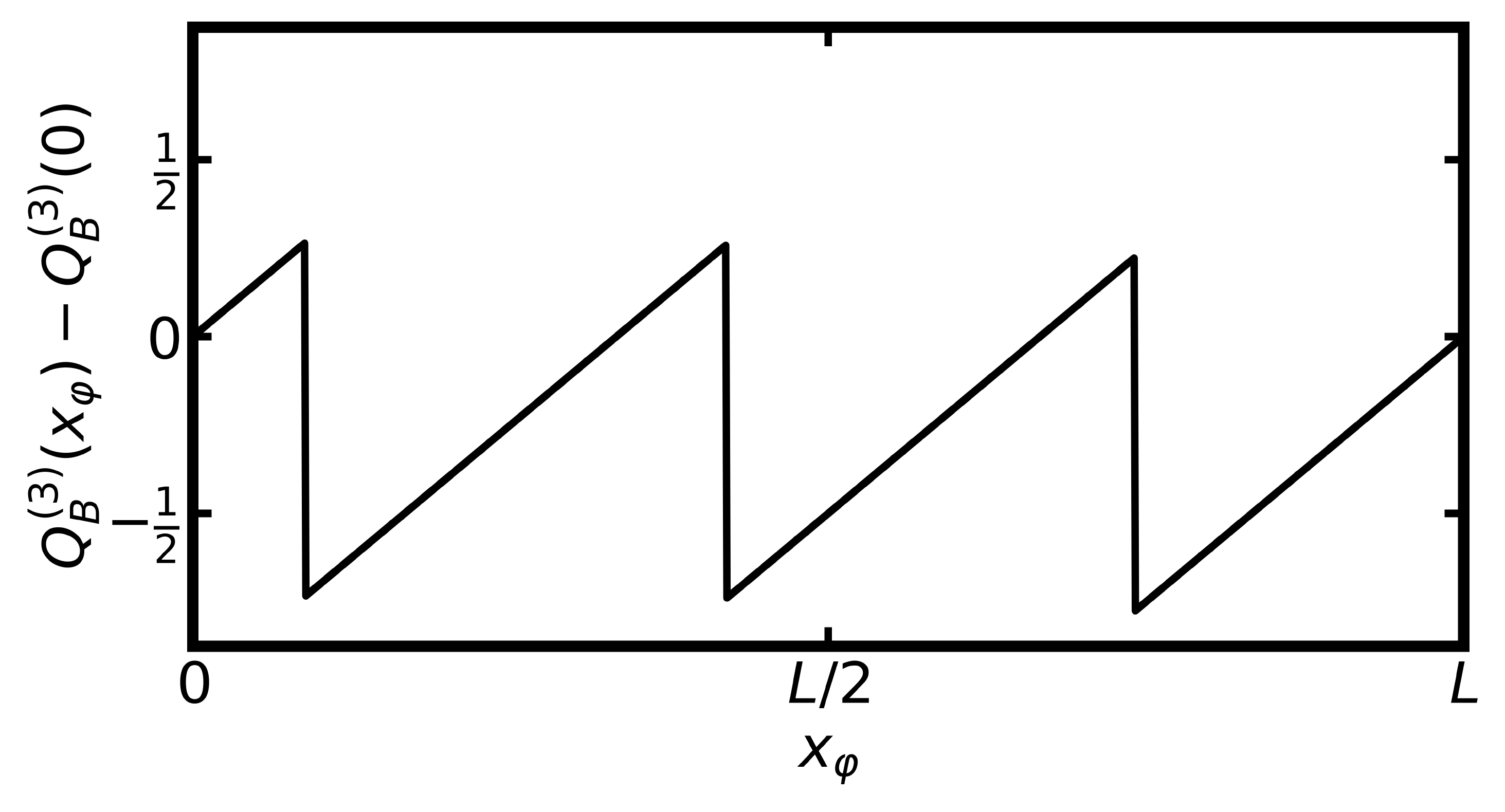}
    \caption{Total boundary charge for $\mu=\epsilon_{\frac{ \pi}{L},3}$. According to Eq. \eqref{Mmp_alpha}, $\nu=3$ jumps downward are observed at the positions of the touching points with the adjacent band edge hence restoring periodicity over a shift $x_\varphi\rightarrow x_\varphi+L$. }
    \label{fig:TotalBC}
\end{figure}

Setting the chemical potential $\mu_{\nu} = \epsilon_{\frac{\nu \pi}{L}, \nu}$ at the top of the uppermost valence band, we sum up all contributions to the total boundary charge, coming from both the bands and the edge states. Thus we get
\begin{align} \label{Tot_BC_Init}
     Q_B^{(\nu)} = \sum_{\alpha=1}^{\nu} Q_{B,\alpha} + \sum_{\alpha=1}^{\nu-1} Q_{\text{e},\alpha}.
\end{align}
The edge state contribution from the gap $\alpha$ equals  $Q_{\text{e},\alpha}=1$, when $\kappa_{\alpha} >0$, and zero otherwise.
Based on the fact that the boundary charge for a single band is periodic in $x_{\varphi}$ (see Sec. \ref{SubSec:UnivBC}) we recognize that this also holds true for the total boundary charge. Furthermore, considering the individual contributions to Eq. \eqref{Tot_BC_Init}, it can be seen that jumps in $ Q_{B,\alpha}$ occurring at the touching points are compensated by edge states themselves, besides those jumps which occur at the topmost band edge below $\mu_{\nu}$. Therefore,
\begin{align} 
    Q_B^{(\nu)} (x_{\varphi}) = Q_B^{(\nu)} (0) + \frac{\nu}{L} x_{\varphi} - \sum_{j=1}^{\nu} \Theta (x_{\varphi,j}^*) ,
    \label{TotBC_Rewrit}
\end{align}
where $\{x_{\varphi,j}^* \}$ is a set of touching points at which an edge state leaves the top edge of the uppermost valence band $\nu$. According to \eqref{Mmp_alpha} there are precisely $\nu$ such points, and this is the number needed to ensure the periodicity $ Q_B^{(\nu)} (x_{\varphi})=  Q_B^{(\nu)} (x_{\varphi}+L)$. Fig.~\ref{fig:TotalBC} illustrates this behaviour which was already observed by the authors of [\onlinecite{thakurati_etal_18}] in a lattice model with quite large $Z=10$ rendering their results to be tractable in terms of the present continuum theory.

Following the previous studies [\onlinecite{weber_etal_prl_20},\onlinecite{wannier_paper}], we quote the expression for the fluctuations of \eqref{TotBC_Rewrit}
\begin{align}
X_2 \, (\Delta Q_B^{(\nu)})^2 
&= \sum_{\alpha,\beta=1}^{\nu} \int_{-\pi/L}^{\pi/L} \frac{d k}{2 \pi} \,\, (\mathcal{Q}_{k})_{ \alpha \beta}.
\label{fluct_def}
\end{align}
Here $(\mathcal{Q}_{k})_{ \alpha \beta}$ is the geometric tensor [see \eqref{geom_tensor}], and the relation
\begin{align}
\frac{1}{X_2} = \int_{-\infty}^{\infty} d x \,\, [f' (x)]^2
\end{align}
quantifies the crossover range of the envelope function $f (x)$. In the narrow-gap limit we re-derive (see Appendix \ref{app:fluct_ident}) the previously established universal low-energy scaling of \eqref{fluct_def} -- the so called surface fluctuation theorem [\onlinecite{weber_etal_prl_20},\onlinecite{wannier_paper}]
\begin{align}
X_2 \, (\Delta Q_B^{(\nu)})^2 \approx  \frac{v_{F,\nu}}{8 E_{g,\nu}},
\label{sft}
\end{align}
expressed in terms of the Fermi velocity $v_{F, \nu}$ and the gap value $E_{g,\nu}$.

\section{Interface charge}
\label{sec:Interface_Charge}

This section discusses the properties of interface charges $Q_I$ that accumulate at interfaces with nontrivial conjunctions of two half-infinite subsystems. Similarly to our proceedings in Sec. \ref{sec:Boundary_Charge} and particularly in Ref.~[\onlinecite{RatBCPlet}] (see Appendix C therein) we determine the scattering eigenstates by modelling the interface microscopically. This construction, in turn, provides the scattering matrix that has already been shown to be determining for the properties at the interface [\onlinecite{fulga_2011},\onlinecite{fulga_2012}]. In addition, we construct interface localized states --- exponentially localized states with bilateral support.  Energetically these states reside in the gaps between bands formed by the scattering eigenstates.

At the interface we allow for a mutual phase mismatch $x_\varphi\neq x_{\varphi\prime}$ of the potentials in the two subsystems on the right and left in addition to a local impurity potential $\lambda \, \delta(x)$. Fixing the value of $x_{\varphi'}$ and cyclically pumping $x_{\varphi}$, we establish  a number of similarities between $Q_I (x_{\varphi})$ and the boundary charge discussed in the previous section. In particular, we  derive for $Q_I (x_{\varphi})$ analogs of the universal expressions \eqref{eq:QBmain} and \eqref{TotBC_Rewrit}. In doing so we identify  a function $\tilde{d}_{k\alpha} (x_{\varphi})$ whose windings and phase singular points play a tremendous determining role in describing a spectral flow of the interface problem, not to speak of the equation $\tilde{d}_{k\alpha} (x_{\varphi})=0$ determining the interface localized states.

It is then recognized that the linear change of $Q_I (x_{\varphi})$ is associated with an incomplete compensation of the dipole-moment contributions \eqref{eq:QP} from the right and left subsystems, which accords with the main motif of Ref.~[\onlinecite{RatBCPlet}]. Other changes to $Q_I (x_{\varphi})$ are found to be discontinuously integer-valued, and this serves as an additional justification for the nearsightedness principle, which was heuristically postulated in Ref.~[\onlinecite{NearSightednessPrin}].

\subsection{Scattering eigenstates}

\begin{figure}[t]
    \centering
    \includegraphics[width=\columnwidth]{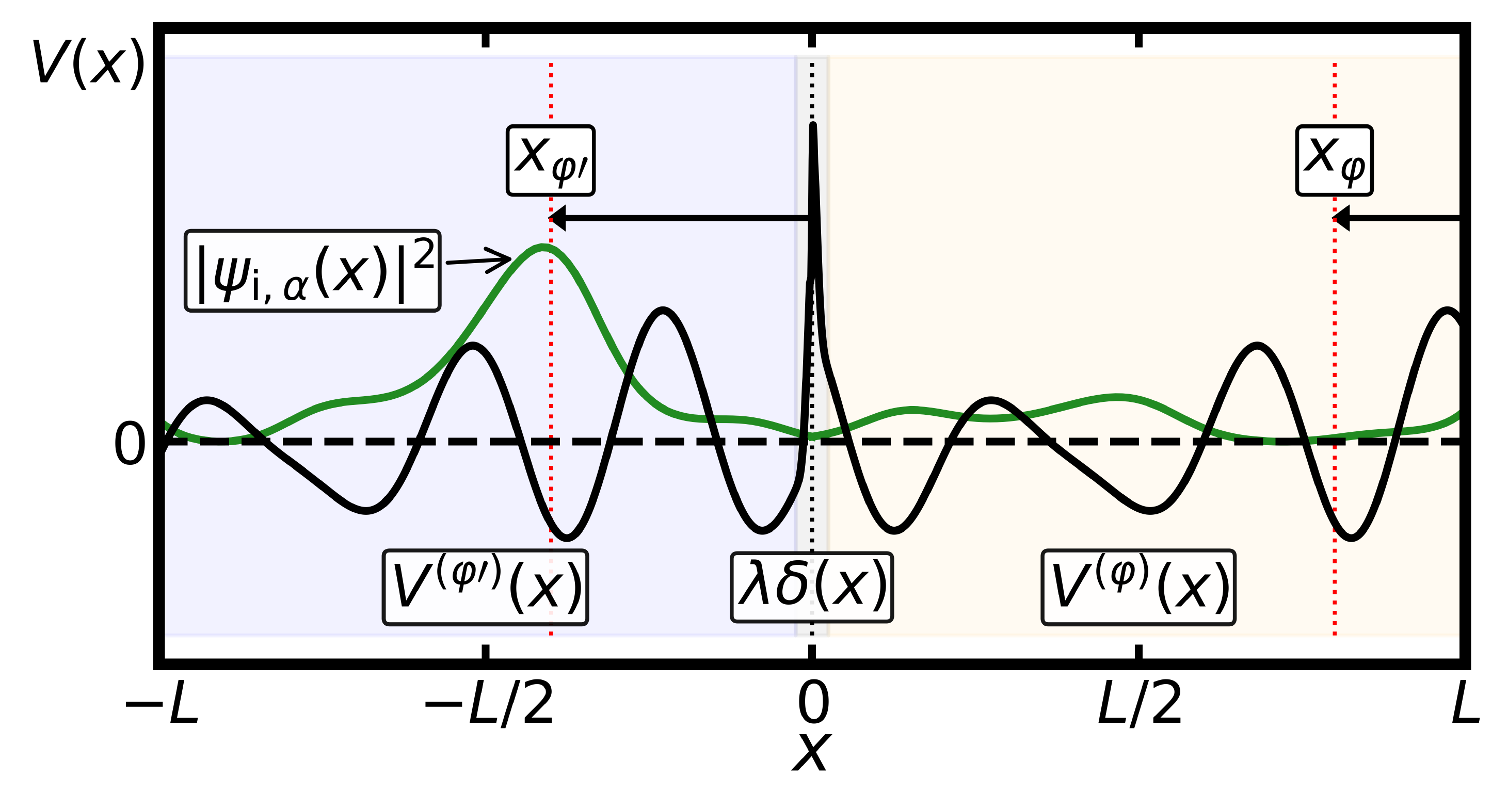}
    \caption{Potential $\Theta (x) \, V (x+x_{\varphi}) + \Theta (-x) \, V (x+x_{\varphi'}) + \lambda \, \delta (x)$ of the interface problem (black) and exemplary interface localized  state (green) belonging to the gap $\alpha=1$ and evaluated for the parameters $x_{\varphi}=0.1 L$, $x_{\varphi'}=0$, and $\lambda=1$.}
    \label{fig:MismatchPotential}
\end{figure}

We consider the periodic potential $V^{(\varphi)} (x) = V (x + x_{\varphi})$ in the right half-space, and the periodic potential $V^{(\varphi')} (x) = V (x + x_{\varphi'})$ in the left half-space. In addition, we add an extra impurity potential at the interface $V_{\text{imp}} (x) = \lambda \, \delta (x)$. The overall interface potential used in numerical calculations is depicted in Fig.~\ref{fig:MismatchPotential}.

In this subsection we construct scattering eigenstates of this model using the two bases, $\Psi_{k \alpha}^{(\varphi)} (x) $ and $\Psi_{k \alpha}^{(\varphi')} (x) $, which are the eigenstates of the bulk eigenvalue problems with the corresponding periodic potentials. Note that these bulk eigenvalue problems are isospectral, since they are mutually related by the unitary transformation --- the translation by $x_{\varphi} - x_{\varphi'}$. Therefore, the bands $\epsilon_{k \alpha}$ of the scattering states in the interface problem also coincide with those of the bulk models.

We make the following ansatz for the scattering eigenstates
\begin{align}
\Psi_{k \alpha}^{(r)} (x) &= \Theta (-x) \,  \left[ \Psi_{k \alpha}^{(\varphi')} (x)  + r_{k \alpha} \, \Psi_{ -k, \alpha}^{(\varphi')} (x)  \right] \nonumber \\
& + \Theta (x) \,\,  t_{k \alpha} \,  \Psi_{k \alpha}^{(\varphi)} (x)  \label{scat_state_r}, \\
\Psi_{k \alpha}^{(l)} (x) &= \Theta (-x) \,\, t'_{k \alpha} \, \Psi_{ -k,\alpha}^{(\varphi')} (x) 
\nonumber \\
&+ \Theta (x)  \left[  \Psi_{ -k,\alpha}^{(\varphi)} (x)  + r'_{k\alpha} \, \Psi_{k\alpha}^{(\varphi)} (x)  \right], \label{scat_state_l}
\end{align}
labeled by $k \in [0, \frac{\pi}{L}]$ and the band index $\alpha$.

To establish the transmission ($t_{k \alpha}, t'_{k \alpha}$) and reflection ($r_{k\alpha},r'_{k \alpha}$) amplitudes, we employ the wavefunction's matching conditions at the interface
\begin{align}
\Psi_{k\alpha}^{(\eta)} (0^-) &= \Psi_{k \alpha}^{(\eta)} (0^+), 
\label{match1} \\
\frac{d  \Psi_{k \alpha}^{(\eta)}}{d x} (0^+) - \frac{d  \Psi_{k \alpha}^{(\eta)}}{d x} (0^-) &= 2 m \lambda \, \Psi_{k \alpha}^{(\eta)} (0),
\label{match2}
\end{align}
where $\eta=r,l$. Thereby we get
\begin{align}
  & t_{k\alpha} \, \psi_{k \alpha}^{(\varphi)} (0)  - r_{k \alpha} \, \psi_{ -k,\alpha}^{(\varphi')} (0)  =  \psi_{k \alpha}^{(\varphi')} (0)   , \\
   & t_{k \alpha} \, [ \psi_{k \alpha}^{(\varphi) \, \prime} (0)  - 2  m \lambda \, \psi_{k \alpha}^{(\varphi)} (0) ] - r_{k \alpha} \, \psi_{ -k,\alpha}^{(\varphi') \, \prime} (0)  =  \psi_{k \alpha}^{(\varphi') \, \prime} (0)   ,
\end{align}
and
\begin{align}
  & t'_{k \alpha} \, \psi_{ -k, \alpha}^{(\varphi')} (0)  - r'_{k \alpha} \, \psi_{k \alpha}^{(\varphi)} (0) = \psi_{ -k, \alpha}^{(\varphi)} (0), \\
  & t'_{k \alpha} \,  [\psi_{ -k , \alpha}^{(\varphi')\, \prime} (0) +2 m \lambda  \,  \psi_{ -k, \alpha}^{(\varphi')} (0) ]  - r'_{k \alpha} \, \psi_{k \alpha}^{(\varphi) \, \prime} (0) = \psi_{ -k, \alpha}^{(\varphi) \, \prime} (0)  ,
\end{align}
where
\begin{align}
 \psi_{k \alpha}^{(\varphi)} (0) &= \psi_{ -k, \alpha}^{(\varphi)} (0) = \frac{\psi_2^{(\varphi)} (L) }{\sqrt{N_{k \alpha}^{(\varphi)}}} , \label{psi_phi_0} \\
 \psi_{k \alpha}^{(\varphi)\, \prime} (0) &= \frac{-\psi_1^{(\varphi)} (L)  +e^{i k L}}{\sqrt{N_{k\alpha}^{(\varphi)}}} .
 \label{psi_phi_prime_0}
\end{align}

The solutions of these linear equations read
\begin{align}
t_{k\alpha} &= \frac{ 2 i \, \sin k L \,\, \psi_{2}^{(\varphi')} (L)}{ d_{k \alpha}} \, \sqrt{\frac{N_{k\alpha}^{(\varphi)}}{N_{k\alpha}^{(\varphi')}}} , \label{tk}\\
-r_{k\alpha} &= 1- \frac{ 2 i \,\sin k L \,\, \psi_{2}^{(\varphi)} (L)}{ d_{k\alpha}} ,
\label{rk}
\end{align}
and
\begin{align}
 t'_{k\alpha} &= \frac{2i \, \sin k L \,\, \psi_{2}^{(\varphi)} (L)  }{  d_{k\alpha} } \, \sqrt{\frac{N_{k\alpha}^{(\varphi')}}{N_{k\alpha}^{(\varphi)}}}, \label{tk_prime} \\
 -r'_{k\alpha} &= 1-  \frac{2i \, \sin k L \,\, \psi_2^{(\varphi')} (L) }{d_{k\alpha}},
 \label{rk_prime}
\end{align}
where
\begin{align}
d_{k \alpha} =& \sqrt{ N_{k \alpha}^{(\varphi)}  N_{k \alpha}^{(\varphi')}} \left\{ - \psi_{k \alpha}^{(\varphi)} (0) \, \psi_{-k, \alpha}^{(\varphi') \, \prime} (0) \right. \nonumber \\
& \left. + [ \psi_{k \alpha}^{(\varphi)\, \prime} (0) -2 m \lambda \, \psi_{k \alpha}^{(\varphi)} (0)] \,\, \psi_{-k, \alpha}^{(\varphi')} (0) \right\}  \\
=&  \psi_2^{(\varphi)} (L) \, [\psi_1^{(\varphi')} (L)  -e^{-i k L} ] \nonumber \\
&- [\psi_1^{(\varphi)} (L)  -e^{i k L} + 2 m \lambda  \, \psi_2^{(\varphi)} (L) ] \, \, \psi_2^{(\varphi')} (L).
\end{align}
This function is expressed in term of two different fundamental systems referring to potentials shifted by $x_{\varphi}$ and $x_{\varphi'}$. Using the transformation laws \eqref{psi1_shift}, \eqref{psi2_shift}, \eqref{c_matr} of a fundamental system under the potential's shift, we express $d_{k \alpha}$ in terms of the unshifted fundamental system.
This gives
\begin{align}
    d_{k \alpha} =& \frac{
    \sqrt{ N_{k \alpha}^{(\varphi)}  N_{k \alpha}^{(\varphi')}}}{N_{k \alpha}} \, \tilde{d}_{k \alpha} \, e^{-i \Phi_{k \alpha} (x_{\varphi})\, + \, i  \Phi_{k \alpha} (x_{\varphi'})} , \label{ddt} \\
\frac{\tilde{d}_{k \alpha}}{N_{k \alpha}}=& - \psi_{k \alpha} (x_{\varphi}) \, \psi'_{-k, \alpha}  (x_{\varphi'}) + \psi'_{k \alpha} (x_{\varphi}) \, \psi_{-k, \alpha}  (x_{\varphi'}) \nonumber \\
& - 2 m \lambda \, \psi_{k \alpha} ( x_{\varphi}) \, \psi_{-k,\alpha} (x_{\varphi'})  .
\label{dt}
\end{align}
The introduced function $\tilde{d}_{k \alpha}$ plays an important role in the forthcoming analysis of the interface properties, which will be elucidated later.

The transmission and reflection coefficients form the scattering matrix
\begin{align}
    S_{k \alpha} = \left( \begin{array}{cc}
         t_{k \alpha} &  r'_{k \alpha}\\
         r_{k \alpha} & t'_{k \alpha}
    \end{array}\right),
    \label{scat_matr}
\end{align}
which is unitary (see Appendix \ref{app:unit_check} for verification). This property implies the relations
\begin{align}
    & |t_{k\alpha}|^2 + |r_{k\alpha}|^2 =1 , \label{scat_d1} \\
    & |t'_{k\alpha}|^2 + |r'_{k\alpha}|^2 =1, \label{scat_d2} \\
    & t_{k \alpha}^* \, r'_{k \alpha} + r_{k \alpha}^* \, t'_{k \alpha} =0, \label{scat_off}
\end{align}
which guarantee the normalization and orthogonality of the states \eqref{scat_state_r} and \eqref{scat_state_l}.

\subsection{Interface localized states} 
\label{subsec:interface_loc}

To find eigenstates which are exponentially localized near the interface, we make the ansatz 
\begin{align}
\psi_{\text{i}, \alpha} (x) = & \Theta (-x) \, a_{\alpha} \,  [- \psi_2^{(\varphi')} (x-L) + \psi_2^{(\varphi')} (x) \, (-1)^{\alpha}\,  e^{ \kappa L}] \nonumber \\
+&  \Theta (x) \, b_{\alpha} \, [- \psi_2^{(\varphi)} (x-L) + \psi_2^{(\varphi)} (x) \, (-1)^{\alpha} \, e^{- \kappa L}] ,
\end{align}
which uses the (unnormalized) Bloch states \eqref{eig_fun4}
\begin{align}
  - \psi_2^{(\varphi)} (x-L) + \psi_2^{(\varphi)} (x) \, (-1)^{\alpha} \, e^{\mp \kappa L} \propto \psi_{\frac{\alpha \pi}{L} \pm i\kappa}^{(\varphi)} (x)
\end{align}
with complex $k= \frac{\alpha \pi}{L}\pm i \kappa$ and $\kappa>0$.
Applying the matching conditions \eqref{match1}, \eqref{match2} we obtain the equations
\begin{align}
& b_{\alpha} \,  \psi_2^{(\varphi)} (L) =  a_{\alpha} \, \psi_2^{(\varphi')} (L) , \label{InterMatchEqn1}\\
& b_{\alpha}  \left[-\psi_1^{(\varphi)} (L)  +(-1)^{\alpha} \, e^{-\kappa L} - 2 m \lambda \,   \psi_2^{(\varphi)} (L) \right] \nonumber \\
&=  a_{\alpha} \,  [-\psi_1^{(\varphi')} (L) +(-1)^{\alpha} \, e^{ \kappa L}] .
\end{align}

\begin{figure*}
    \centering
    \includegraphics[width=\textwidth]{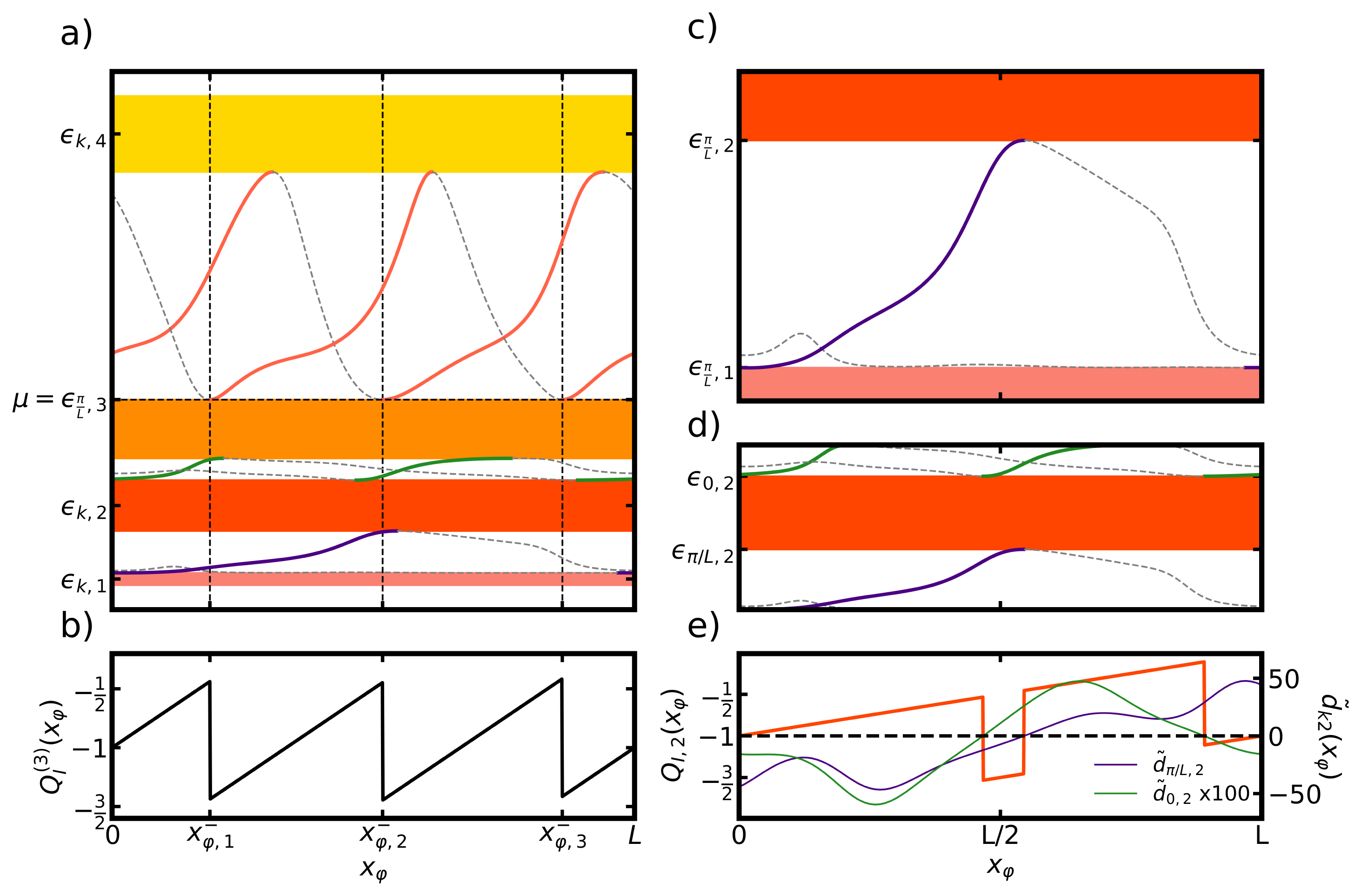}
    \caption{a) Energy spectrum of the interface eigenvalue problem for the potential shown in Fig.~\ref{fig:MismatchPotential} with $\lambda=1$. The solid lines in the bandgaps represent energies of the interface localized states. b) Total interface charge for a system with the chemical potential $\mu=\epsilon_{\frac{\pi}{L},3}$, i.e. lying on the top of the third band. Downward jumps occur whenever the localized states leave the topmost occupied band edge. c) Close-up of the first gap: The continuous curve connecting both physical ($\kappa>0$) and unphysical ($\kappa<0$) solutions of \eqref{tilde_d_eq} exhibits the double period $2 L$. d) Close-up of the second band. e) Second band's contribution to the interface charge. Jumps occur in both downward and upward directions at the touching points (which are the roots of the functions $\tilde{d}_{\frac{\pi}{L},2} (x_{\varphi})=0$ and $\tilde{d}_{0,2} (x_{\varphi})=0$) depending on the direction of the localized state's spectral flow.}
    \label{fig:L1PlotQuadruple}
\end{figure*}

The condition for the existence of a nontrivial solution yields the equation 
\begin{align}
   & \psi_2^{(\varphi')} (L)  \left[-\psi_1^{(\varphi)} (L)  +(-1)^{\alpha} \, e^{-\kappa L} - 2 m \lambda  \, \psi_2^{(\varphi)} (L) \right] \nonumber \\
&=   \psi_2^{(\varphi)} (L) \,  [-\psi_1^{(\varphi')} (L)  +(-1)^{\alpha} \, e^{ \kappa L}].
\label{eq_kappa_inter}
\end{align}
Noticing that by virtue of \eqref{psi_phi_0}, \eqref{psi_phi_prime_0}, and \eqref{phase_elim_rel} it holds
\begin{align}
    \frac{-\psi_1^{(\varphi)} (L)  +(-1)^{\alpha} \, e^{\mp\kappa L} }{\psi_2^{(\varphi)} (L)} = \frac{\psi'_{k \alpha} (x_{\varphi})}{\psi_{k \alpha} (x_{\varphi})} \bigg|_{k = \frac{\alpha \pi}{L} \pm i\kappa},
    \label{def_log_der}
\end{align}
we cast \eqref{eq_kappa_inter} to the form
\begin{align}
    \left[ \frac{\psi'_{k \alpha} (x_{\varphi})}{\psi_{k \alpha} (x_{\varphi})}  - \frac{\psi'_{-k, \alpha} (x_{\varphi'})}{\psi_{-k, \alpha} (x_{\varphi'})}\right]_{k = \frac{\alpha \pi}{L} + i\kappa} - 2 m \lambda =0,
    \label{loc_inter2}
\end{align}
which is equivalent to 
\begin{align}
\tilde{d}_{\frac{\alpha \pi}{L} + i\kappa, \alpha} =0.
\label{tilde_d_eq}
\end{align}
In addition, we have the relations between $\kappa$ and the localized state's energy $\epsilon_{\text{i},\alpha}$
\begin{align}
    (-1)^{\alpha} \, \cosh \kappa L = D^{(\varphi)} (\epsilon_{\text{i},\alpha})=D^{(\varphi')} (\epsilon_{\text{i},\alpha}) = D (\epsilon_{\text{i},\alpha}) ,
\end{align}
where $D^{(\varphi)} (E)$ is defined in \eqref{DE_phi}, and in \eqref{DE_phi_indep} it is shown to be independent of $x_{\varphi}$. Note that to solve \eqref{loc_inter2} for $\kappa >0$ we have to replace $e^{\pm \kappa L}$ with $|D| \pm \sqrt{D^2-1}$.

\begin{figure*}
    \centering
    \includegraphics[width=\textwidth]{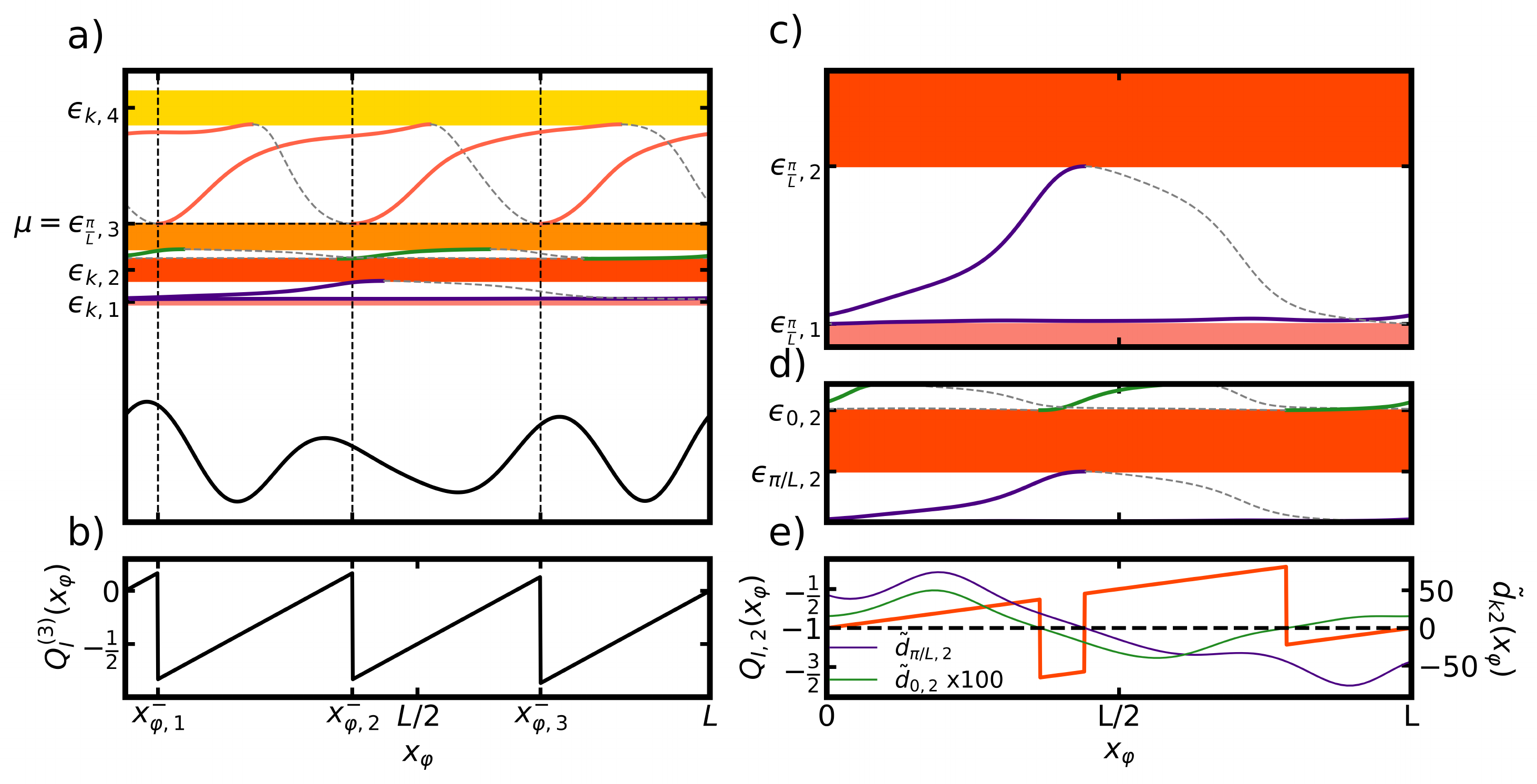}
    \caption{Same as in Fig.~\ref{fig:L1PlotQuadruple} for $\lambda=-1$.}
    \label{fig:LM1PlotQuadruple}
\end{figure*}

From the normalization condition 
\begin{align}
    \int_{-\infty}^{\infty}dx \,\,|\psi_{\text{i}, \alpha} (x)|^2  =1
\end{align}
we derive the equation for $|a_{\alpha}|^2$:
\begin{align}
    \frac{1}{|a_{\alpha}|^2} &=\sum_{n=1}^{\infty} e^{- 2 n \kappa L} \\
    & \times \int_{0}^L dx \left[-\psi_2^{(\varphi')}(x-L)+\psi_2^{(\varphi')}(x) \, (-1)^\alpha \, e^{\kappa L}\right]^2  \nonumber \\ 
    &  + \left(\frac{\psi_2^{(\varphi')}(L)}{\psi_2^{(\varphi)}(L)}\right)^2  \sum_{n=0}^{\infty} e^{- 2 n \kappa L} \nonumber \\ 
    & \times \int_0^{L} d x \left[-\psi_2^{(\varphi)}(x-L)+\psi_2^{(\varphi)}(x) \, (-1)^\alpha \, e^{-\kappa L}\right]^2 .\nonumber
\end{align}
Using the identities \eqref{pr_cross}, \eqref{pr1}, and \eqref{pr2} we then find
\begin{align}
   &  \frac{1}{|a_{\alpha} \, \psi_2^{(\varphi')}(L)|^2}  =  \frac{1}{|b_{\alpha} \, \psi_2^{(\varphi)}(L)|^2} \nonumber \\
     = &  \frac{1}{2 m} \, \frac{d}{d E} \left[  \frac{-\psi_1^{(\varphi)} (L)+(-1)^{\alpha} \,  e^{-\kappa L}}{\psi_2^{(\varphi)} (L)} \right. \nonumber \\
     & \left. \qquad \qquad - \frac{-\psi_1^{(\varphi')} (L)+(-1)^{\alpha}  \, e^{ \kappa L} }{\psi_2^{(\varphi')} (L)}   \right].  \label{InterNormal}
\end{align}
Note that the function under the derivative in the right-hand side is the same --- up to the constant $-2 m \lambda$ --- as the one in Eq. \eqref{loc_inter2}, whose roots determine interface localized states. This implies that a second-order root is impossible (since otherwise the denominators in the left-hand side of \eqref{InterNormal} becomes infinite, and the corresponding state is unnormalizable).  Hence, any two physical roots of \eqref{tilde_d_eq} may not coalesce, and on this basis we conclude that interface localized states are always nondegenerate.

The wavefunction of an exemplary interface localized state lying in the first gap is shown in Fig.~\ref{fig:MismatchPotential}. An important question is for which parameter values such a state is present, and how many of them can be accommodated in each gap? To systematically study this, we fix $x_{\varphi'}=0$ and vary the parameter $x_{\varphi}$. There are two characteristic cases: $\lambda >0$ (Fig.~\ref{fig:L1PlotQuadruple}(a,c,d)) and $\lambda<0$ (Fig.~\ref{fig:LM1PlotQuadruple}(a,c,d)). The bands of the scattering eigenstates energetically coincide with those of the bulk eigenvalue problem. The localized states dispersions in $x_{\varphi}$ are depicted by solid lines lying in the bandgaps. They are continuously prolonged by grey dashed lines which represent unphysical solutions (that is with $\kappa<0$) of the equation \eqref{tilde_d_eq}. We remark that the whole continuous curve has a double period $2L$ (see particularly close-ups of the first gap in Figs.~\ref{fig:L1PlotQuadruple}(c) and \ref{fig:LM1PlotQuadruple}(c) for a confirmation that this property holds in each gap). This is qualitatively different from the behavior of an edge state dispersion in the boundary problem [Fig.~\ref{fig:BoundaryCharge}(a,b)], which has a period $L$. The only exception from the double-period rule is the localized state residing beneath the lowest band: It has a period $L$, and there is no counterpart for this state in the boundary problem. Note that this state may occur not only for negative $\lambda$, but also for positive $\lambda$ (see additionally Figs.~\ref{fig:my_labelP}(a,c) and \ref{fig:my_labelM}(a,c) in Appendix \ref{app:add_fig} for other values of $\lambda$). It is also remarkable that, unlike in the boundary problem, there are physical dispersions which both enter and leave a band at the same band edge (as shown e.g. in the first gap in Fig.~\ref{fig:my_labelP}(c)). Overall, we may have either zero, one, or two localized states in each gap for a fixed value of $x_{\varphi}$, as well as either zero or one state beneath the lowest band.

\begin{figure}
    \centering
    \includegraphics[width=0.45\textwidth]{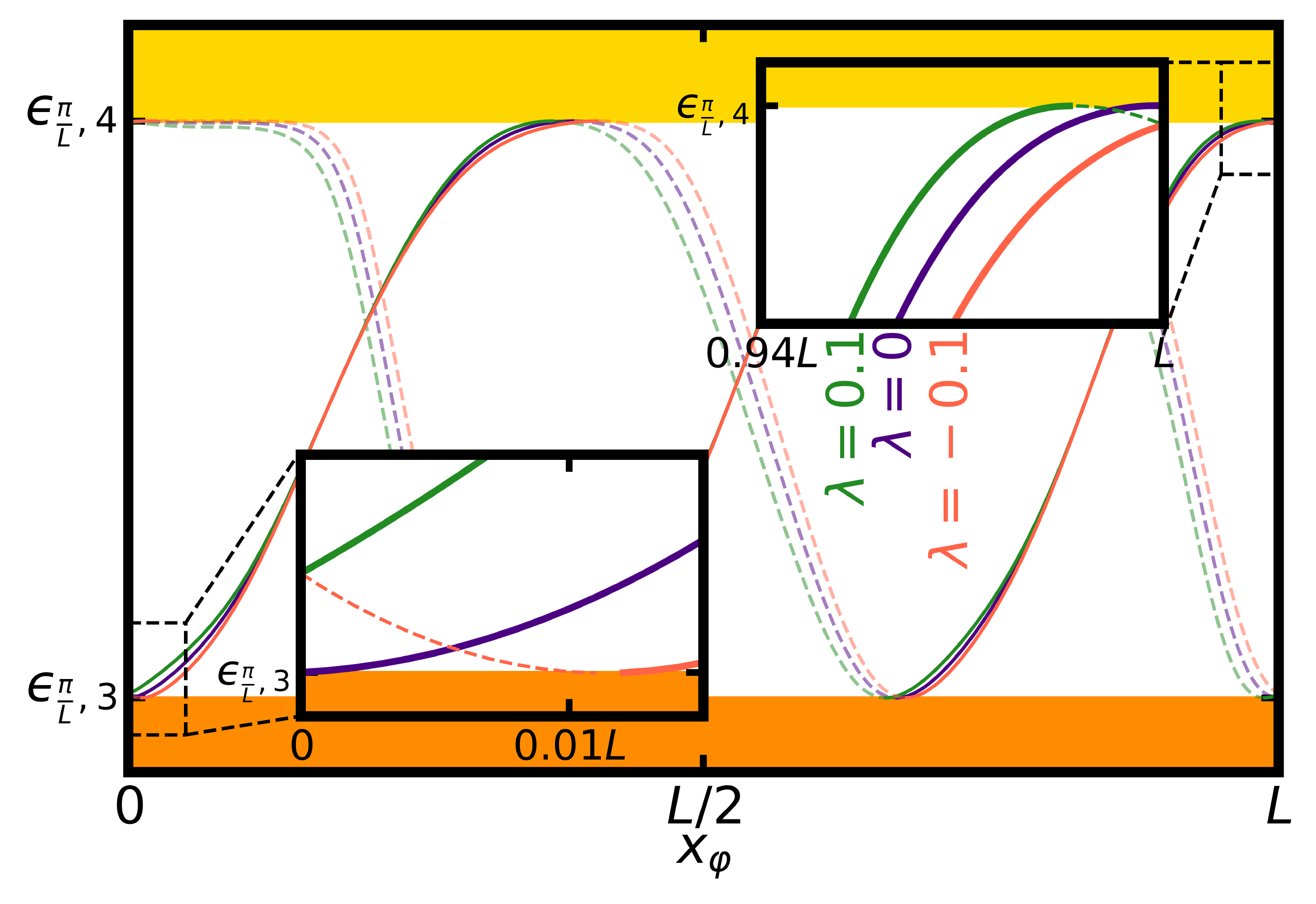}
    \caption{Localized states dispersions for $\lambda=-0.1,0,0.1$: Illustration of the nontrivial limits $\lambda \to 0^{\pm}$  close to the translationally invariant point $x_{\varphi}=x_{\varphi'}=0$ (shown is the third gap). In particular, for $\lambda \to 0^+$ there is a state in the gap at $\epsilon_{\frac{\pi}{L},3} +0^+$, while for $\lambda \to 0^-$ there is a state in the gap at $\epsilon_{\frac{\pi}{L},4} -0^+$. At $\lambda=0$ these states touch the bands.}
    \label{fig:TripleLambdaMkII}
\end{figure}

The case $\lambda =0$ is realized by the  limits $\lambda \to 0^{\pm}$, as shown in  Fig.~\ref{fig:TripleLambdaMkII}. Note that these limits are especially nontrivial close to the translationally invariant point $x_{\varphi}=x_{\varphi'}=0$. In particular, for $\lambda>0$ the localized state detaches from the top of the lower band, while for $\lambda<0$ it detaches from the bottom of the upper band.

\subsection{Interface charge: Band's contribution} 
\label{SubSec:UnivInterface}

The $\alpha$th band's contribution to the interface charge is defined by
\begin{align}
    Q_{I, \alpha} =& \int_{-\infty}^{\infty} d x \,\, f (x) \nonumber \\
    & \times \left[\int_0^{\pi/L} d k \left( |\Psi_{k\alpha}^{(r)} (x)|^2 + |\Psi_{k\alpha}^{(l)} (x)|^2\right) - \bar{\rho}_{\alpha}\right],
    \label{QI_def}
\end{align}
with a symmetric envelope function $f(-x)=f(x)$. Due to the unitarity of the scattering matrix and the property \eqref{tteq}, the average bulk charge density $\bar{\rho}_{\alpha}$ is cancelled out, and \eqref{QI_def} can be represented as a sum $Q_{I, \alpha}=Q_{I,\alpha}^R  +Q_{I,\alpha}^L$ of the two charge contributions integrated over the right and left half-space, respectively. They amount to
\begin{align}
Q_{I,\alpha}^R 
=& L \,\, \text{Re}\, \int_{-\pi/L}^{\pi/L} \frac{d k }{2 \pi} \int_0^{L} d x   \,\, r'_{k\alpha} \, \psi_{k \alpha}^{(\varphi)\, 2} (x)   \nonumber  \\
& \times \sum_{n=1}^{\infty}  e^{2 i  (k L+i 0^+) (n-1)} +Q_{P,\alpha} (x_{\varphi})
\end{align}
and
\begin{align}
Q_{I,\alpha}^L =& L \,\, \text{Re} \, \int_{-\pi/L}^{\pi/L} \frac{d k }{2 \pi} \int_{0}^{L} d x  \,\, r^*_{k \alpha} \, \psi_{k \alpha}^{(\varphi')\, 2} (x)   \nonumber \\
& \times \sum_{n=-\infty}^0 e^{2 i ( k L -i 0^+)(n-1)} - Q_{P,\alpha} (x_{\varphi'}),
\end{align}
where the polarization charge
\begin{align} 
   Q_{P,\alpha} (x_{\varphi}) =& - \frac{1}{L} \int_0^L  dx \,\, x \left( L \int_{-\pi/L}^{\pi/L} \frac{d k}{2 \pi} \,|u^{(\varphi)}_{k \alpha} (x) |^2 - \frac{1}{L} \right) \nonumber \\
   =& \frac{x_{\varphi}}{L} + Q_{P,\alpha}
   \label{QPMismOneSide}
\end{align}
is an analog of \eqref{eq:QP}. Note that the last equality follows from \eqref{u_shift} and the periodicity of $u^{(\varphi)}_{k \alpha} (x)$ in $x$.

Taking into account the special structure of the reflection coefficients \eqref{rk} and \eqref{rk_prime}, the relation \eqref{eq:QBLQBR}, and the boundary charge formula \eqref{eq:QBmain} we deduce
\begin{align} 
    Q_{I,\alpha}=& -1 + \frac{x_{\varphi} - x_{\varphi'}}{L} - \text{wn} \, [e^{i \Phi_{k \alpha} (x_{\varphi})}] + \text{wn} \,[e^{i \Phi_{k \alpha} (x_{\varphi'})}] \nonumber \\
    &+\bar{Q}_{I,\alpha},
    \label{QI_res}
\end{align}
where
\begin{align}
 \bar{Q}_{I,\alpha}  =& - L \,\, \text{Re} \, \int_{-\pi/L}^{\pi/L} \frac{d k}{2 \pi} \int_{0}^{L} d x   \nonumber \\
 \times &\frac{ \psi_{2}^{(\varphi')} (L) \, \psi_{k \alpha}^{(\varphi)\, 2} (x) \,  e^{- i k L} +\psi_{2}^{(\varphi)} (L) \,  \psi_{-k,\alpha}^{(\varphi')\, 2} (x) \,  e^{i k L}}{d_{k \alpha}}  .
 \label{QI_bar_res}
\end{align}
It can be shown that $\bar{Q}_{I,\alpha}$ is an integer contribution given in terms of the winding number of $d_{k \alpha}$:
\begin{align}
\bar{Q}_{I,\alpha} =  i\int_{-\pi/L}^{\pi/L} \frac{d k }{2 \pi} \,  \frac{1}{d_{k\alpha}} \, \frac{d}{d k} d_{k \alpha}= - \text{wn}\,  [d_{k \alpha}],
\label{QI_bar_win}
\end{align}
see Appendix \ref{app:inter_charge} for details of this derivation. By virtue of \eqref{ddt}, \eqref{dt} we  similarly obtain Eq. \eqref{QI_res} in terms of $\tilde{d}_{k\alpha}$ as
\begin{align} 
      Q_{I,\alpha}=& -1 + \frac{x_{\varphi} - x_{\varphi'}}{L} - \text{wn}\,  [\tilde{d}_{k \alpha}].
    \label{QI_res_red}
\end{align}
This is the first  main result of this section. It establishes the universal form of a  band's contribution to the interface charge. This form is similar to the one obtained for the boundary charge \eqref{eq:QBmain}. The phase of the function $\tilde{d}_{k \alpha}$ plays here a role similar to that of the phase $e^{i \Phi_{k \alpha}}$ in the boundary problem.

Without loss of generality we set $x_{\varphi'}=0$.  Then, taking into account \eqref{eig_fun4} as well as \eqref{psi_phi_0}, \eqref{psi_phi_prime_0}, we conclude 
\begin{align}
    \tilde{d}_{k\alpha} (x_{\varphi}) =&  - [-\psi_2 (x_{\varphi}-L) + \psi_2 (x_{\varphi}) e^{i k L}] \nonumber \\
    & \times [-\psi_1 (L) + e^{-i k L} +2 m \lambda \, \psi_2 (L)] \nonumber \\
&  + [-\psi'_2 (x_{\varphi}-L) + \psi'_2 (x_{\varphi}) e^{i k L}] \, \psi_2 (L).
\label{dxp_def}
\end{align}
Considering $Q_{I,\alpha}$ as a function of $x_{\varphi}$, we notice its periodicity $Q_{I,\alpha} (x_{\varphi}) = Q_{I,\alpha} (x_{\varphi}+L)$, which is due to the properties $\tilde{d}_{k\alpha} (x_{\varphi}+L) = \tilde{d}_{k\alpha} (x_{\varphi}) e^{i k L}$ and $\text{wn}\, [\tilde{d}_{k\alpha} (x_{\varphi}+L)] =\text{wn}\, [\tilde{d}_{k\alpha} (x_{\varphi})]+1$. Note that the latter follow from the definitions \eqref{dxp_def} and \eqref{wn_def} of $\tilde{d}_{k \alpha} (x_{\varphi})$ and of the winding number, respectively.

The behavior $Q_{I,\alpha=2} (x_{\varphi})$ is illustrated in Figs.~\ref{fig:L1PlotQuadruple}(e) (for $\lambda=1$) and \ref{fig:LM1PlotQuadruple}(e) (for $\lambda=-1$) as well as in the analogous panels in Appendix \ref{app:add_fig} (for other values of $\lambda$). For a detection of the touching points we plot the real-valued functions $\tilde{d}_{\frac{\pi}{L}, 2} (x_{\varphi})$ (for the bottom band edge) and $\tilde{d}_{0, 2} (x_{\varphi})$ (for the top band edge) and determine their roots. Their roots give the band's touching points at the corresponding edge. Whenever the localized state enters/leaves the band, the value of $Q_{I,\alpha=2} (x_{\varphi})$ jumps by $\pm 1$. A relation between the winding number change and the spectral flow of the localized states into/out of the bands is discussed in greater detail in the next subsection.

\subsection{Total interface charge}
\label{subsec:total_int_chrge}

The total interface charge in the system with the chemical potential $\mu_{\nu}$ at the top of the $\nu$th band is given by a sum of contributions from all states below $\mu_{\nu}$, 
 \begin{align}
     Q_I^{(\nu)} (x_{\varphi}) = \sum_{\alpha=1}^{\nu} Q_{I,\alpha} (x_{\varphi}) + \sum_{\alpha=0}^{\nu-1} Q_{\text{i}, \alpha} (x_{\varphi}).
     \label{QI_total}
 \end{align}
Here $Q_{\text{i}, \alpha} (x_{\varphi})$ is a number of the interface localized states in the gap $\alpha$. Note that in contrast to \eqref{Tot_BC_Init} the summation in the second term of \eqref{QI_total} begins from $\alpha=0$ ("zeroth gap", i.e. the energy range beneath the lowest band), since an  interface localized can occur there as well.

To establish the $x_{\varphi}$ dependence of $Q_I^{(\nu)}$, we first derive $ Q_I^{(\nu)} (0)$ at finite $\lambda$, that is the zero phase mismatch result. Then, exploiting the continuity of the spectral flow with $x_{\varphi}$, we address the sought-after dependence.

In the zero mismatch limit $x_{\varphi}=0$, the bands' contributions are evaluated in terms of
\begin{align}
    \tilde{d}_{k\alpha} (0) =   2 \, \psi_2 (L)  \, [  i \sin kL - m \lambda \, \psi_2 (L)] .
    \label{dxp_zero}
\end{align}
This expression immediately follows from \eqref{dxp_def} after applying the defining relations of the functions $\psi_1$ and $\psi_2$, stated after Eq.~\eqref{lin_comb}, and the properties \eqref{LminusL1}, \eqref{LminusL2}.
Since the function $\psi_2 (L)= \psi_2 (L, \epsilon_{k \alpha})$ does not change its sign within the band $\alpha$ [see \eqref{sign_psi2}], the complex function $\tilde{d}_{k \alpha} (0)$ does not wind with $k$ around the origin, leaving $\text{wn}\,  [\tilde{d}_{k \alpha} (0)]=0$. Then from \eqref{QI_res_red} it follows that at $x_{\varphi} = x_{\varphi'}=0$ each band provides the $-1$ contribution to the interface charge, that is $Q_{I,\alpha} (0)=-1$.

To compute $Q_I^{(\nu)} (0)$ by means of \eqref{QI_total}, we also need to establish the number of localized states. At $x_{\varphi} = x_{\varphi'}=0$ they obey the equation
\begin{align}
            -(-1)^{\alpha} m \lambda \, \psi_2 (L) =\sinh \kappa L,
            \label{loc_inter_zm}
\end{align}
which follows from $\tilde{d}_{k \alpha} (0)=0$ [note that $\psi_2 (L)=0$ does not give a nontrivial solution inside a gap away from bands' edges]. It is convenient to rewrite \eqref{loc_inter_zm} as
\begin{align}
            -m \lambda \, \frac{\psi_2 (L)}{D(E)} =\frac{\sqrt{D^2 (E)-1}}{|D (E)|}.
            \label{inter_state_no_mismatch}
\end{align}
According to \eqref{sign_psi2} and \eqref{sign_D_dpsi2}, the function in the left-hand side is monotonically decreasing across the gap from a positive to a negative value for $\lambda>0$, and it is monotonically increasing from a negative to a positive value for $\lambda<0$. In turn, the function in the right-hand side is positive within the gap and vanishes at its ends. Since both sides of \eqref{inter_state_no_mismatch} are continuous in $E$, they have a single intersection point for any $\lambda \neq 0$, which gives the energy of the interface localized state. Thus, each gap hosts one localized state.

It is necessary to additionally inspect the "zeroth gap", i.e. the energy range below the first band. The right-hand side of \eqref{inter_state_no_mismatch} is monotonically decreasing
 from $+1$ (at $E=-\infty$) down to $0^+$. Remarking that
 \begin{align} \label{zero_gap_sol_form}
     \psi_2 (L) \stackrel{E \to -\infty}{\approx} \frac{\sinh (L \sqrt{2 m |E|})}{\sqrt{2 m |E|}},
 \end{align}
 we conclude that the left-hand side of \eqref{inter_state_no_mismatch} for $\lambda>0$ is monotonically decreasing from $0^-$ (at $E=-\infty$) to a negative value, and for $\lambda<0$ it is monotonically increasing from $0^+$ (at $E=-\infty$) to a positive value. Thus, only in the latter case we obtain an additional localized solution. It is a remnant of the bound state in the attractive ($\lambda <0$) delta-potential and the flat background potential, which has the energy $E=-\frac{m \lambda^2}{2}$ [note that this result is easily recovered from \eqref{inter_state_no_mismatch} with help of \eqref{zero_gap_sol_form}].

Summarizing the above results in the zero mismatch limit $x_{\varphi}=0$, we obtain
 \begin{align} \label{TotICNoMism}
     Q_I^{(\nu)} (0) = \begin{cases} -1, & \lambda> 0, \\ 0, & \lambda <0. \end{cases}
 \end{align}
Next, we investigate how these universal values change with $x_{\varphi}$.

\begin{figure}[t]
    \centering
    \includegraphics[width=\columnwidth]{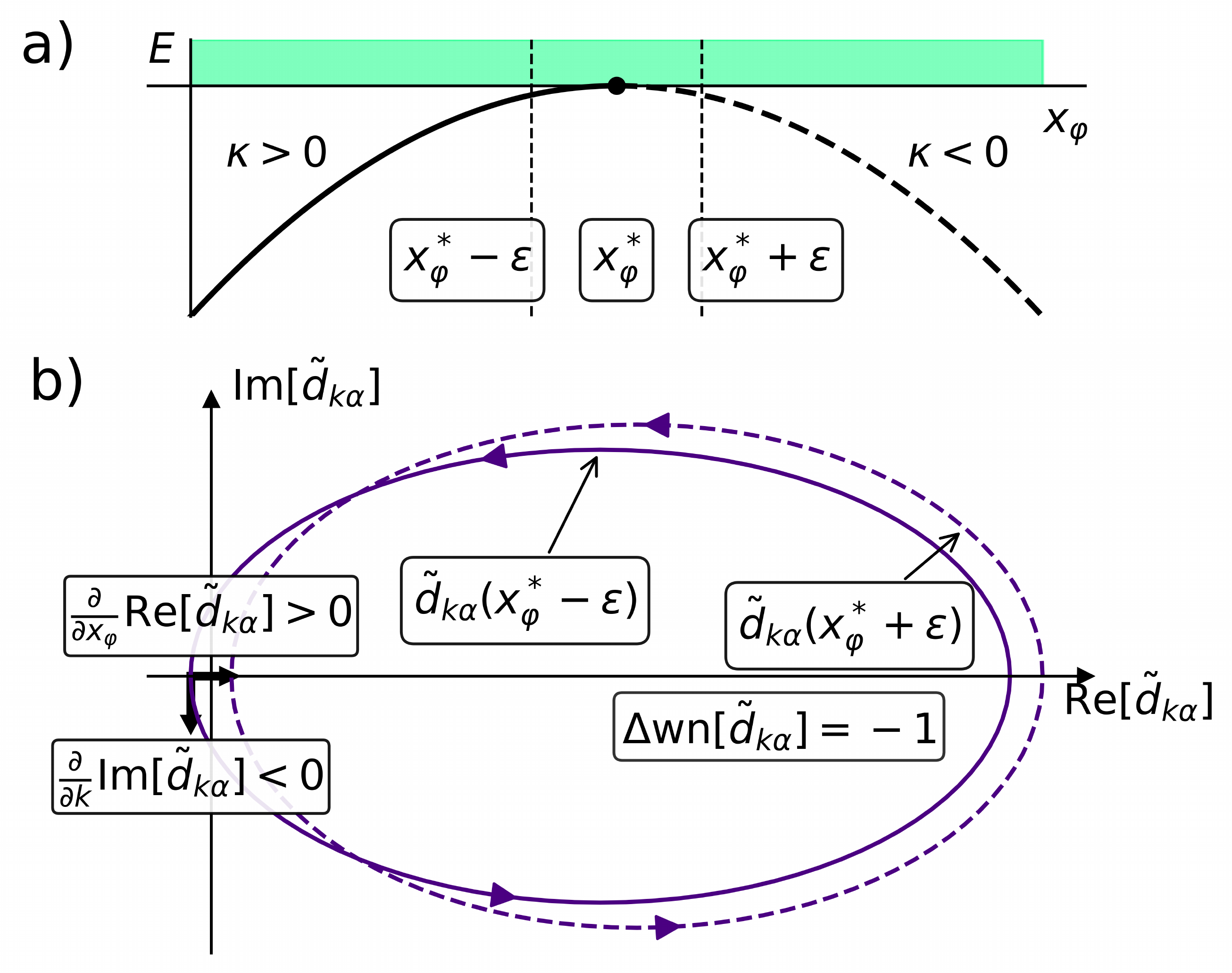}
    \caption{A sketch of the proof of \eqref{delta_wn}. a) Fixing notations for the present and absent localized state before and after the touching point, respectively. b) The winding number changes upon traversing the touching point:  The contour of $\tilde{d}_{k\alpha}(x_\varphi)$ is swept across the origin. The precise value of the change ($+1$ or $-1$) depends on a mutual orientation of the normal and the tangent vectors to the contour in the vicinity of the origin.  }
    \label{fig:delta_wn}
\end{figure}

Moving along the localized state dispersion in $x_{\varphi}$ [Fig.~\ref{fig:delta_wn}(a)], we establish that at the touching point $x_{\varphi}^*$ [obeying the equation $\tilde{d}_{k \alpha}  (x_{\varphi}^*)=0$ with $\text{Re} \, k =0, \pi/L$ and $\kappa=0$] 
it holds 
\begin{align}
   \frac{d E (x_{\varphi}^*)}{d x_{\varphi}} = (-1)^{\alpha} L \, \sinh \kappa L \, \frac{d \kappa (x_{\varphi}^*)}{d x_{\varphi}} =0 ,
\end{align}
 as well as
 \begin{align}
     &\frac{d \kappa (x_{\varphi}^*)}{d x_{\varphi}} = - \frac{\frac{\partial \tilde{d}_{k \alpha}}{\partial x_{\varphi}}}{\frac{\partial \tilde{d}_{k \alpha}}{\partial \kappa}} \nonumber \\
     &=   \frac{(-1)^{\alpha} \frac{\partial \tilde{d}_{k \alpha}}{\partial x_{\varphi}}}{ L \psi_2 (L) [\psi_1 (x_{\varphi}^*)+ \psi'_2 (x_{\varphi}^*) -2 m \lambda \, \psi_2 (x_{\varphi}^*)]}.
     \label{vort_touch_inter}
 \end{align}
 
On the other hand, looking at the touching point from the band's perspective, we notice that the winding number of $\tilde{d}_{k \alpha}$ associated with the band $\alpha$ changes upon going across the touching point by
\begin{align}
    \Delta \, \text{wn} \, [\tilde{d}_{k \alpha}] &= \text{wn} \, [\tilde{d}_{k \alpha} (x_{\varphi}^* + \varepsilon)]-\text{wn} \, [\tilde{d}_{k \alpha} (x_{\varphi}^*-\varepsilon)] \nonumber \\
    &=  \text{sign} \, \left[ \frac{\partial \tilde{d}_{k \alpha}}{\partial x_{\varphi}} \, \frac{\partial \, \text{Im} \, \tilde{d}_{k \alpha}}{\partial k}\right],
    \label{delta_wn}
\end{align}
where
\begin{align}
    \frac{\partial \, \text{Im} \, \tilde{d}_{k \alpha}}{\partial k}  =& (-1)^{\alpha} L \psi_2 (L) \nonumber \\
    & \times [\psi_1 (x_{\varphi}^*)+ \psi'_2 (x_{\varphi}^*) -2 m \lambda \, \psi_2 (x_{\varphi}^*)].
    \label{dd_dkappa}
\end{align}
The derivation of \eqref{delta_wn} is sketched in Fig.~\ref{fig:delta_wn}(b). For $k =0, \pi/L$ the function $\tilde{d}_{k \alpha}$ becomes real. These values correspond to points on the real axis, where the parametric plot of $\tilde{d}_{k\alpha}$ (or the loop) intersects the horizontal axis. Moving one of these points (say, the left one, as shown on the sketch) across the origin (where the touching point condition is satisfied) by means of tuning $x_{\varphi}$ leads to a change in the winding number. This change (by either $+1$ or $-1$) depends on the orientation of the tangent vector to the loop in the vicinity of the origin. A careful inspection of all possible scenarios leads us to the expression \eqref{delta_wn}.

By analogy with \eqref{vort_def} we introduce the \textit{interface vorticity} as a sign of \eqref{vort_touch_inter}. It quantifies whether the localized interface state leaves the band (the value $+1$ indicates the change of $\kappa$ from negative to positive) or enters the band (the value $-1$ indicates the change of $\kappa$ from positive to negative). Comparing \eqref{vort_touch_inter} with \eqref{delta_wn}, \eqref{dd_dkappa} we show that this vorticity equals
\begin{align}
    \text{sign} \, \left[ \frac{d \kappa_{\alpha} (x_{\varphi}^*)}{d x_{\varphi}} \bigg|_{\alpha} \right] = \Delta \, \text{wn} \, [\tilde{d}_{k \alpha}].
    \label{bbc_if1}
\end{align}
Here in the left-hand side we additionally mark that the localized state in the gap $\alpha$ (encoded in $\kappa_{\alpha}$) touches the band $\alpha$ (encoded in the lowest subscript $\alpha$). Analogously we show that upon touching the upper band it holds
\begin{align}
    \text{sign} \, \left[ \frac{d \kappa_{\alpha} (x_{\varphi}^*)}{d x_{\varphi}} \bigg|_{\alpha+1} \right] = \Delta \, \text{wn} \, [\tilde{d}_{k, \alpha+1}].
    \label{bbc_if2}
\end{align}

The expressions \eqref{bbc_if1} and \eqref{bbc_if2} have a transparent physical meaning: When the localized state enters/leaves the band, the band's contribution to the total interface charge changes accordingly, so that there is no net contribution to $Q_I^{(\nu)}$. The only uncompensated changes occur at the topmost occupied band edge (since we put there the chemical potential and therefore do not count the localized states touching that band edge). Hence we obtain the following expression for the total interface charge
\begin{align} \label{ICTotFormula}
    Q_I^{(\nu)} (x_{\varphi}) =& Q_I^{(\nu)} (0)+ \frac{\nu}{L} x_{\varphi}\nonumber \\
    &- \sum_{j=1}^{\nu+l} \Theta (x_{\varphi,j}^{-}) + \sum_{j=1}^{l} \Theta (x_{\varphi,j}^{+}).
\end{align}
This is the second main result of this section: It states the universal dependence of the total interface charge on the pumping parameter $x_{\varphi}$.
It is similar to \eqref{TotBC_Rewrit}, the main difference being that the localized state energy may now both enter and leave a band at the same band edge. Here $\{x_{\varphi,j}^{\pm} \}$ is a (model-specific) set of points, where the interface localized state enters/leaves the topmost occupied band edge. The difference between the numbers of $x_{\varphi,j}^-$ and $x_{\varphi,j}^+$ must be exactly equal to $\nu$: This is needed to ensure the periodicity $Q_I^{(\nu)} (x_{\varphi}+L)=Q_I^{(\nu)} (x_{\varphi})$. 

The behavior $Q_{I}^{(3)} (x_{\varphi})$ for the three occupied bands is illustrated in Figs.~\ref{fig:L1PlotQuadruple}(b) (for $\lambda=1$) and \ref{fig:LM1PlotQuadruple}(b) (for $\lambda=-1$) as well as in the analogous panels in Appendix \ref{app:add_fig} for the values $\lambda = \pm 0.1$, which additionally give an idea about the behavior of $Q_{I}^{(3)} (x_{\varphi})$ close to $\lambda=0$.

\section{Summary}
\label{sec:Summary}

In this work, we have studied universal properties of the boundary change $Q_B (x_{\varphi})$ in one-dimensional single-channel continuum insulator models. We have found a rigorous proof of the linear dependence of $Q_B$ on the phase $x_{\varphi}$ of the periodic potential modulation. In continuum models, this result appears in an undisguised form: All nonuniversal $\frac{2\pi}{Za}$-periodic contributions previously found in the lattice model considerations [\onlinecite{pletyukhov_etal_prb_20}] are suppressed in the continuum limit $Z \to \infty$ (where $Z$ is a number of sites in a unit cell, and $a$ is a lattice spacing). 

The total boundary charge consists of both band contributions and potentially edge states  residing in the band gaps. We have extended the earlier findings [\onlinecite{thakurati_etal_18},\onlinecite{pletyukhov_etal_prb_20}] for both the individual band contributions $Q_{B,\alpha}$ and the total boundary charge to the class of continuum models studied here. In particular, in the first case the slope of the linear dependence is given by the first Chern index $C_{1,\alpha}$ of band $\alpha$, which acquires the same value $C_{1,\alpha} =1$ for each band. This value is in tune with the difference between the numbers $M_{\alpha}^{(\mp)}$ of the touching points, at which the edge state dispersions (evolved in $x_{\varphi}$) leave and enter the band. We have shown that over one pumping cycle $x_{\varphi} \to x_{\varphi}+L$ (i.e. pushing the periodic potential towards the boundary over one period), in continuum models, there are exactly $\alpha$ edge states which leave the band $\alpha$ from its top side, and $\alpha-1$ edge states which enter the band from its bottom side. At the corresponding touching points, $Q_{B,\alpha}$ acquires discontinuous jump contributions $\pm 1$ in such a way that the periodicity $Q_{B, \alpha} (x_{\varphi}+L) = Q_{B, \alpha} (x_{\varphi})$ is maintained. Thus, the equality $C_{1,\alpha} = M_{\alpha}^{(-)}- M_{\alpha}^{(+)}$ is not only a manifestation of the bulk-boundary correspondence, but also an expression for charge conservation.

An analogous result has been obtained for the total boundary charge $Q_{B}^{(\nu)}$ in a system with $\nu$ fully occupied bands: The slope $\nu$ (with respect to $\frac{x_{\varphi}}{L}$)  of the linear dependence is compensated by the occurrence of exactly $\nu$ edges states removing a single-electron charge during one pumping cycle. 

Studying the boundary charge fluctuations, we have confirmed the earlier stated surface fluctuation theorem [\onlinecite{weber_etal_prl_20},\onlinecite{wannier_paper}], which expresses the universal low-energy $1/E_g$ scaling of these fluctuations with the gap size $E_g$.

The other interesting facet of the excess charges studied in the present work is the interface charge $Q_I$ and its universal properties. Bringing two semi-infinite systems with different phases $x_{\varphi} \neq x_{\varphi'}$ of the periodic potential modulation into contact and introducing an additional delta-potential barrier between them, we have revealed similar universal dependencies of $Q_I$ on $x_{\varphi}$ at fixed $x_{\varphi'}$, i.e., a slope 1 for the single band contribution and a slope $\nu$ for the total interface charge (in a system with $\nu$ occupied bands). To restore the periodicity $Q_I (x_{\varphi})= Q_I (x_{\varphi}+L)$, the linear growth is accompanied by discontinuous jumps $\pm 1$ at the points where the interface localized state enters/leaves the band. Thereby, the required charge conservation is respected. However, in interface models there is no analog of the bulk-boundary correspondence, since the spectral flow of the interface localized states is in accord with changes in winding numbers of a novel function called $\tilde{d}_{k\alpha}$, and not with those of the phases of the Bloch states. The function $\tilde{d}_{k \alpha}$ which we introduced not only represents a specific combination of the Bloch states in the bulk of both right and left subsystems, but also contains information about their interface. It connects to the denominator of the scattering matrix, and thus inherits the analytic features of the reflection coefficient. We have elucidated the tremendous role of the function $\tilde{d}_{k \alpha}$ in describing the universal properties of the interface charge; in particular, the contribution from its winding number to $Q_I$ and the equation $\tilde{d}_{k \alpha}=0$ to determine interface localized states. The integer-valued changes of $Q_I$ analysed in our study, which are associated with the winding number changes and an emergence/cease of localized states, serve as a manifestation of the nearsightedness principle [\onlinecite{NearSightednessPrin}].

A perspective extension of the present work consists in studying a multichannel generalization of  the continuum one-dimensional insulator models and revealing similar dependencies for both boundary and interface charges [\onlinecite{mueller_etal_2021}].

\section{Acknowledgments}
\label{sec:Acknowledgments}

We appreciate a lasting fruitful exchange of ideas on the boundary charge subject with J. Klinovaja and D. Loss. The work was supported by the Deutsche Forschungsgemeinschaft via RTG 1995.

\begin{appendix}

\section{Properties of the Hill equation}
\label{app:hill_eq}

To prove the validity of  \eqref{psi2L}, we notice that both left- and right-hand sides  this relation satisfy \eqref{schr_eq} and have the same initial conditions at $x=L$
\begin{align}
\psi_1 (L) \, \psi_2 (L) - \psi_1 (L) \, \psi_2 (L) &= 0 = - \psi_2 (0), \\
\psi'_1 (L) \, \psi_2 (L) - \psi_1 (L) \, \psi'_2 (L) &= -1 = - \psi'_2 (0).
\end{align}
Here the second identity holds on the basis of \eqref{wron}.
Since for given initial conditions a solution of a differential equation is unique, the functions on the both sides of \eqref{psi2L} appear to be  identical.

Analogously we prove the relations
\begin{align}
\psi_1 (x-L)  &=  \psi_2 (x) \, \psi'_1 (-L)  +  \psi_1 (x) \, \psi_1 (-L) , \label{psi1L} \\
\psi_1 (x+L) &=  \psi_2 (x) \, \psi'_1 (L)  +  \psi_1 (x) \, \psi_1 (L), 
\label{psi1bL} \\
\psi_2 (x+L) &=  \psi_2 (x)\, \psi'_2 (L)  +  \psi_1 (x) \, \psi_2 (L), 
\label{psi2bL}
\end{align}
and remark that together with \eqref{psi2L} they imply the following consequences
\begin{align}
\psi_2 (-L) &= - \psi_2 (L), \label{LminusL1} \\
\psi'_2 (-L) &=  \psi_1 (L), \label{LminusL2} \\
\psi_1 (-L) &=  \psi'_2 (L),  \label{LminusL3} \\
\psi'_1 (-L) &= - \psi'_1 (L).\label{LminusL4}
\end{align}

To prove \eqref{psi_int}, we consider the following equations
\begin{align}
- \frac{1}{2m} \frac{\partial \psi''_2 (x)}{\partial E} + [V (x) - E] \frac{\partial \psi_2 (x)}{\partial E} &= \psi_2 (x), \label{eqdE} \\
- \frac{1}{2m} \frac{\partial \psi''_2 (x-L)}{\partial E} + [V (x) - E] \frac{\partial \psi_2 (x-L)}{\partial E} &= \psi_2 (x-L).
\end{align}
Multiplying the first equation with $\frac{\partial \psi_2 (x-L)}{\partial E}$, the second equation with $\frac{\partial \psi_2 (x)}{\partial E}$, and subtracting the obtained results from each other, we find
\begin{align}
& -2 m \, F (x) \nonumber \\
&= \frac{\partial \psi_2 (x-L)}{\partial E} \, \frac{\partial \psi''_2 (x)}{\partial E} - \frac{\partial \psi_2 (x)}{\partial E} \, \frac{\partial \psi''_2 (x-L)}{\partial E} .
\end{align}
Integrating both sides over $x$ from $0$ to $L$ and applying the integration by parts, we obtain
\begin{align}
& -2 m\int_0^L d x \,\, F (x) \nonumber \\
& =\left[ \frac{\partial \psi_2 (x-L)}{\partial E} \, \frac{\partial \psi'_2 (x)}{\partial E} - \frac{\partial \psi_2 (x)}{\partial E} \, \frac{\partial \psi'_2 (x-L)}{\partial E} \right]_0^L  .
\end{align}
But the right-hand side equals zero, since
\begin{align}
 \frac{\partial \psi_2 (0)}{\partial E}  = \frac{\partial \psi'_2 (0)}{\partial E}  =0
\end{align}
(we recall that $\psi_2 (0) =0$ and $\psi'_2 (0) =1$ are constants). This proves  the identity \eqref{psi_int}.

Below we list further properties of the Hill's equation.

Multiplying \eqref{eqdE}  with $\psi_2 (x-L)$ and integrating  over $x$ from $0$ to $L$ (also using the integration by parts), we obtain
\begin{align}
 & \frac{1}{2m} \, \psi'_2 (x-L) \, \frac{\partial \psi_2 (x)}{\partial E}  \, \bigg|_0^L - \frac{1}{2m} \, \psi_2 (x-L) \, \frac{\partial \psi'_2 (x)}{\partial E} \, \bigg|_0^L \nonumber \\
 &= \int_0^L d x \,\, \psi_2 (x-L) \, \psi_2 (x) .
\end{align}
It follows
\begin{align}
 \frac{1}{2m}\, \frac{\partial \psi_2 (L)}{\partial E} = \int_0^L d x \,\, \psi_2 (x-L) \, \psi_2 (x) .
 \label{pr_cross}
\end{align}

In the analogous manner we derive
\begin{align}
 \frac{1}{2m} \left[ \psi'_2 (x) \, \frac{\partial \psi_2 (x)}{\partial E}  -  \psi_2 (x) \, \frac{\partial \psi'_2 (x)}{\partial E} \right]_0^L = \int_0^L d x \,\, \psi_2^2 (x) 
\end{align}
and
\begin{align}
  & \frac{1}{2m} \left[\psi'_2 (x-L) \, \frac{\partial \psi_2 (x-L)}{\partial E}  - \psi_2 (x-L) \, \frac{\partial \psi'_2 (x-L)}{\partial E} \right]_0^L 
  \nonumber \\
  &= \int_0^L d x \,\, \psi_2^2 (x-L) ,
\end{align}
giving
\begin{align}
 \frac{1}{2m} \left[ \psi'_2 (L) \frac{\partial \psi_2 (L)}{\partial E}   -   \psi_2 (L) \frac{\partial \psi'_2 (L)}{\partial E} \right] &= \int_0^L d x \,\, \psi_2^2 (x) , \label{pr1}
\end{align}
and
\begin{align}
 \frac{1}{2m} \left[ \psi_1 (L) \frac{\partial \psi_2 (L)}{\partial E}  -  \psi_2 (L) \frac{\partial \psi_1 (L)}{\partial E} \right] = \int_0^L d x \, \psi_2^2 (x-L) , \label{pr2}
\end{align}
respectively.

Summing up \eqref{pr1} and \eqref{pr2}, we obtain
\begin{align}
& \frac{1}{m} \left[ D(E) \, \frac{\partial \psi_2 (L)}{\partial E}   -   \psi_2 (L) \, \frac{\partial D(E)}{\partial E} \right] \nonumber \\
&= \int_0^L d x \,\, \psi_2^2 (x) +  \int_0^L d x \,\, \psi_2^2 (x-L).
\label{sign_D_dpsi2}
\end{align}

On the basis of \eqref{pr_cross} and \eqref{sign_D_dpsi2} we state the Schwarz inequality
\begin{align}
 D(E) \, \frac{\partial \psi_2 (L)}{\partial E}   -   \psi_2 (L) \,\frac{\partial D(E)}{\partial E} \geq \bigg|  \frac{\partial \psi_2 (L)}{\partial E} \bigg| .
 \label{schwarz1}
\end{align}
Note that from this inequality it follows that strictly inside a band, i.e. for $-1 < D (E) < 1$, the function $D (E)$ may not have extrema, that is $\frac{\partial D(E)}{\partial E} \neq 0$, since otherwise we get the contradiction
\begin{align}
 \bigg|  \frac{\partial \psi_2 (L)}{\partial E} \bigg|  > D(E) \, \frac{\partial \psi_2 (L)}{\partial E}    \geq \bigg|  \frac{\partial \psi_2 (L)}{\partial E} \bigg| .
 \label{contradict1}
\end{align}

Similar relations can be derived for the function $\psi_1$:
\begin{align}
 &- \frac{1}{2m} \, \frac{\partial \psi'_1 (L)}{\partial E} = \int_0^L d x \,\,  \psi_1 (x-L) \, \psi_1 (x) , \\
 &\frac{1}{2m} \left[ \psi'_1 (L) \frac{\partial \psi_1 (L)}{\partial E}   -  \psi_1 (L) \frac{\partial \psi'_1 (L)}{\partial E} \right] = \int_0^L d x \, \psi_1^2 (x) , \label{pr3} \\
 & \frac{1}{2m} \left[ \psi'_1 (L) \frac{\partial \psi'_2 (L)}{\partial E} -  \psi'_2 (L) \frac{\partial \psi'_1 (L)}{\partial E}  \right]= \int_0^L d x \, \psi_1^2 (x-L) .  \label{pr4}
\end{align}

Summing up \eqref{pr3} and \eqref{pr4} we obtain
\begin{align}
 & \frac{1}{m} \left[ \psi'_1 (L) \, \frac{\partial D(E)}{\partial E}   -  D (E) \, \frac{\partial \psi'_1 (L)}{\partial E} \right] \nonumber \\
 &= \int_0^L d x \,\, \psi_1^2 (x) + \int_0^L d x \,\, \psi_1^2 (x-L) .
\end{align}

Analogously to \eqref{schwarz1} we state the other Schwarz inequality
\begin{align}
\psi'_1 (L) \, \frac{\partial D(E)}{\partial E}   -  D (E) \, \frac{\partial \psi'_1 (L)}{\partial E} \geq \bigg| \frac{\partial \psi'_1 (L)}{\partial E} \bigg| .
\label{schwarz2}
\end{align}

\section{Some properties of the Bloch states}
\label{app:sign_psi2}

Differentiating \eqref{schr_eq} for $\psi_{k\alpha} (x)$ with respect to $k$, then multiplying it with $\psi^*_{k\alpha} (x)$, and integrating over $x$ from $0$ to $L$ (also using the integration by parts), we get 
\begin{align}
2 m \, \frac{d \epsilon_{k\alpha}}{dk} =& -\psi_{k\alpha}^* (L) \, \frac{d \psi'_{k\alpha} (L)}{d k} + \psi_{k\alpha}^* (0) \, \frac{d \psi'_{k\alpha} (0)}{d k} \nonumber \\
&+ \psi^{\prime \, *}_{k\alpha} (L) \, \frac{d \psi_{k\alpha} (L)}{d k} - \psi^{\prime \, *}_{k\alpha} (0) \, \frac{d \psi_{k\alpha} (0)}{d k} \nonumber \\
=& \, 2 L \,  \text{Im}\, [\psi_{k\alpha}^* (0) \, \psi'_{k\alpha} (0) ],
\end{align}
where the last equality follows from \eqref{bc1} and \eqref{bc2}.

Expressing from \eqref{eig_fun4} 
\begin{align}
\psi_{k\alpha} (0) &=  \frac{\psi_2 (L)}{\sqrt{N_{k\alpha}}}, \\ 
\psi'_{k\alpha} (0) &= \frac{1}{\sqrt{N_{k\alpha}}} \, [e^{i k L}-\psi_1 (L)],
\end{align}
we establish 
\begin{align}
  \frac{d \epsilon_{k\alpha}}{dk}= \frac{L}{m} \, \frac{\psi_2 (L)}{N_{k\alpha}} \, \sin k L.
  \label{dEdk}
\end{align}
Differentiating \eqref{dr} with respect to $k$
\begin{align}
    -L \, \sin k L = \frac{\partial D (E)}{\partial E} \, \frac{d \epsilon_{k\alpha}}{d k},
    \label{dEdk_sin}
\end{align}
we also get an alternative form of \eqref{dEdk}
\begin{align}
N_{k\alpha} = - \frac{\psi_2 (L)}{m}\frac{\partial D (E)}{\partial E} = - \frac{\psi_2 (L)}{2m} \left( \frac{\partial \psi_1  (L)}{\partial E} + \frac{\partial \psi'_2 (L)}{\partial E} \right).
\label{Nk_alt}
\end{align}
Comparing it with \eqref{sign_dD} we conclude that within the band $\alpha$ the following sign
\begin{align}
    \text{sign} \, [\psi_2 (L, \epsilon_{k\alpha})] = - (-1)^{\alpha}
    \label{sign_psi2}
\end{align}
is constant and solely determined by the band index $\alpha$.

\section{Evaluation of the Chern index}
\label{app:chern_eval}

To evaluate \eqref{chern_def}, we represent
\begin{align}
& \int_0^L d x \,\,  \text{Im} \, \frac{d \psi_{k \alpha}^{(\varphi)\, *} (x)}{d k} \, \frac{d \psi_{k \alpha}^{(\varphi)} (x)}{d x_{\varphi}} \nonumber \\
=& \int_0^L d x \,\,  \text{Im} \, \frac{d \psi_{k \alpha}^{ *} (x+ x_{\varphi})}{d k} \, \psi'_{k \alpha} (x+x_{\varphi}) \label{c1_chern} \\
&-\frac{d \Phi_{k \alpha} (x_{\varphi})}{d x_{\varphi}} \, \frac12 \,  \frac{d}{d k} \int_0^L d x \,\, |\psi_{k \alpha} (x+x_{\varphi})|^2 \label{c2_chern} \\
&+\frac{d \Phi_{k \alpha} (x_{\varphi})}{d k} \, \frac12 \,\int_0^L d x \,\,  \frac{d}{d x} | \psi_{k \alpha} (x+x_{\varphi}) |^2, \label{c3_chern}
\end{align}
where we used \eqref{psi_shift} and the periodicity of $|\psi_{k \alpha} (x)|^2 = |u_{k\alpha} (x)|^2$ in $x$. Apparently, the contributions \eqref{c2_chern} and \eqref{c3_chern} identically vanish.

Transforming the contribution \eqref{c1_chern}
\begin{align}
    & \int_0^L d x \,\,  \text{Im} \, \frac{d \psi_{k \alpha}^{*} (x+x_{\varphi})}{dk} \, \psi'_{k \alpha} (x+x_{\varphi}) \nonumber \\
    =& \frac{d}{dk} \int_0^L d x \,\,  \text{Im} \,  \psi_{k \alpha}^{*} (x+x_{\varphi}) \, \psi'_{k \alpha} (x+x_{\varphi}) \nonumber \\
    &- \int_0^L d x \,\,  \text{Im} \,  \psi_{k \alpha}^{*} (x+x_{\varphi}) \, \frac{d\psi'_{k \alpha} (x+x_{\varphi}) }{dk } \\
     =& \frac{d }{dk} \int_0^L d x \,\,  \text{Im}  \,  \psi_{k \alpha}^{*} (x+x_{\varphi}) \, \psi'_{k \alpha} (x+x_{\varphi}) \nonumber \\
    &- \left[ \text{Im} \,  \psi_{k \alpha}^{*} (x+x_{\varphi}) \, \frac{d\psi_{k \alpha} (x+x_{\varphi}) }{dk} \right]_0^L \nonumber \\
    &+ \int_0^L d x \,\,  \text{Im} \,  \psi_{k \alpha}^{\prime \, *} (x+x_{\varphi}) \, \frac{d\psi_{k \alpha} (x+x_{\varphi}) }{dk},
\end{align}
we establish
\begin{align}
    & \int_0^L d x \,\,  \text{Im} \, \frac{d \psi_{k \alpha}^{*} (x+x_{\varphi})}{d k} \, \psi'_{k \alpha} (x+x_{\varphi}) \nonumber \\
    =&\frac12 \, \frac{d }{dk} \int_0^L d x \,\,  \text{Im}  \,  \psi_{k \alpha}^{*} (x+x_{\varphi}) \, \psi'_{k \alpha} (x+x_{\varphi})
    \label{c4_chern} \\
    &-\frac{L}{2} \, 
    |u_{k \alpha} (x_{\varphi})|^2. \label{c5_chern}
\end{align}
Since $\psi_{k \alpha} (x)$ is periodic in $k$, the term \eqref{c4_chern} drops out under the $k$-integration, and we get
\begin{align}
    C_{1,\alpha} =&  L \, \int_0^L d x_{\varphi} \int_{-\pi/L}^{\pi/L} 
    \frac{d k}{2 \pi} \,\, |u_{k\alpha} (x_{\varphi})|^2 =1 .
\end{align}

\section{One edge state per gap}
\label{app:one_per_gap}

Consider odd gap $\alpha$, that is $E \in [\epsilon_{\frac{\pi}{L}, \alpha}, \epsilon_{\frac{\pi}{L}, \alpha +1}] $, characterized by  $D (E) \leq -1$. According to \eqref{sign_psi2}, the function $\psi_2 (L,E)$ must change its sign across this gap from positive to negative value, and this guarantees the existence of at least one edge state per gap.

\begin{figure}
    \centering
    \includegraphics[width=0.45\textwidth]{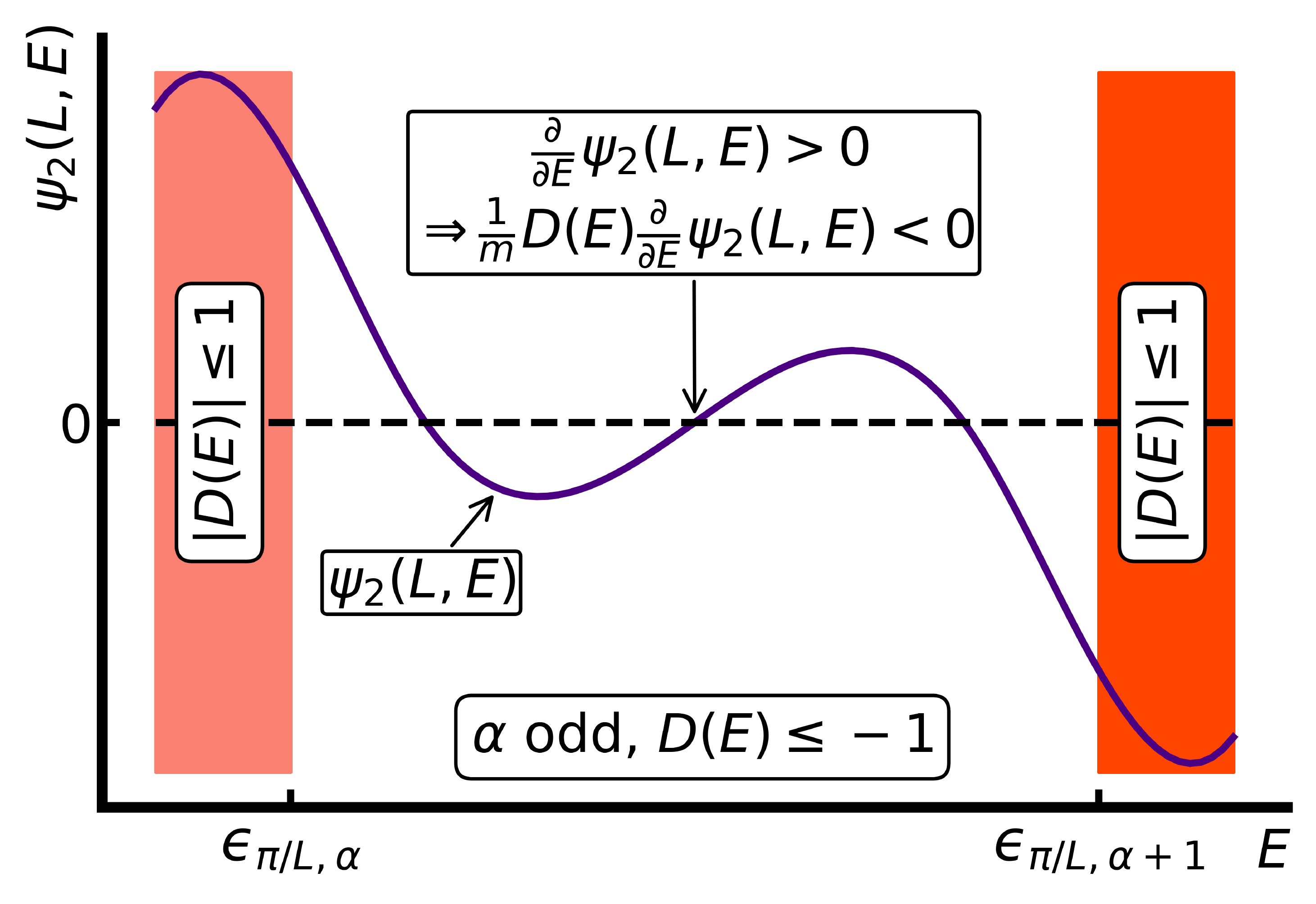}
    \caption{Illustration of the impossibility to have multiple roots of $\psi_2(L,E)$ within gaps. Assume an odd gap index $\alpha$ ($D(E)\leq-1$) which hosts an odd number of roots of $\psi_2(L,E)$. If the latter number is greater than one, there must necessarily be one root that satisfies both $\psi_2(L,E)=0$ and $\frac{\partial}{\partial E}\psi_2(L,E)>0$. Since this implies $\text{sign}[D(E)\frac{\partial}{\partial E}\psi_2(L,E)]<0$ we obtain a contradiction to \eqref{sign_D_dpsi2}. A similar argumentation holds for even gap index $\alpha$. }
    \label{fig:AppHRootIllustration}
\end{figure}

Let us show that it is impossible to have more than one edge state per gap. Suppose that we have multiple roots of $\psi_2 (L,E)$ on the indicated above interval, see Fig.~\ref{fig:AppHRootIllustration}. Because of the sign-changing property, their number must be odd, degenerate roots (that is, with $\psi_2 (L,E)= \frac{\partial}{\partial E} \psi_2 (L,E)=0$) being excluded on the basis of \eqref{sign_D_dpsi2}. This entails that at every even root we should have $\frac{\partial}{\partial E} \psi_2 (L,E) > 0$. But this contradicts to \eqref{sign_D_dpsi2}, since its left-hand side appears to be negative at even roots. Therefore, even roots are impossible, and thus we have only a single edge state per gap.

Analogously, we prove the same property for even gaps.

\section{Zeros of (anti)periodic solutions of \eqref{schr_eq}}
\label{app:zeros}

Consider a family of Hamiltonians
\begin{align}
H_{\xi} = - \frac{1}{2 m} \frac{d^2}{d x^2} + \zeta \, V (x), \quad V(x) = V (x+L),
\label{schr_xi}
\end{align}
where $\zeta$ is continuously varied on the interval $(0,1]$. Equipping the eigenvalue problem \eqref{schr_xi} with the periodic boundary conditions, we obtain periodic eigenfunctions $\psi_{l;\zeta}^{(p)} (x)= \psi_{l;\zeta}^{(p)} (x+L)$ labelled by $l=0,1,2,\ldots$, which are continuous in $\zeta$. The corresponding eigenenergies are denoted by $\epsilon_{l; \zeta}^{(p)}$. In addition, we consider the antiperiodic boundary conditions leading to the solutions $\psi_{l;\zeta}^{(a)} (x)= -\psi_{l;\zeta}^{(a)} (x+L)$ of \eqref{schr_xi} with eigenenergies $\epsilon_{l;\zeta}^{(a)}$, labelled by $l=1,2,\ldots$. 

The energies $\epsilon_{l;\zeta}^{(p)}$ and $\epsilon_{l;\zeta}^{(a)}$ are identified with the bands' edges of the eigenvalue problem \eqref{schr_eq}:
\begin{align}
      \epsilon_{k=0, \alpha;\zeta} &=\epsilon_{\alpha-1;\zeta}^{(p)} , \\
      \epsilon_{k=\frac{\pi}{L},\alpha;\zeta} &=\epsilon_{\alpha;\xi}^{(a)},
\end{align}
where the band index $\alpha \geq 1$. Analogously we identify the corresponding eigenfunctions
\begin{align}
      \psi_{k=0, \alpha;\zeta} (x) &=\psi_{\alpha-1;\zeta}^{(p)} (x) , 
      \label{rel1} \\
      \psi_{k=\frac{\pi}{L},\alpha;\zeta} (x) &=\psi_{\alpha;\zeta}^{(a)} (x).
      \label{rel2}
\end{align}

The main observation is that a number of zeros of either $\psi_{l;\zeta}^{(p)} (x)$ or $\psi_{l;\zeta}^{(a)} (x)$ on the interval $x \in [ 0, L)$   does not depend on the value of $\zeta$. Indeed, to change a number of zeros of a continuous eigenfunction we would need to make its two nearby roots coalesce at some value $\zeta^*$ and point $x^*$. Thereby we get a second-order root with $\psi_{l;\zeta^*}^{(p,a)} (x^*)=\psi^{(p,a) \, \prime}_{l;\zeta^*} (x^*)=0$. But this implies that $\psi_{l;\zeta^*}^{(p,a)} (x) \equiv 0$. This is impossible, since there must be a nontrivial solution corresponding to the energy $\epsilon_{l; \zeta^*}^{(p,a)}$. 

Let us find numbers of zeros of the periodic functions $\psi_{\alpha;\zeta}^{(p)} (x)$
 and $\psi_{\alpha;\zeta}^{(a)} (x)$ for infinitesimal $\zeta \ll 1$. 
 
 In this limit, the eigenstates $\psi_{\alpha-1;\zeta}^{(p)} (x)$ and $\psi_{\alpha; \zeta}^{(p)} (x)$ with even $\alpha$ become nearly degenerate with the energy $\epsilon_{\alpha}^{(0)} = \frac{\alpha^2 \pi^2}{2 m L^2}$. Using $\frac{1}{\sqrt{L}} e^{\pm i \frac{\pi \alpha}{L} x}$ as a basis in this two-dimensional degenerate subspace, we find in the first-order degenerate perturbation theory $\epsilon_{\alpha}^{(p)} =\epsilon_{\alpha}^{(0)} + \zeta |\tilde{V}_{\alpha}|$ and $\epsilon_{\alpha-1}^{(p)} =\epsilon_{\alpha}^{(0)} - \zeta |\tilde{V}_{\alpha}|$ with the corresponding eigenstates
 \begin{align}
     \psi_{\alpha;\zeta \ll 1}^{(p)} (x) &\approx \sqrt{\frac{2}{L}} \cos \left(\frac{\pi \alpha}{L} x + \frac{\varphi_{\alpha}}{2} \right), 
     \label{psi_p_even} \\
     \psi_{\alpha-1;\zeta \ll 1}^{(p)} (x) &\approx  \sqrt{\frac{2}{L}} \sin \left(\frac{\pi \alpha}{L} x + \frac{\varphi_{\alpha}}{2} \right),
     \label{psi_m_even}
 \end{align}
 where $\tilde{V}_{\alpha} = |\tilde{V}_{\alpha}| e^{i \varphi_{\alpha}} = \frac{1}{L} \int_0^L d x \, V (x)\,  e^{-i \frac{2 \pi \alpha}{L} x}$ is the $\alpha$th Fourier component of the periodic potential. Both function \eqref{psi_p_even} and \eqref{psi_m_even} have $\alpha$ zeros on the interval $x \in [0,L)$.
An exception in this consideration is $\alpha=0$: the eigenfunction $\psi_{\alpha=0;\zeta \ll 1}^{(p)} (x) \approx \frac{1}{\sqrt{L}}$ remains nondegenerate. It has no zeros.

Analogously we consider $\psi_{\alpha+1;\zeta}^{(a)} (x)$ and $\psi_{\alpha; \zeta}^{(a)} (x)$ with odd $\alpha$, and find them to be of the form  \eqref{psi_p_even} and  \eqref{psi_m_even}, respectively. This implies that they also have $\alpha$ zeros on the interval $x \in [0,L)$.
 
 Thus, by virtue of \eqref{rel1} and \eqref{rel2}, the functions $\psi_{k=0, \alpha+1} (x)$ and $\psi_{k=0, \alpha} (x)$ with even $\alpha$ have $\alpha$ zeros, as well as the functions $\psi_{k=\frac{\pi}{L}, \alpha+1} (x)$ and $\psi_{k=\frac{\pi}{L}, \alpha} (x)$ with odd $\alpha$, have $\alpha$ zeros. These properties ensure the relation \eqref{Mmp_alpha}.

\section{Universal scaling of the boundary charge fluctuations}
\label{app:fluct_ident}

The geometric tensor appearing in \eqref{fluct_def} is defined by
\begin{align} 
(\mathcal{Q}_{k})_{ \alpha \beta} &=\langle \frac{d u_{k \alpha}}{dk} | \frac{d u_{k \beta}}{d k} \rangle \, \delta_{\alpha \beta} - |\langle u_{k \alpha}| \frac{d u_{k \beta}}{dk} \rangle |^2.
\label{geom_tensor}
\end{align}
Here we employ the shorthand notation $\langle F | G \rangle = \int_0^L d x \, F^* (x) G (x)$. Using the completeness relation $\sum_{\beta} |u_{k \beta} \rangle \langle u_{k \beta}| = \hat{1}$, we alternatively express
\begin{align}
    \sum_{\alpha,\beta=1}^{\nu} (\mathcal{Q}_k )_{\alpha \beta} = \sum_{\alpha=1}^{\nu} \sum_{\beta=\nu+1}^{\infty} |\langle u_{k \alpha} | \frac{d u _{k \beta}}{d k}\rangle |^2 .
\end{align}
In the obtained double sum the index $\alpha$ runs through the valence bands, while the index $\beta$ runs through the conduction bands. 

For $\alpha \neq \beta$ we establish 
\begin{align}
    &-i \, \langle \psi_{k \alpha}| \frac{d \psi_{k \beta}}{dk} \rangle = -i \, \langle u_{k \alpha}| \frac{d u_{k \beta}}{dk} \rangle + \langle u_{k \alpha}| \, x \, | u_{k \beta}  \rangle \label{qb_id1}\\
    =& \frac{L}{2m (\epsilon_{k \alpha} - \epsilon_{k \beta})}  \left[ \psi_{k \alpha}^* (0) \, \psi'_{k \beta} (0) - \psi_{k \alpha}^{* \, \prime} (0) \, \psi_{k \beta} (0)\right] \label{qb_id2} \\
     =&\frac{1}{2m (\epsilon_{k \alpha} - \epsilon_{k \beta})} \int_0^L d x \,\, [\psi_{k \alpha}^* (x) \, \psi'_{k \beta} (x) -\psi_{k \alpha}^{* \, \prime} (x) \, \psi_{k \beta} (x)] \nonumber \\
     &+ \langle u_{k \alpha}| \, x \, | u_{k \beta}  \rangle . \label{qb_id3}
\end{align}

To obtain \eqref{qb_id2} from \eqref{qb_id1}, we consider the equation
\begin{align}
    - \frac{1}{2m}  \frac{d \psi''_{k \beta} (x)}{d k} + [V (x) - \epsilon_{k \beta}] \, \frac{d\psi_{k \beta} (x) }{dk} =\frac{d \epsilon_{k \beta}}{dk} \, \psi_{k \beta} (x) ,
\end{align}
and project it onto the state $\langle \psi_{k \alpha}|$ with $\alpha \neq \beta$, such that $\langle \psi_{k \alpha} | \psi_{k \beta} \rangle =0$. It follows
\begin{align}
 & \frac{1}{2m} \int_0^L dx  \left[\psi^*_{k \alpha} (x) \frac{d \psi''_{k \beta} (x)}{d k} - \psi^{* \, \prime \prime}_{k \alpha} (x)  \frac{d\psi_{k \beta} (x) }{dk} \right] \label{part_int_ident1} \\
 =&   (\epsilon_{k \alpha} - \epsilon_{k \beta}) \, \langle \psi_{k \alpha}|  \frac{d\psi_{k \beta} }{dk} \rangle.
\end{align}
Integrating \eqref{part_int_ident1} by parts, we obtain \eqref{qb_id2}.

To derive \eqref{qb_id3} from \eqref{qb_id2}, we multiply the Schr{\"o}dinger equation for $| \psi_{k\beta} \rangle$ with $x$ and then project the result onto the state $\langle \psi_{k \alpha}|$. This gives
\begin{align}
    & \frac{1}{2 m} \int_0^L d x \,\left[ x \, \psi_{k \alpha}^* (x) \,  \psi''_{k \beta} (x) - \, x \, \psi_{k \alpha}^{* \, \prime \prime} (x)  \psi_{k \beta} (x) \right]\label{part_int_ident2} \\
    &=  (\epsilon_{k \alpha} - \epsilon_{k \beta}) \, \langle \psi_{k \alpha} | \, x \, |  \psi_{k \beta} \rangle.
\end{align}
Integrating \eqref{part_int_ident2} by parts and using $\langle \psi_{k \alpha} |x|  \psi_{k \beta} \rangle = \langle u_{k \alpha} |x|  u_{k \beta} \rangle$ as well as $\psi_{k\alpha} (L) = e^{i k L} \psi_{k \alpha} (0)$, $\psi'_{k\alpha} (L) = e^{i k L} \psi'_{k \alpha} (0)$, we obtain \eqref{qb_id3}.

To further simplify \eqref{qb_id3} we notice that
\begin{align}
     & \int_0^L d x \,\, [\psi_{k \alpha}^* (x) \, \psi'_{k \beta} (x) -\psi_{k \alpha}^{* \, \prime} (x) \, \psi_{k \beta} (x)] \nonumber \\
    =& \frac{2}{\epsilon_{k \alpha} - \epsilon_{k \beta}} \int_0^L d x \,\, V (x) \, \frac{d}{dx} \, [\psi_{k \alpha}^* (x) \, \psi_{k \beta} (x)].
    \label{qb_id4}
\end{align}
Thus, we get
\begin{align}
    &-i \, \langle u_{k \alpha}| \frac{d u_{k \beta}}{dk} \rangle \nonumber \\
    =&  \frac{1}{m(\epsilon_{k \alpha} - \epsilon_{k \beta})^2} \int_0^L d x \,\, V (x) \, \frac{d}{dx} \, [\psi_{k \alpha}^* (x) \, \psi_{k \beta} (x)].
    \label{der_V_int}
\end{align}

The validity of \eqref{qb_id4} follows from the two equations
\begin{align}
    &-\frac{1}{2m} \int_0^L d x \,\, \psi_{k \alpha}^{* \, \prime} (x) \, \psi''_{k \beta} (x) \nonumber \\
    &+ \int_0^L d x \,\, \psi_{k \alpha}^{* \, \prime} (x) \, [V (x) - \epsilon_{k \beta}] \, \psi_{k \beta} (x) =0, \\
    &-\frac{1}{2m} \int_0^L d x \,\, \psi_{k \alpha}^{* \, \prime \prime} (x) \, \psi'_{k \beta} (x) \nonumber \\
    &+ \int_0^L d x \,\, \psi_{k \alpha}^{* \, } (x) \, [V (x) - \epsilon_{k \alpha}] \, \psi'_{k \beta} (x) =0.
\end{align}
Adding them up, we establish
\begin{align}
    &\int_0^L d x \,\, V(x) \, \frac{d}{dx} [\psi_{k \alpha}^{*} (x) \, \psi_{k \beta} (x)]  \nonumber \\
    =& \epsilon_{k \alpha} \int_0^L d x \,\, \psi_{k \alpha}^{* \, } (x)  \, \psi'_{k \beta} (x)+ \epsilon_{k \beta}\int_0^L d x \,\, \psi_{k \alpha}^{* \, \prime} (x) \, \psi_{k \beta} (x) \nonumber \\
    =& \frac{\epsilon_{k \alpha}-\epsilon_{k \beta}}{2} \int_0^L d x \,\, [\psi_{k \alpha}^{* \, } (x)  \, \psi'_{k \beta} (x) - \psi_{k \alpha}^{* \, \prime} (x) \, \psi_{k \beta} (x)] \nonumber \\
    &+ \frac{\epsilon_{k \alpha}+\epsilon_{k \beta}}{2} \int_0^L d x \,\, \frac{d}{d x} [\psi_{k \alpha}^{* \, } (x)  \, \psi_{k \beta} (x)].
\end{align}
Due to the periodicity of $[\psi_{k \alpha}^{* \, } (x) \, \psi_{k \beta} (x)]$ the last term vanishes, and we obtain \eqref{qb_id4}.

In the narrow-gap limit we find the low-energy approximation for \eqref{fluct_def} (cf. [\onlinecite{wannier_paper}] and Appendix \ref{app:zeros}): The leading contribution is received from the valence band  $\epsilon_{k \nu} \approx \epsilon_{\nu}^{(0)} - \sqrt{ (v_{F,\nu} k)^2 + |\tilde{V}_{\nu}|^2}$ and the conduction band $\epsilon_{k ,\nu+1} \approx \epsilon_{\nu}^{(0)} + \sqrt{ (v_{F,\nu} k)^2 + |\tilde{V}_{\nu}|^2}$, which are adjacent to the chemical potential $\mu_{\nu} = \epsilon_{\nu}^{(0)} = \frac{\nu^2 \pi^2}{2 m L^2}$ (assuming even $\nu$ and performing the expansion near $k=0$; for odd $\nu$ one first has to shift $k \to k - \pi/L$). Here  $v_{F,\nu}= \frac{\sqrt{2 m \mu_{\nu}}}{m} = \frac{\nu \pi}{m L}$ is the Fermi velocity, and  the $\nu$th Fourier component $\tilde{V}_{\nu}= \frac{1}{L} \int_0^L d x \, V(x) \, e^{-i \frac{2 \pi \nu}{L} x}$ of the potential $V (x)$ determines the $\nu$th gap's size $E_{g,\nu} =2 |\tilde{V}_{\nu}|$. To approximate \eqref{der_V_int} we use \eqref{rel1}, \eqref{rel2} as well as \eqref{psi_p_even}, \eqref{psi_m_even}, and obtain
\begin{align}
     & \int_0^L d x \,\, V (x) \, \frac{d}{dx} \, [\psi_{k \alpha}^* (x) \, \psi_{k \beta} (x)] \nonumber \\
     & \approx \int_0^L d x \,\, V (x) \, \frac{d}{dx} \, \frac{1}{L} \sin \left( \frac{2 \pi \nu}{L} x + \varphi_{\nu} \right) \\
     & = \frac{2 \pi \nu}{L} |\tilde{V}_{\nu}| = 2 m v_{F,\nu}  |\tilde{V}_{\nu}|.
\end{align}

Combining all approximations made together,  we establish
\begin{align}
& X_2 \, (\Delta Q_B^{(\nu)})^2  \approx \frac{1}{2 \pi} \int_{-\infty}^{\infty} dk  \,\, |\langle u_{k \nu} | \frac{d u _{k , \nu+1}}{d k}\rangle |^2 \nonumber \\
& \approx \frac{1}{8 \pi} \int_{-\infty}^{\infty} dk \,\, \frac{v_{F,\nu}^2 |\tilde{V}_{\nu}|^2}{[(v_{F,\nu} k)^2 + |\tilde{V}_{\nu}|^2]^2} = \frac{v_{F,\nu}}{16 |\tilde{V}_{\nu}|} ,
\end{align}
which is equivalent to \eqref{sft}.

\section{Unitarity of the scattering matrix \eqref{scat_matr}}
\label{app:unit_check}

To check the unitarity of \eqref{scat_matr}, we first observe that
by virtue of the identity
\begin{align}
\frac{\psi_2^{(\varphi)} (L)}{N_{k \alpha}^{(\varphi)}} = \frac{\psi_2^{(\varphi')} (L)}{N_{k \alpha}^{(\varphi')}} = - \frac{m}{\frac{\partial D (E)}{\partial E}},
\label{ratNpsi}
\end{align}
following from \eqref{Nk_alt}, it holds
\begin{align}
t_{k \alpha} = t'_{k \alpha} = - \frac{m}{\frac{\partial D (E)}{\partial E}} \, \frac{ 2 i \,\sin k L }{ d_{k \alpha}} \, \sqrt{N_{k \alpha}^{(\varphi)}N_{k \alpha}^{(\varphi')}}.
\label{tteq}
\end{align}

Next, we establish that
\begin{align}
|r_{k \alpha}|^2 =& 1-  \frac{ 4 \sin k L \,\, \psi_{2}^{(\varphi)} (L)}{ |d_{k \alpha}|^2 } \, \text{Im} \,  d_{k\alpha} \nonumber \\
&+  \frac{ 4 \sin^2 k L \,\, [\psi_{2}^{(\varphi)} (L)]^2}{ |d_{k \alpha}|^2}, \\
|r'_{k \alpha}|^2 =&  1-  \frac{ 4 \sin k L \,\, \psi_{2}^{(\varphi')} (L)}{ |d_{k \alpha}|^2 } \, \text{Im} \,  d_{k \alpha} \nonumber \\
&+  \frac{ 4 \sin^2 k L \,\, [\psi_{2}^{(\varphi')} (L)]^2}{ |d_{k \alpha}|^2}.
\end{align}
Since
\begin{align}
 \text{Im} \,  d_{k \alpha} = [\psi_{2}^{(\varphi)} (L)+\psi_{2}^{(\varphi')} (L)] \, \sin k L,
 \label{imDgen}
\end{align}
we get
\begin{align}
 |r_{k \alpha}|^2= |r'_{k \alpha}|^2 = 1-  \frac{ 4 \sin^2 k L \,\, \psi_{2}^{(\varphi)} (L)\, \psi_{2}^{(\varphi')} (L)}{|d_{k \alpha}|^2} .
\end{align}

To demonstrate the validity of \eqref{scat_d1} and \eqref{scat_d2} it suffices to notice that
\begin{align}
\left[  \frac{m}{\frac{\partial D (E)}{\partial E}}\right]^2 N_{k \alpha}^{(\varphi)} N_{k \alpha}^{(\varphi')} = \psi_{2}^{(\varphi)} (L)\, \psi_{2}^{(\varphi')} (L),
\end{align}
which holds true by virtue of \eqref{ratNpsi}.

The relation \eqref{scat_off}  requires
\begin{align}
  & \frac{ 1}{ d_{k \alpha}^*} \left[1-  \frac{  2i \sin k L \,\, \psi_2^{(\varphi')} (L) }{d_{k \alpha}} \right] \nonumber \\
  -& \frac{ 1 }{ d_{k \alpha}}  \left[1+ \frac{ 2 i \sin k L \,\, \psi_{2}^{(\varphi)} (L)}{ d_{k \alpha}^*} \right] =0,
\end{align}
which indeed follows from \eqref{imDgen}.

Thus, the unitarity of \eqref{scat_matr} is verified.

\begin{figure*}[t]
    \centering
    \includegraphics[width=\textwidth]{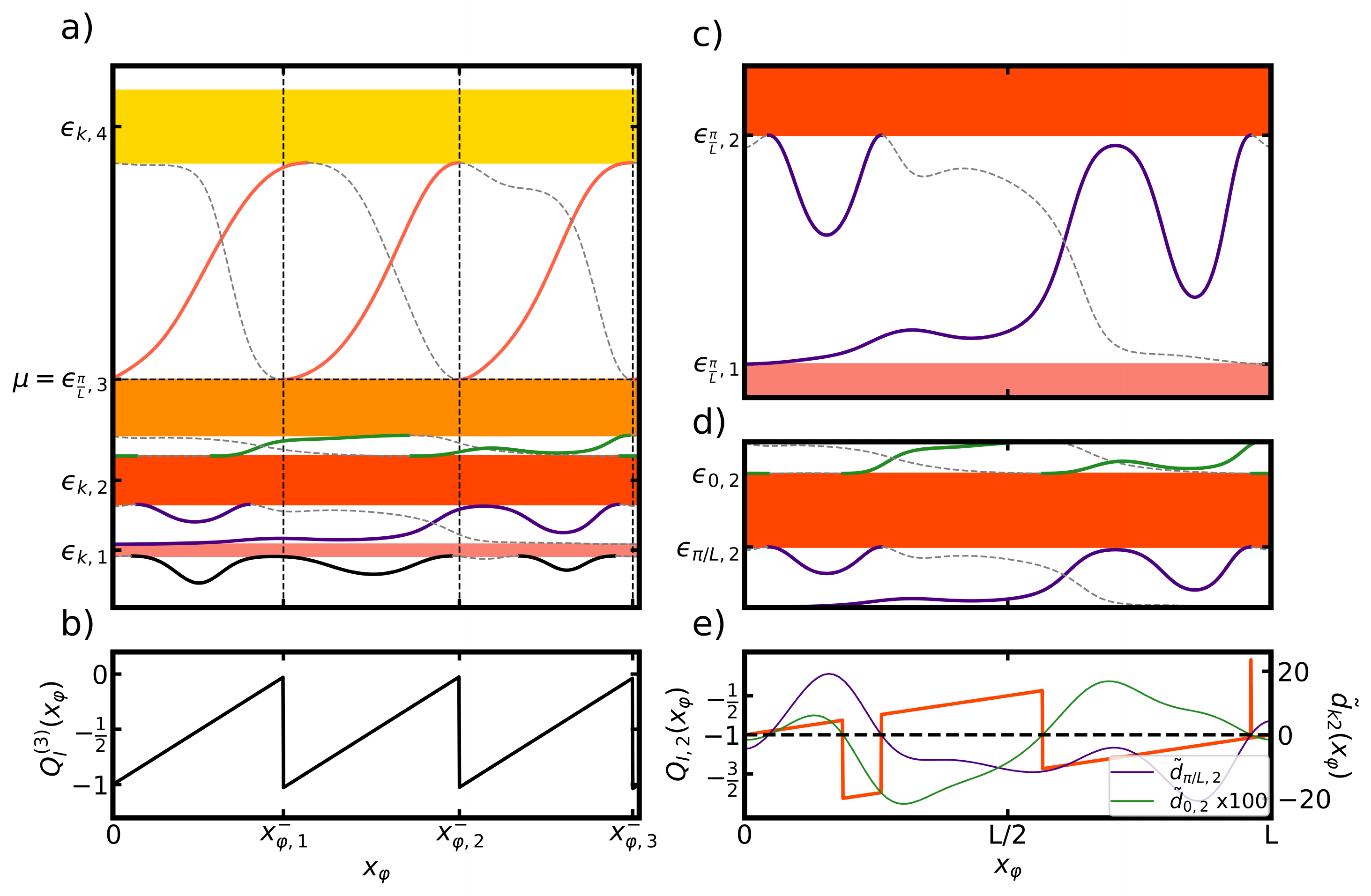}
    \caption{Same as in Fig.~\ref{fig:L1PlotQuadruple} (besides a separate first gap close-up) for $\lambda=0.1$.}
    \label{fig:my_labelP}
\end{figure*}
\begin{figure*}
    \centering
    \includegraphics[width=\textwidth]{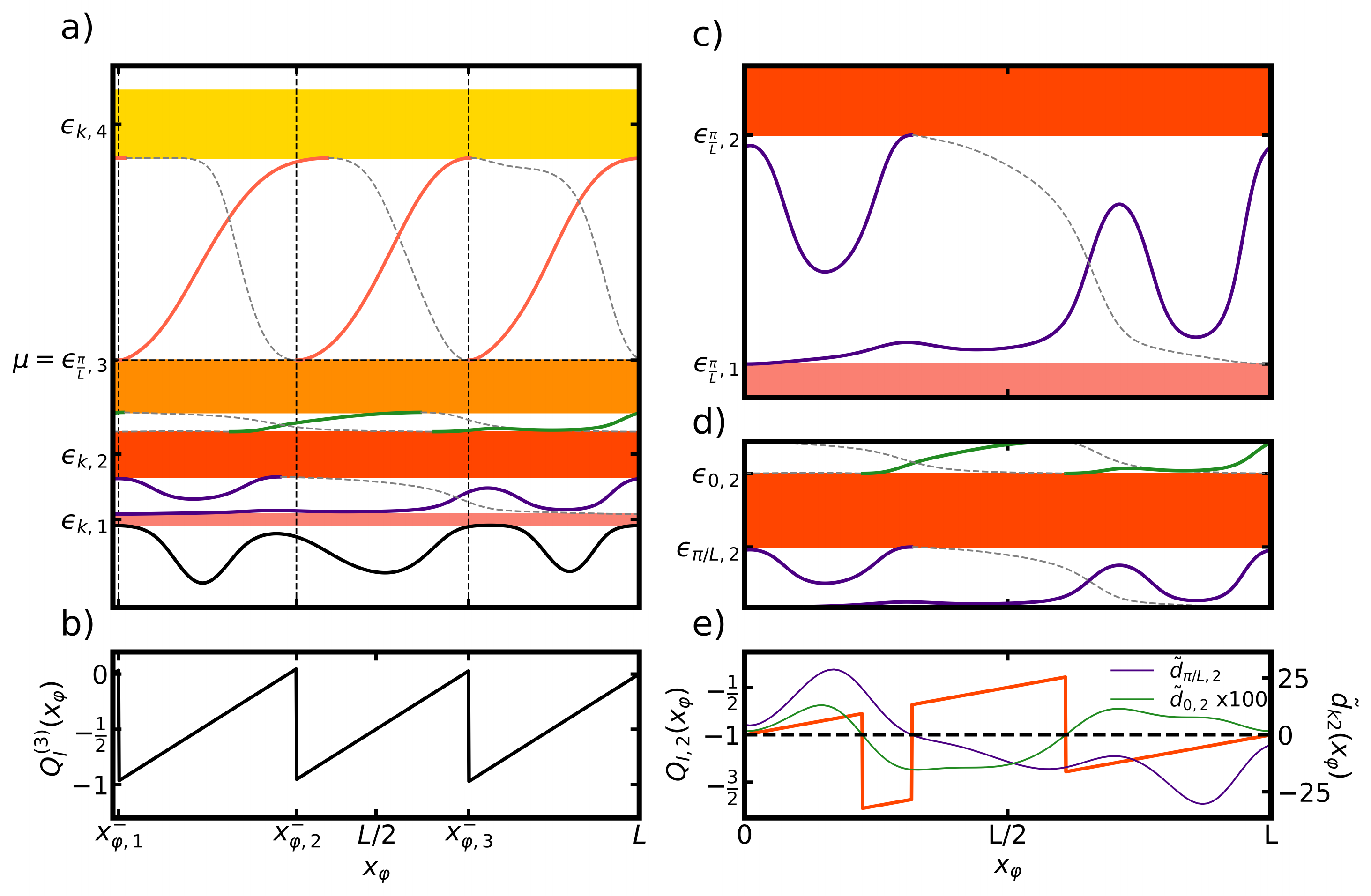}
    \caption{Same as in Fig.~\ref{fig:L1PlotQuadruple} (besides a separate first gap close-up) for $\lambda=-0.1$.}
    \label{fig:my_labelM}
\end{figure*}

\section{Additional illustrations}
\label{app:add_fig}

This section contains additional Figures \ref{fig:my_labelP} and \ref{fig:my_labelM} to highlight the previously discussed properties of energy spectrum of the interface eigenvalue problem. The chosen parameters $\lambda = \pm 0.1$ serve to illustrate the limit of the absent impurity potential $\lambda \to 0^{\pm}$, which is most  nontrivially achieved close to the translationally invariant point $x_{\varphi}=x_{\varphi'}=0$.

\section{Derivation of \eqref{QI_bar_win}}
\label{app:inter_charge}

Let us consider
\begin{align}
& \frac{d}{d k} d_{k \alpha} =   \frac{d\psi_{ 2}^{(\varphi)} (L)}{dk} \, [\psi_{1}^{(\varphi')} (L)- e^{-i k L}]  \nonumber \\
&-   [\psi_{ 1}^{(\varphi)} (L) - e^{i k L} + 2 m \lambda  \, \psi_2^{(\varphi)} (L)] \,  \frac{d \psi_{ 2}^{(\varphi')} (L)}{dk} \nonumber \\
&+  \psi_{ 2}^{(\varphi)} (L) \, \frac{d \psi_{1}^{(\varphi')} (L)}{dk}  \nonumber \\
& -  \left[\frac{d  \psi_{ 1}^{(\varphi)} (L) }{dk}  +2 m  \lambda  \, \frac{d \psi_2^{(\varphi)} (L)}{dk} \right]  \psi_{ 2}^{(\varphi')} (L)  \nonumber \\
&+  i  L \left\{ \psi_{ 2}^{(\varphi)} (L) \, e^{-ik L} +   \psi_{ 2}^{(\varphi')} (L) \, e^{i k L} \right\} , \label{dddk}
\end{align}
and compare it with
\begin{align}
P_{k \alpha} =& L \int_0^L d x \,\, \psi_{2}^{(\varphi')} (L) \, \psi_{k \alpha}^{(\varphi)\, 2} (x) \,  e^{- i k L} \nonumber \\
+& L \int_0^L d x \,\,  \psi_{2}^{(\varphi)} (L) \, \psi_{-k,\alpha}^{(\varphi')\, 2} (x) \, e^{i k L},
\end{align}
which appears in \eqref{QI_bar_res}, that is
\begin{align}
    \bar{Q}_{I,\alpha} = - \text{Re} \, \int_{-\pi/L}^{\pi/L} \frac{dk}{2 \pi} \, \frac{P_{k \alpha}}{d_{k\alpha}}.
\end{align}
With help of \eqref{pr_cross}, \eqref{pr1} and \eqref{dEdk}
we evaluate
\begin{align}
    & \int_0^L d x \,\, \psi_{k \alpha}^{(\varphi)\, 2} (x)\nonumber \\
    &= 1+\int_0^L d x \,\, \psi_{k \alpha}^{(\varphi)} (x) \left[ \psi_{k \alpha}^{(\varphi)} (x) -\psi_{-k, \alpha}^{(\varphi)} (x)\right] \nonumber \\
    &= 1- \frac{ i }{L} \,  \frac{d \ln \psi_2^{(\varphi)} (L)}{d k}   \nonumber \\
    & + \frac{ i }{L} \, \left[ \psi^{(\varphi) \, \prime}_2 (L) \, \frac{d \ln \psi_2^{(\varphi)} (L)}{d k}   -    \frac{d \psi^{(\varphi) \, \prime}_2 (L)}{d k}\right] \, e^{i k L}.
\end{align}
Thus
\begin{align}
    P_{k \alpha} &= L \, \left\{ \psi_{ 2}^{(\varphi)} (L) \, e^{ik L} +  \psi_{ 2}^{(\varphi')} (L) \,  e^{-i k L} \right\} \nonumber \\
    & - i \, e^{- i k L} \,  \psi_{2}^{(\varphi')} (L)  \, \frac{d \ln \psi_2^{(\varphi)} (L)}{d k}  \nonumber \\
& + i \, \psi_{2}^{(\varphi')} (L) \, \left[ \psi^{(\varphi) \, \prime}_2 (L) \, \frac{d \ln \psi_2^{(\varphi)} (L)}{d k}   -    \frac{d \psi^{(\varphi) \, \prime}_2 (L)}{d k}\right] \nonumber  \\
 & + i\, \psi_{2}^{(\varphi)} (L) \, \frac{d \ln \psi_2^{(\varphi')} (L)}{d k}  \,  e^{ i k L} \nonumber \\
& -i\, \psi_{2}^{(\varphi)} (L) \, \left[ \psi^{(\varphi') \, \prime}_2 (L) \, \frac{d \ln \psi_2^{(\varphi')} (L)}{d k}   -    \frac{d \psi^{(\varphi') \, \prime}_2 (L)}{d k}\right] \nonumber
\end{align}
\begin{align}
 &= L \, \left\{ \psi_{ 2}^{(\varphi)} (L) \, e^{-ik L} +   \psi_{ 2}^{(\varphi')} (L)\, e^{i k L} \right\} \nonumber \\
    & - i\, e^{- i k L} \, \psi_{2}^{(\varphi')} (L) \, \frac{d \ln \psi_2^{(\varphi)} (L)}{d k}  \nonumber \\
& + i \, \psi_{2}^{(\varphi')} (L)\, \left[ \psi^{(\varphi) \, \prime}_2 (L) \, \frac{d \ln \psi_2^{(\varphi)} (L)}{d k}   +   \frac{d \psi^{(\varphi)}_1 (L)}{d k} \right] \nonumber  \\
 & + i \, \psi_{2}^{(\varphi)} (L) \, \frac{d \ln \psi_2^{(\varphi')} (L)}{d k}  \,  e^{ i k L} \nonumber \\
& -i \, \psi_{2}^{(\varphi)} (L) \, \left[ \psi^{(\varphi') \, \prime}_2 (L) \, \frac{d \ln \psi_2^{(\varphi')} (L)}{d k} +\frac{d \psi^{(\varphi') }_1 (L)}{d k} \right].
\end{align}

Replacing in \eqref{dddk} $ [\psi_{1}^{(\varphi')} (L)- e^{-i k L} ]$ by
\begin{align}
   \frac{d_{k \alpha} +[\psi_{ 1}^{(\varphi)} (L) - e^{i k L} + 2 m \lambda  \, \psi_2^{(\varphi)} (L)] \, \psi_{ 2}^{(\varphi')} (L)}{\psi_{ 2}^{(\varphi)} (L)},
\end{align}
and $[\psi_{ 1}^{(\varphi)} (L) - e^{i k L} + 2 m \lambda  \, \psi_2^{(\varphi)} (L)]$ by
\begin{align}
    \frac{[\psi_{1}^{(\varphi')} (L)- e^{-i k L}] \, \psi_{ 2}^{(\varphi)} (L) - d_{k \alpha}}{\psi_{ 2}^{(\varphi')} (L)},
\end{align}
we evaluate
\begin{align}
 \frac{d}{d k} d_{k \alpha} - i \, P_{k \alpha} =  d_{k \alpha} \, \frac{d }{dk}  \ln [ \psi_{ 2}^{(\varphi)} (L) \, \psi_{ 2}^{(\varphi')} (L)],
\end{align}
and
\begin{align}
  & \text{Im} \,  \int_{-\pi/L}^{\pi/L} \frac{d k }{2 \pi} \,\, \frac{1}{ d_{k \alpha}} \frac{d}{d k} d_{k \alpha} -  \text{Re} \, \int_{-\pi/L}^{\pi/L} \frac{d k }{2 \pi} \,\,  \frac{P_{k \alpha}}{ d_{k \alpha}} =0.
\end{align}
Hence,
\begin{align}
\bar{Q}_{I,\alpha} =  - \text{Im}\, \int_{-\pi/L}^{\pi/L} \frac{d k }{2 \pi} \,\,  \frac{1}{d_{k\alpha}} \, \frac{d}{d k} d_{k \alpha},
\end{align}
which is equivalent to \eqref{QI_bar_win}.

\end{appendix}

\end{document}